\documentclass[12pt,onecolumn]{article}
\usepackage{color}
\usepackage{amsmath}
\usepackage{amssymb}
\usepackage{amsfonts}
\usepackage{geometry}
\usepackage{setspace}
\usepackage{amsmath,bm}
\usepackage{graphicx}
\usepackage[section]{placeins}% avoid figures cross the section
\usepackage{subfigure}
\usepackage[sort&compress]{natbib}
\usepackage{booktabs}  
\usepackage{soul}
\usepackage{makecell}
\usepackage{caption}% caption
\usepackage[colorlinks, citecolor=blue]{hyperref}% link to the superlink
\usepackage{lineno} %line number
\usepackage{float} % fix the figure position

\author{Shuwei Zhou$^{1,2}$, Xiaoying Zhuang$^{1,2}$,  Jiaming Zhou$^{3}$, Fang Liu$^{1*}$}
\title {Phase field characterization of rock fractures in Brazilian splitting test specimens containing voids and inclusions}
\begin{document}

%\linenumbers
%\captionsetup[figure]{labelfont={bf},name={Fig.},labelsep=space,justification=raggedright, singlelinecheck = false}
\captionsetup[figure]{labelfont={bf},name={Fig.},labelsep=space}
% referecne style
%\bibliographystyle{unsrtnat}
%\bibliographystyle{plainnat}
\bibliographystyle{apa}
\setcitestyle{authoryear,round,aysep={},yysep={,}}
% referecne style
%\bibliographystyle{unsrtnat}
%\setcitestyle{numbers,square,aysep={},yysep={,}}
% delte the date
% Increase the section level (declaration}
%\setcounter{tocdepth}{4} 
%\setcounter{secnumdepth}{4}

\date{}
\maketitle

\spacing {1.2}
\noindent
1. Department of Geotechnical Engineering, College of Civil Engineering, Tongji University, Shanghai 200092, P.R. China\\
2. Institute of Continuum Mechanics, Leibniz University Hannover, Hannover 30167, Germany\\
3. School of Mechanical Engineering, Beijing Institute of Technology, Beijing 100081, P.R. China\\
* Corresponding author: Fang Liu, Email: liufang@tongji.edu.cn
%1 Division of Computational Mechanics, Ton Duc Thang University, Ho Chi Minh City, Viet Nam\\
%2 Faculty of Civil Engineering, Ton Duc Thang University, Ho Chi Minh City, Viet Nam\\ 
%3 Department of Geotechnical Engineering, College of Civil Engineering, Tongji University, Shanghai 200092, P.R. China\\
%4 Institute of Continuum Mechanics, Leibniz University Hannover, Hannover 30167, Germany\\
%5 School of Mechanical Engineering, Beijing Institute of Technology, Beijing 100081, P.R. China\\
%* Corresponding author: Xiaoying Zhuang, Division of Computational Mechanics, Ton Duc Thang University, Ho Chi Minh City, Viet Nam; Email: Xiaoying Zhuang\\

%\begin{spacing}{2.0}
\begin{abstract}
\noindent The Brazilian splitting test is a widely used testing procedure for characterizing the tensile strength of natural rock or rock-like material due to the fact. However, the results of Brazilian tests on specimens with naturally existing voids and inclusions are strongly influenced by size effects and boundary conditions, while numerical modeling can assist in explaining and understanding the mechanisms. On the other hand, the potential of utilizing Brazilian test to characterize inhomogeneous deformation of rock samples with voids and inclusions of dissimilar materials still awaits to be explored. In the present study, fracture mechanisms in Brazilian discs with circular voids and filled inclusions are investigated by using the phase field model (PFM). The PFM is implemented within the framework of finite element method to study the influence of diameter, eccentricity, and quantity of the voids and inclusions on the fracture patterns and stress-strain curves. The phase field simulations can reproduce previous experimental phenomena and furthermore it deepens the understanding of the influence of inclusion and voids on the fracture pattern, overall strength and deformation behavior of inhomogeneous rock. The findings in the study highlight the potential of characterizing inhomogeneous rock through combining Brazilian tests and numerical modeling. 
\end{abstract}
%\end{spacing}


%\twocolumn
\section {Introduction}\label{Introduction}

Fracture initiation and propagation is of great importance in rock engineering and rock mechanics \citep{lee2017mechanism, gou2019propagation}. Natural rocks contain a large number of inclusions \citep{ni2016displacement}, voids \citep{xiao2018finite}, and cracks \citep{han2018quantifying}, which in turn affect the overall stability and strength of the rocks \citep{xu2018review, yang2019estimation}. For most cases, cracks initiate and propagate from the natural flaws of the rocks, and then decrease the stiffness and bearing capacity of the rocks. Fracture propagation therefore becomes a governing factor in many situations in petroleum geomechanics, rock engineering and mining such as hydraulic fracturing and rock cutting \citep{marji2015simulation, abdollahipour2016time, abdollahipour2016numerical, cao2018mechanical, zhou2018phase2}. It is necessary to investigate the crack initiation, propagation and coalescence during loading process, which is helpful for predicting and assessing the mechanical performance of rock. 

In the past decades, many efforts have been devoted to develop effective and efficient methods either analytically or numerically for predicting fractures in rock \citep{bobet1998fracture, wong2001analysis, sagong2002coalescence, jongpradist2015high, xia2015strength, zhang2015crack, oterkus2017fully, xu2017digital, yang2017experimental, zhou2018phase3, le2018effect, pakzad2018numerical}. Many novel tests were developed to investigate fracture patterns under compression. Popular tests include notched semi-circular bending test (NSCB) \citep{chong1984new}, cracked chevron notched semi-circular bending test (CCNSCB) \citep{dai2011determination}, and Brazilian splitting tests \citep{vardar1975analysis, zhou2016numerical, zhou2019propagation}. Recently, \citet{wei2016experimental, wei2017experimental} prove that the CCNSCB method is more suitable to measure the fracture toughness of rocks than some of the ISRM-suggested methods.

The Brazilian disc test \citep{vardar1975analysis} has been extensively adopted to assess fracture performance of rock and rock-like materials \citep{villeneuve2011effects, chen2018verification, masoumi2018scale}. For example, the static and dynamic fracture toughness of a rock is commonly determined by compressing a Brazilian disc specimen with a central pre-existing notch \citep{wei2016stress, wei2017fracture}. In addition, a cracked straight through Brazilian disk (CSTBD) specimen is used for achieving the mixed Mode I/II fracture toughness. More new knowledge on fractures in Brazilian discs with pre-existing notches has been achieved during recent years. These efforts are made mainly based on some advanced numerical techniques because corresponding experimental tests are difficult to perform. For instance, \citet{marji2013use, marji2014numerical, haeri2015experimental} used the displacement discontinuity method (DDM) and \citet{sarfarazi2017fracture} applied the particle flow code (PFC). However, the knowledge on the fracture mechanisms in the Brazilian discs with multiple circular voids and inclusions is still insufficient. More specifically, in this type of test how the dimension, quantity, and spatial distribution of voids and filled inclusions affect the crack pattern and indirect tensile strength is unknown.

Therefore, to better understand the strength characteristic of rock-like materials and the fracture initiation and propagation phenomenon in them \citep{wang2017temperature, zhang2017longwall}, this study investigates the fracture behavior of rock-like discs containing multiple circular voids and inclusions under diametral compression. In this paper, an effective fracture modeling approach--the phase field method (PFM) is used due to the difficulties in actual experimental tests such as physical property discreteness of rock-like samples. Although to date there have been many PFMs such as the elasto-plastic phase field crack method \citep{duda2015phase} and finite-deformation phase field method \citep{hesch2014thermodynamically}, the anisotropic PFM for brittle crack \citep{miehe2010phase, miehe2010thermodynamically, zhou2018phase} is applied in our work. The used PFM has clear physical meaning and obeys the thermodynamic consistency. The PFM can predict complex fracture patterns such as coalescence and branching, and the propagation path is automatically achieved without any additional tracking algorithms \citep{zhou2019phase1, zhou2019phase2, zhuang2020hydraulic}. The PFM is implemented using a finite element method and the influence of diameter, eccentricity, and quantity of the voids and inclusions on the fracture patterns and stress-strain curves of the Brazilian discs are investigated. The PFM with some previous experimental tests are also compared to show the effectiveness and prediction capability of the PFM.

This paper is outlined as follows. Section \ref{Phase field method} describes the fundamental theory of the phase field method to fracture and Section \ref{Numerical implementation and parameter adjustment} presents the detailed finite element implementation of the PFM. Section \ref{Fracture patterns in Brazilian discs with voids} discusses the fracture mechanism of Brazilian discs with multiple voids while the results on Brazilian discs with multiple inclusions are presented in Section \ref{Fracture patterns in Brazilian discs with inclusions}. Finally, some concluding remarks are given in Section \ref{Conclusions}.

%section 2
\section {Phase field model for rock fracture}\label{Phase field method}

In this section, the theory framework of phase field model for fracture initiation and propagation is described and the governing equations on the multi-field fracture modeling are achieved.

\subsection {Variational method to fracture}\label{Variational method to fracture}

\begin{figure}[htbp]
	\centering
	\includegraphics[width = 6cm]{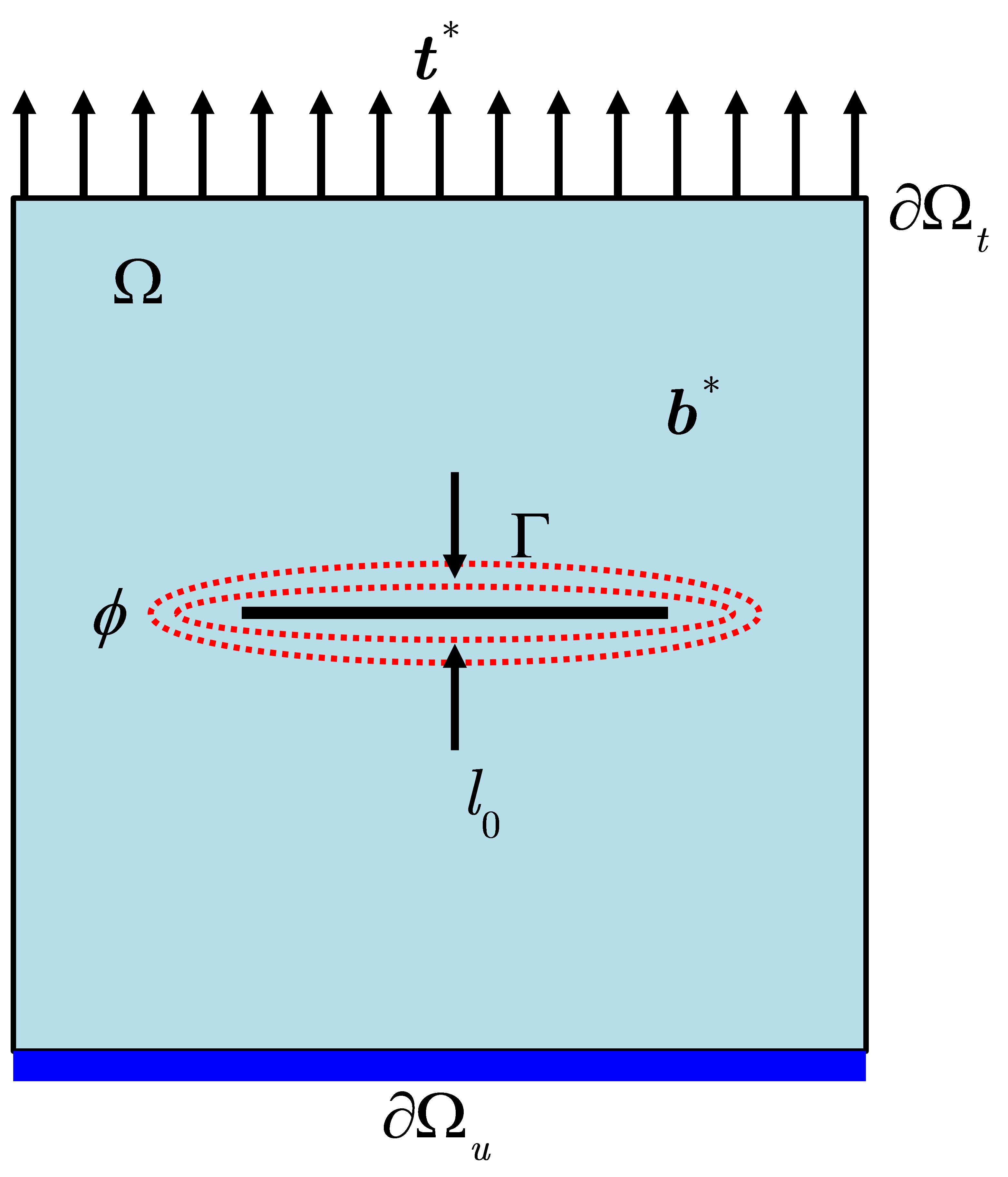}
	\caption{Phase field approximation of the crack surface}
	\label{Phase field approximation of the crack surface}
	\end{figure}

Let $\Omega\subset \mathbb R^d$ ($d\in \{1,2,3\} $) be a cracked solid body in Fig. \ref{Phase field approximation of the crack surface} and $\bm x$ be the position vector. The external boundary is denoted as $\partial \Omega$ and divided into two disjointed parts $\partial \Omega_t$ and $\partial \Omega_u$. That is, $\partial \Omega_t \cap \partial \Omega_u = \emptyset$ and $\overline{\partial \Omega_t \cup \partial \Omega_u} = \partial \Omega$. In addition, the outward unit normal vector of the boundary is denoted as $\bm n$. An internal discontinuity $\Gamma$ is in the body and the time-dependent Dirichlet boundary condition $\bm {u^*}(\bm x)$ and Neumann condition $\bm {t^*}(\bm x)$ are applied on $\partial \Omega_u$ and $\partial \Omega_t$, respectively. 

In a cracked solid, three leading types of energy are considered in this paper, namely, the elastic energy $\Psi_E$, fracture energy $\Psi_f$, and the external work $W_{ext}$. The classic Griffith's fracture theory \citep{anderson2005fracture} is used and it is considered that a crack initiates after the stored elastic energy increases to be large enough and overcome the fracture resistance of the solid. Therefore, crack propagation is considered as a process to minimize the total energy functional $L$, which is a combination of the elastic energy $\Psi_E$, fracture energy $\Psi_f$ and external work $W_{ext}$:
	\begin{equation}
 	L = \Psi_E+\Psi_f-W_{ext}
	\label{L}
	\end{equation} 

The variational approach to fracture is used and then the zero first variation of energy functional $L$ is applied:
	\begin{equation}
 	\delta L = \delta \Psi_E+\delta \Psi_f-\delta W_{ext}=0
	\label{delta L}
	\end{equation} 

\subsection{Phase field approximation for the fracture energy}\label{Phase filed approximation for the fracture energy}

In the context of phase field methods \citep{miehe2010phase, miehe2010thermodynamically, borden2012phase}, the sharp crack $\Gamma$ is smeared by a scalar field over a narrow transition band in the domain $\Omega$; see also Fig. \ref{Phase field approximation of the crack surface}. The scalar field, i.e. phase field $\phi(\bm x,t)\in[0,1]$ (similar to the damage factor \citep{cao2018use}) uses a length scale parameter $l_0$ to control the shape of transition region and also to reflect the width of the crack. The setting of the phase field is assumed to satisfy the following conditions:
	\begin{equation}
	\phi = 
	\begin{cases}
	0,\hspace{1cm}\text{if material is intact}\\1,\hspace{1cm}\text{if material is cracked}
	\end{cases}
	\end{equation}

It should be noted that the crack region will have a larger transition width as $l_0$ increases and when the $\Gamma-$convergence condition holds the phase field will recover to a sharp crack with $l_0$ tending to zero. Additionally, we follow \citet{miehe2010phase} and calculate the fracture energy $\Psi_f$ and its variation as
	\begin{equation}
	\Psi_f=\int_{\Gamma}G_c \mathrm{dS} \approx \int_{\Omega}G_c  \gamma \mathrm{d}\Omega = \int_{\Omega}G_c\left(\frac{\phi^2}{2l_0}+\frac{l_0}2|\nabla\phi|^2 \right)\mathrm{d}{\Omega}
	\label{fracture energy}
	\end{equation}

\noindent and
	\begin{equation}
	\delta \Psi_f=\int_{\Omega}G_c  \delta \gamma \mathrm{d}\Omega = \int_{\Omega}G_c \delta \phi \left( \frac{\phi}{l_0}-l_0\nabla^2\phi \right)\mathrm{d}{\Omega}
	\label{variation of fracture energy}
	\end{equation}

\noindent where $G_c$ is the critical energy release rate.

\subsection{Strain decomposition}\label{Strain decomposition}

The elastic energy $\Psi_E$ is calculated by  $\Psi_E=\int_{\Omega}\psi_{\varepsilon}(\bm \varepsilon) \mathrm{d}{\Omega}$ with $\psi_{\varepsilon}(\bm \varepsilon)$ being the elastic energy density and $\bm \varepsilon$ being the linear strain tensor of the solid. According to the variational approach \citep{bourdin2000numerical}, the elastic energy drives the phase field evolution in a phase field method. Thus, to capture cracks only under tension, the elastic energy must be divided into tensile and compressive parts based on the strain decomposition \citep{miehe2010thermodynamically}:
	\begin{equation}
	\bm\varepsilon_{\pm}=\sum_{a=1}^d \langle\varepsilon_a\rangle_{\pm}\bm n_a\otimes\bm n_a
	\end{equation}

\noindent where $\bm\varepsilon_+$  and $\bm\varepsilon_-$  are the tensile and compressive strain tensors, respectively. $\varepsilon_a$ is the principal strain with  $\bm n_a$  being their directions. The operators $\langle\centerdot\rangle_{\pm}$  are defined as : $\langle\centerdot\rangle_{\pm}=(\centerdot \pm |\centerdot|)/ 2$. The positive and negative elastic energy densities are then calculated by
	\begin{equation}
	\psi_{\varepsilon}^{\pm}(\bm \varepsilon) = \frac{\lambda}{2}\langle \mathrm{tr}(\bm\varepsilon)\rangle_{\pm}^2+\mu \mathrm{tr} \left(\bm\varepsilon_{\pm}^2\right) 
	\end{equation}

\noindent where $\lambda>0$ and $\mu>0$ are the Lam\'e constants. 

The phase field is assumed to affect only the positive elastic energy density and the elastic energy density is calculated by
	\begin{equation}
	\psi_{\varepsilon}(\bm\varepsilon)=w(\phi)\psi_{\varepsilon}^+(\bm \varepsilon)+\psi_{\varepsilon}^-(\bm \varepsilon)
	\label{elastic energy}
	\end{equation}

\noindent where $w(\phi)$ is a degradation function to model the energy and stiffness reduction in the solid.

Now the variation of the elastic energy $\delta \Psi_E$ is rewritten as
	\begin{equation}
	\delta \Psi_E = 	\int_{\Omega} \frac{\partial \psi_{\varepsilon}}{\partial \bm \varepsilon}: \nabla^{\rm{sym}} \delta \bm u \mathrm{d}{\Omega}+\int_{\Omega}w'(\phi)\psi_{\varepsilon}^+\rm{d}{\Omega} 
	\label{variation of elastic energy}
	\end{equation}

\noindent where $w'(\phi)$ = $\mathrm{d}w(\phi)/\rm{d}\phi$.

\subsection{Governing equations}\label{Governing equations}

To date, there are many forms of degradation function $w(\phi)$ such as those in \citet{wu2017unified}. Among these types of functions, the most used one \citep{borden2012phase} is adopted as
	\begin{equation}
	w(\phi)=(1-k)(1-\phi)^2+k
	\label{W}
	\end{equation}

\noindent where $0<k\ll1$ is a stability parameter for avoiding numerical singularity when the phase field $\phi$ tends to 1. Note that the stability parameter $k$ may be different in different PFMs. For example, in the phase field models of \citet{fang2019phase, natarajan2019phase}, the stability parameter $k=0$.

The variation of the external work $\delta W_{ext}$ is expressed in terms of the body force $\bm {b^*}$ and the traction $\bm {t^*}$:
	\begin{equation}
	\delta W_{ext} = \int_{\Omega} \bm {b^*}\cdot \delta \bm  u \mathrm{d}\Omega + \int_{\partial \Omega_t} \bm {t^*}\cdot \delta \bm u \mathrm{d}S
	\label{delta W_ext}
	\end{equation}

By substituting Eqs. \eqref{variation of fracture energy}, \eqref{variation of elastic energy}, \eqref{W}, and \eqref{delta W_ext} into Eq. \eqref{delta L}, the zero first variation of the energy functional $L$ produces the initial strong form of the system \citep{borden2012phase}: 	\begin{equation}
	  \left\{
	   \begin{aligned}
	\text{Div}(\bm\sigma)+\bm {b^*}=\bm 0
	\\ \left[\frac{2l_0(1-k)\psi_{\varepsilon}^+}{G_c}+1\right]\phi-l_0^2\nabla^2\phi=\frac{2l_0(1-k)\psi_{\varepsilon}^+}{G_c}
	   \end{aligned}\right.
	\label{governing equation1}
	\end{equation}

\noindent where $\bm\sigma$ is Cauchy stress tensor.

For a solid under loading or unloading, the crack irreversibility condition $\Gamma(\bm x,s)\in\Gamma(\bm x,t)(s<t)$ must be fulfilled. That is, a crack cannot be healed and the phase field must be monotonically increasing. We enforce the irreversibility condition by introducing a history field $H(\bm x,t)$ \citep{miehe2010phase, miehe2010thermodynamically}. The history field $H$ records the maximum value of the positive elastic energy in the time domain:
	\begin{equation}
	H(\bm x,t) = \max \limits_{x\in[0,t]}\psi_\varepsilon^+\left(\bm\varepsilon(\bm x,s)\right)
	\end{equation}

By replacing $\psi_\varepsilon^+$  by  $H(\bm x,t)$  in Eq. \eqref{governing equation1}, the final strong form of the system is expressed as follows,
	\begin{equation}
	  \left\{
	   \begin{aligned}
	\text{Div}(\bm\sigma)+\bm {b^*}=\bm 0
	\\ \left[\frac{2l_0(1-k)H}{G_c}+1\right]\phi-l_0^2\nabla^2\phi=\frac{2l_0(1-k)H}{G_c}
	   \end{aligned}\right.
	\label{governing equation2}
	\end{equation}

\section {Numerical implementation and parameter adjustment}\label{Numerical implementation and parameter adjustment}

This section shows the implementation approach of the phase field model and obtains the mechanical parameters used in the Brazilian splitting tests by experimental calibration.
\subsection{Numerical implementation}
The governing equations \eqref{governing equation2} is solved within the framework of Bubnov-Galerkin finite element method. Firstly, we obtain the weak form of \eqref{governing equation2} as follows,
	\begin{subequations}
	\begin{equation}
	\int_{\Omega}\left(-\bm\sigma:\delta \bm {\varepsilon}\right) \mathrm{d}\Omega +\int_{\Omega}\bm {b^*} \cdot \delta \bm u  \mathrm{d}\Omega +\int_{\partial \Omega_{t}}\bm {t^*} \cdot \delta \bm u  \mathrm{d}S=0
	\label{weak form 1}
	\end{equation}
	\begin{equation}
	\int_{\Omega}-2(1-k)H(1-\phi)\delta\phi\mathrm{d}\Omega+\int_{\Omega}G_c\left(l_0\nabla\phi\cdot\nabla\delta\phi+\frac{1}{l_0}\phi\delta\phi\right)\mathrm{d}\Omega=0
	\label{weak form 2}
	\end{equation}
	\end{subequations}

Standard vector-matrix notation is used and the nodal values of the displacement and phase field are defined as $\bm u_i$ and $\phi_i$. The discretized forms of the displacement and phase field then read
\begin{equation}
\bm u = \bm N_u \bm d,\hspace{0.5cm} \phi = \bm N_{\phi} \hat{\bm\phi}
\end{equation}

\noindent where node values $\bm u_i$ and $\phi_i$ sequentially construct the vectors $\bm d$ and $\hat{\bm\phi}$ to be solved for. $\bm N_u$ and $\bm N_{\phi}$ are the shape function matrices defined as
	\begin{equation}		
			\bm N_u = \left[ \begin{array}{ccccccc}
			N_{1}&0&0&\dots&N_{n}&0&0\\
			0&N_{1}&0&\dots&0&N_{n}&0\\
			0&0&N_{1}&\dots&0&0&N_{n}
			\end{array}\right], \hspace{0.5cm}
			\bm N_\phi = \left[ \begin{array}{cccc}
			N_{1}&N_{2}&\dots&N_{n}
			\end{array}\right]
	\end{equation}

\noindent where $n$ is the node number in one element and $N_i$ is the shape function of node $i$. The same discretization is applied to the test functions and we obtain
	\begin{equation}
	\delta \bm u = \bm N_u \delta \bm d,\hspace{0.5cm} \delta \phi = \bm N_{\phi} \delta \hat{\bm\phi}
	\end{equation}

\noindent where $\delta \bm d$ and $\delta \hat{\bm\phi}$ are the vectors consisting of node values of the test functions.

The gradients are thereby calculated by
	\begin{equation}
	\bm \varepsilon =  \bm B_u \bm d,\hspace{0.5cm} \nabla\phi = \bm B_\phi \hat{\bm\phi}, \hspace{0.5cm}\bm \delta \varepsilon =  \bm B_u \delta \bm d,\hspace{0.5cm} \nabla\phi = \bm B_\phi \delta \hat{\bm\phi}
	\end{equation}

\noindent where $\bm B_u$ and $\bm B_\phi$ are the derivatives of the shape functions defined by
	\begin{equation}
	\bm B_u=\left[
		\begin{array}{ccccccc}
		N_{1,x}&0&0&\dots&N_{n,x}&0&0\\
		0&N_{1,y}&0&\dots&0&N_{n,y}&0\\
		0&0&N_{1,z}&\dots&0&0&N_{n,z}\\
		N_{1,y}&N_{1,x}&0&\dots&N_{n,y}&N_{n,x}&0\\
		0&N_{1,z}&N_{1,y}&\dots&0&N_{n,z}&N_{n,y}\\
		N_{1,z}&0&N_{1,x}&\dots&N_{n,z}&0&N_{n,x}
		\end{array}\right],\hspace{0.2cm}
		\bm B_\phi=\left[
		\begin{array}{cccc}
		N_{1,x}&N_{2,x}&\dots&N_{n,x}\\
		N_{1,y}&N_{2,y}&\dots&N_{n,y}\\
		N_{1,z}&N_{2,z}&\dots&N_{n,z}
		\end{array}\right]
		\label{BiBu}
	\end{equation}

The equations of weak form \eqref{weak form 1} and \eqref{weak form 2} are further expressed as
	\begin{subequations}
\begin{equation}
	-(\delta\bm d)^\mathrm{T} \int_{\Omega} \bm B_u^\mathrm{T} \bm D_e \bm B_u \mathrm{d}\Omega \bm d + (\delta\bm d)^\mathrm{T} \left[\int_{\Omega}\bm N_u^\mathrm{T}\bm {b^*} \mathrm{d}\Omega+\int_{\Omega_{h_i}} \bm N_u^\mathrm{T} \bm {t^*} \mathrm{d}S \right]=0
	\label{discrete equation 1}
	\end{equation}
	\begin{equation}
	-(\delta\hat{\bm \phi})^{\mathrm{T}} \int_{\Omega}\left\{\bm B_\phi^{\mathrm{T}} G_c l_0 \bm B_\phi +\bm N_\phi^{\mathrm{T}} \left [ \frac{G_c}{l_0} + 2(1-k)H \right ]  \bm N_\phi \right \} \mathrm{d}\Omega \hat{\bm \phi}+ (\delta\hat{\bm \phi})^\mathrm{T} \int_{\Omega}2(1-k)H\bm N_\phi^{\mathrm{T}} \mathrm{d}\Omega = 0
	\label{discrete equation 2}      
	\end{equation}
\end{subequations}

\noindent where $\bm D_e$ is the degraded elasticity matrix, which is originated from the fourth order elasticity tensor $\bm D$:
	\begin{equation}
		\begin{aligned}
		\bm D &= \frac {\partial \bm\sigma}{\partial \bm\varepsilon}\\
	&=\lambda\left\{ \left[(1-k)(1-\phi)^2+k \right]H_\varepsilon(tr(\bm\varepsilon))+H_\varepsilon(-tr(\bm\varepsilon))\right\}\bm J +2\mu\left\{\left[(1-k)(1-\phi)^2+k \right]\bm P^++\bm P^- \right \}
		\end{aligned}
	\end{equation}

Moreover, $J_{ijkl}=\delta_{ij}\delta_{kl}$, $\delta_{ij}$ being the Kronecker and $P_{ijkl}^\pm= \sum_{a=1}^3\sum_{b=1}^3 H_\varepsilon(\varepsilon_a)\delta_{ab}n_{ai}n_{aj}n_{bk}n_{bl}+\sum_{a=1}^3\sum_{b\neq a}^3 \frac 1 2 \frac {\langle \varepsilon_a\rangle_\pm - \langle \varepsilon_b\rangle_\pm}{\varepsilon_a-\varepsilon_b}n_{ai}n_{bj}(n_{ak}n_{bl}+n_{bk}n_{al})$ with $n_{ai}$ the $i$-th component of vector $\bm n_a$. $H_\varepsilon \langle x \rangle$  is the Heaviside function defined as $H_\varepsilon \langle x \rangle=\mathrm{max} (x,0)$. 

Eqs. \eqref{discrete equation 1} and \eqref{discrete equation 2} must always hold for all admissible arbitrary test functions. Therefore, it follows that
	\begin{subequations}
	\begin{equation}
	\bm R^u = \bm K_u \bm d - \bm F_u^{ext} = \bm 0
	\label{displacement solution}
	\end{equation}
	\begin{equation}
	\bm R^\phi = \bm K_\phi  \hat{\bm\phi} - \bm F_\phi^{ext} = \bm 0
	\label{phase field solution}
	\end{equation}
	\end{subequations}

\noindent where $\bm K_u$ and $\bm K_\phi$ are the respective stiffness matrices of the displacement  and phase fields expressed as
	\begin{subequations}
		\begin{equation}
	\bm K_u = \int_{\Omega} \bm B_u^\mathrm{T} \bm D_e \bm B_u \mathrm{d}\Omega
	\end{equation}
	\begin{equation}
		\bm K_\phi = \int_{\Omega}\left\{\bm B_\phi^{\mathrm{T}} G_c l_0 \bm B_\phi +\bm N_\phi^{\mathrm{T}} \left [ \frac{G_c}{l_0} + 2(1-k)H \right ] \bm N_\phi \right \} \mathrm{d}\Omega
		\end{equation}
	\end{subequations}

In addition, the external force vectors of the displacement and phase field are defined as
	\begin{subequations}
		\begin{equation}
	\bm F_u^{ext} = \int_{\Omega}\bm N_u^\mathrm{T}\bm {b^*} \mathrm{d}\Omega+\int_{\partial\Omega_{t}} \bm N_u^\mathrm{T} \bm {t^*}\mathrm{d}S
	\end{equation}
	\begin{equation}
		\bm F_\phi^{ext} = \int_{\Omega}2(1-k)H\bm N_\phi^{\mathrm{T}} \mathrm{d}\Omega
		\end{equation}
	\end{subequations}

The system of nonlinear equations \eqref{displacement solution} and \eqref{phase field solution} is solved in an incremental manner. The time interval of interest $(0,T]$ is discretized into small intervals of $[t_n,t_{n+1}]$ and thereby the variables at time $t_{n+1}$ is solved with all the variables known at time ${t_n}$. We use the implicit time integration scheme in this work and a staggered scheme is used to solve the coupled equations. The solving order is fixed as $\bm d$, $H$, and $\hat{\bm \phi}$ for the best convergence rate. We implement the numerical model and time integration scheme in a powerful simulation environment--COMSOL Multiphysics, and detailed descriptions about the COMSOL implementations of phase field modeling can be referred to our previous work \citep{zhou2018phase3}. Note that the maximum iteration number is set as 50 in the simulations. The code can be seen in the website "https://sourceforge.net/projects/phasefieldmodelingcomsol/".

\subsection{Parameter adjustment}\label{Parameter adjustment}

We first simulate the fracture propagation in an intact Brazilian disc without voids and inclusions. During the simulations, we adjust the used parameters to fit the load-deformation curve of a real experimental test \citep{chang2018mechanical}, and then these parameters are applied as the base parameters for prediction of the fracture mechanism in Brazilian discs with circular voids and inclusions. The diameter of the specimens is $D=50$ mm and the length scale parameter $l_0$ is fixed to 1 mm in this study. The calculation domain is discretized using the linear triangular elements with the maximum element size $h=0.5$ mm. The parameter $k=1\times10^{-9}$ is used. We apply an increasing downwards displacement to a finite width of $0.05D$ on the upper boundary of the Brazilian disc (this width is $0.12D$ in \citet{zhu2006numerical}), to ensure the specimen failure inside the disc rather than local compression failure on the upper and bottom boundaries. The symmetrical part on the bottom boundary is fixed in the normal displacement. In addition, the displacement increment is set as $\Delta u=5\times10^{-5}$ mm in each time step. In all cases, the 2D plane stress model is used. Although the 3D Brazilian splitting test does not perfectly satisfy the plane stress assumption, the comparative study in \citet{saksala2013numerical} and other numerical simulations on Brazilian splitting tests \citep{zhu2006numerical, haeri2015experimental} have shown that the 2D model can achieve satisfactory results. Note that all of the phase field simulations in this study apply the same numerical settings in this section.

\begin{figure}[htbp]
	\centering
	\subfigure[PFM]{\includegraphics[height = 5cm]{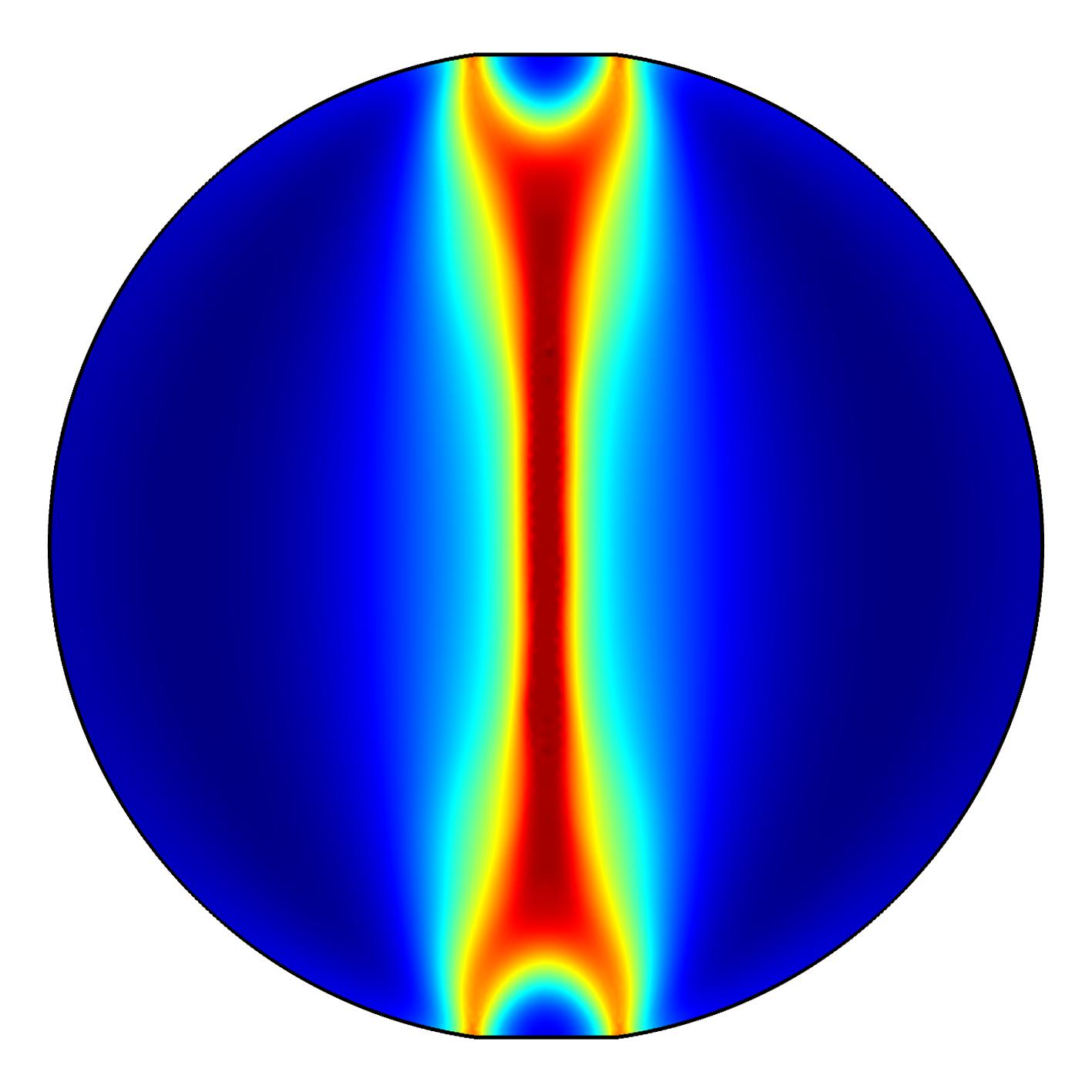}}
	\subfigure{\includegraphics[height = 4cm]{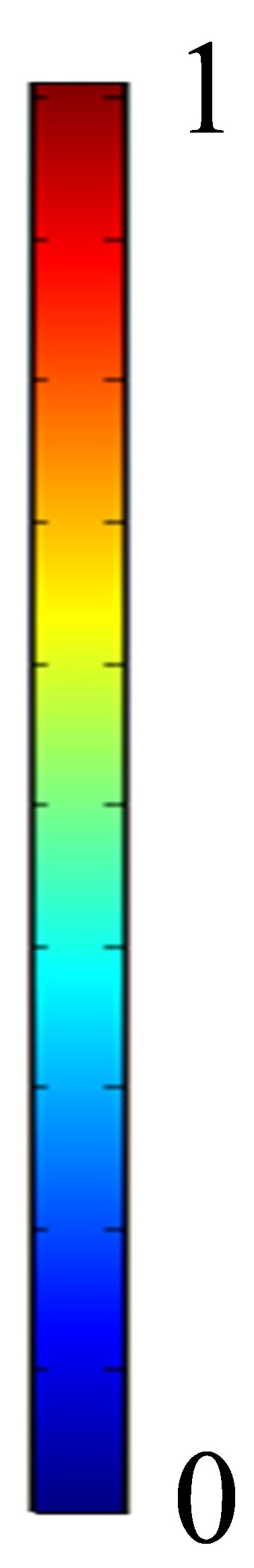}}
	\subfigure[Experimental test \citep{chang2018mechanical}]{\includegraphics[height = 5cm]{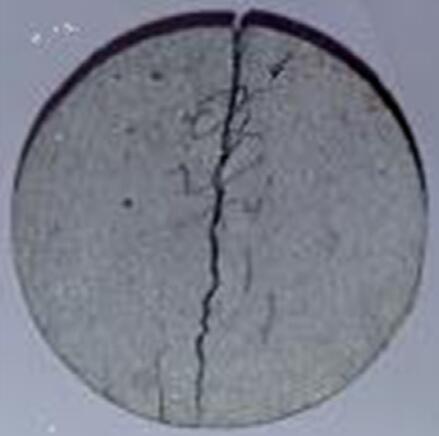}}
	\subfigure[RFPA simulation \citep{chang2018mechanical}]{\includegraphics[height = 5cm]{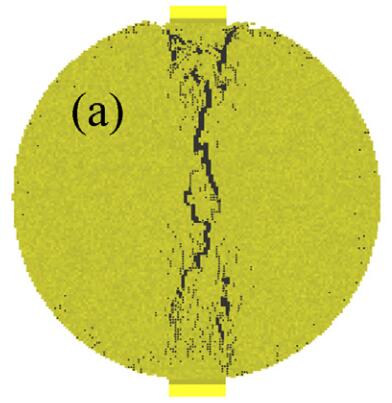}} 
	\caption{Comparison of the fracture patterns in the Brazilian disc without voids and inclusion}
	\label{Comparison of the fracture patterns in the Brazilian disc without voids and inclusion}
	\end{figure}

\begin{figure}[htbp]
	\centering
	\includegraphics[width = 8cm]{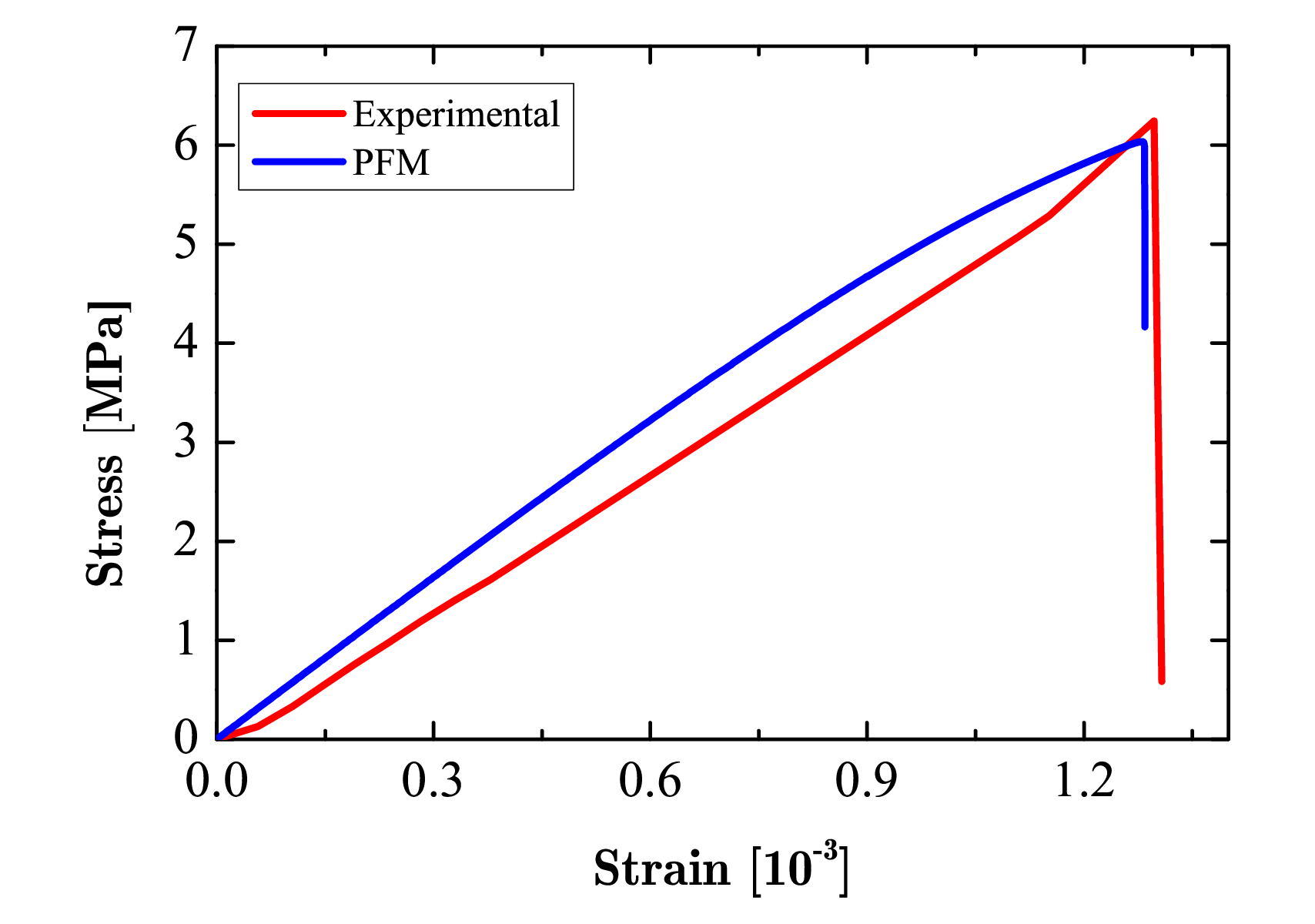}
	\caption{Comparison of the stress-strain curves in the Brazilian disc without voids and inclusion}
	\label{Comparison of the stress-strain curves in the Brazilian disc without voids and inclusion}
	\end{figure}

Figure \ref{Comparison of the fracture patterns in the Brazilian disc without voids and inclusion} shows a comparison of the fracture patterns obtained by the phase field modeling, experimental tests \citep{chang2018mechanical}, and RFPA simulation \citep{chang2018mechanical}. As can be seen from this figure,  the simulated fracture is consistent with those experimental results and RFPA simulation; in fact, the phase field model achieves a more obvious fracture than RFPA, which collects the fracture pattern using scattered damage regions. In addition to the fracture patterns, a good agreement is also observed in the stress-strain curves of the disc in Fig. \ref{Comparison of the stress-strain curves in the Brazilian disc without voids and inclusion}. Note that the stress $S$ and strain $\varepsilon_d$ in the figure are calculated as $S=2P/(\pi D t)$ and $\varepsilon_d=\Delta D/D$ with $P$, $t$, and $\Delta D$ being the loads on the disc end, thickness of the disc, and the relative deformation between the two disc ends, respectively. Therefore, these parameters are used for the Brazilian disc in the further simulations: Young's modulus $E = 26.1$ GPa, critical energy release rate $G_c = 28.2$ N/m, and Poisson's ratio $\nu=0.3$.

\section {Fracture patterns in Brazilian discs with voids} \label{Fracture patterns in Brazilian discs with voids}

This section describes the fractures pattern in Brazilian discs with voids by using the phase field method. The parameters for calculation are identical to those in the previous section.

\subsection{Brazilian discs with a centered void}

\begin{figure}[htbp]
	\centering
	\subfigure[D = 0]{\includegraphics[height = 5cm]{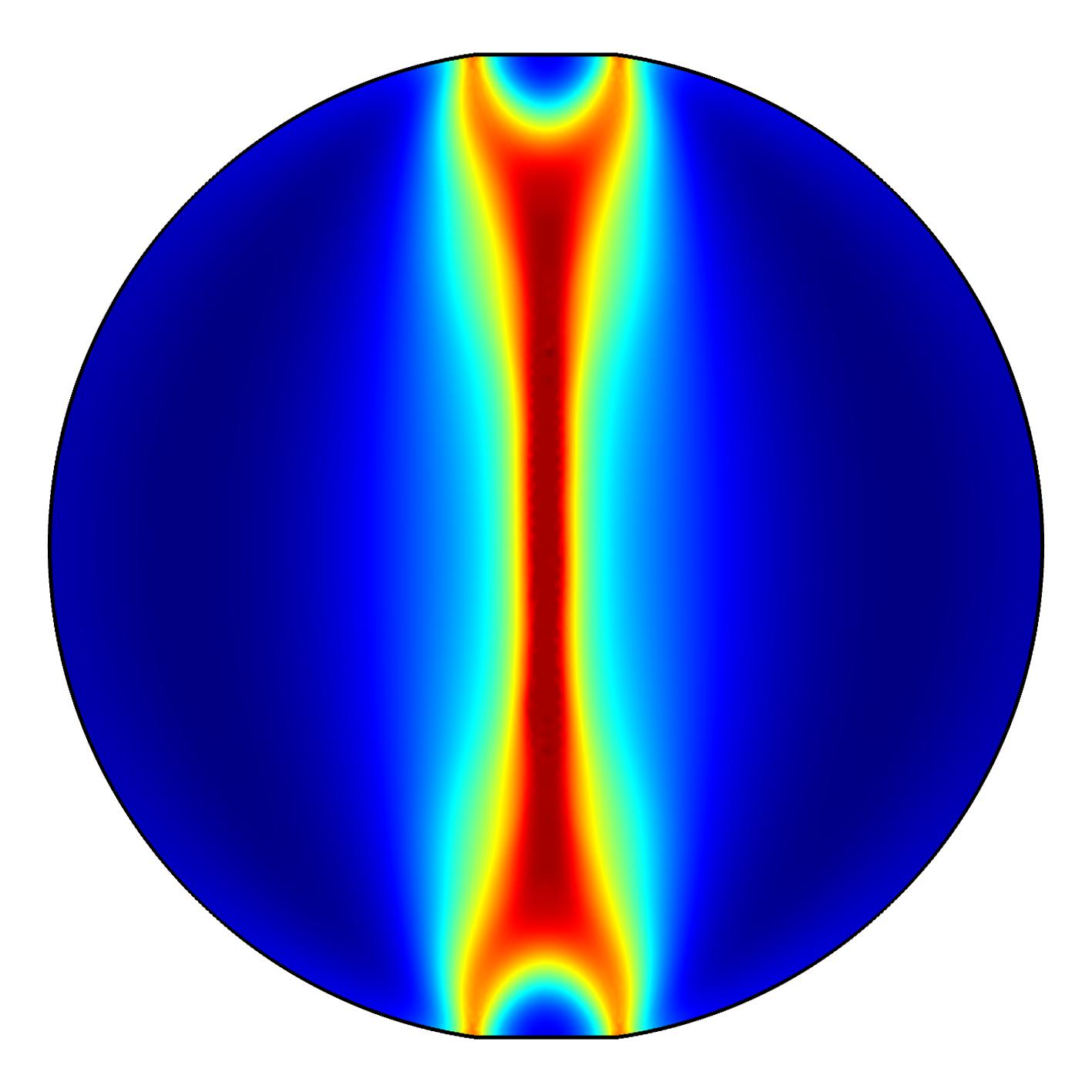}}
	\subfigure[D = 10 mm]{\includegraphics[height = 5cm]{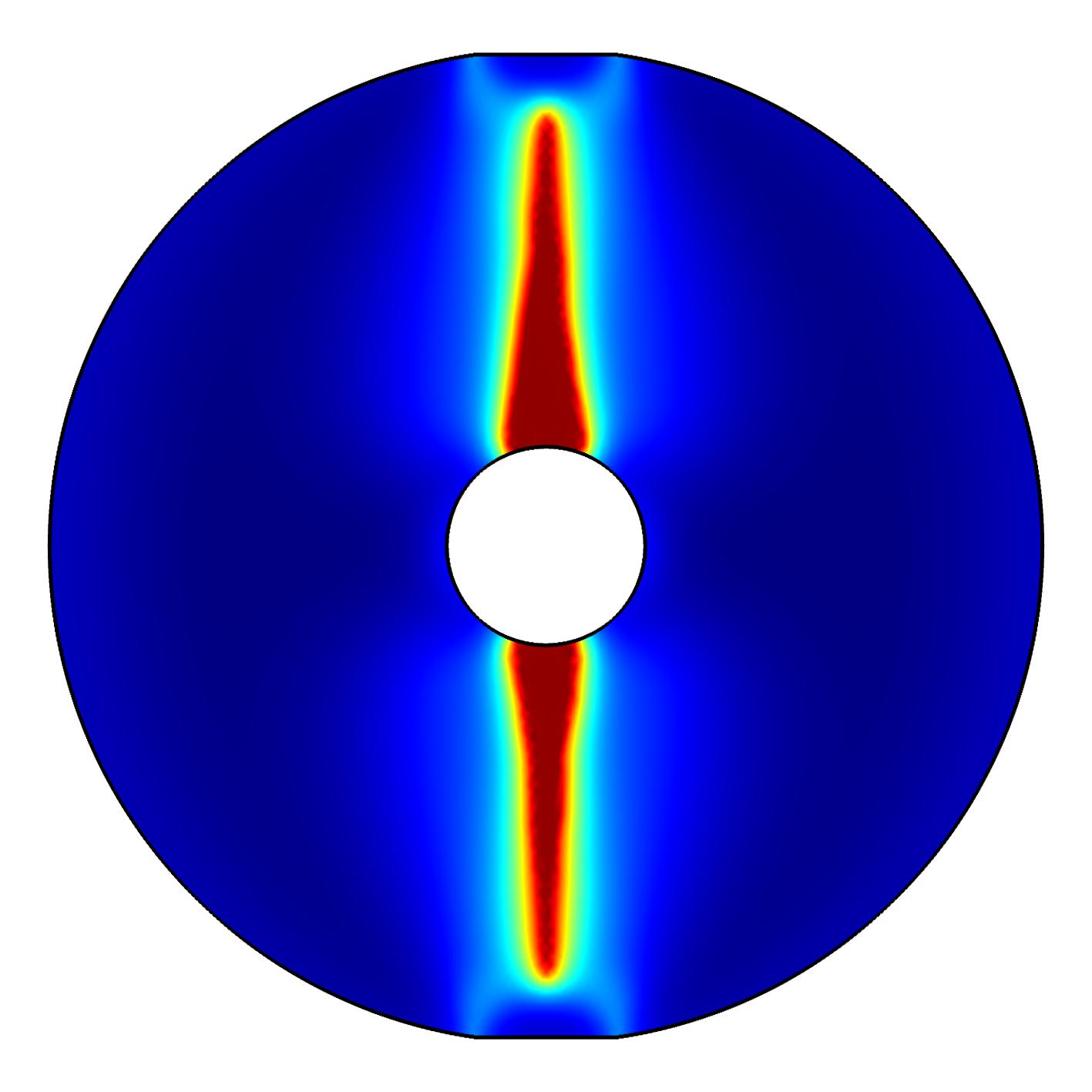}}\\
 	\subfigure[D = 20 mm]{\includegraphics[height = 5cm]{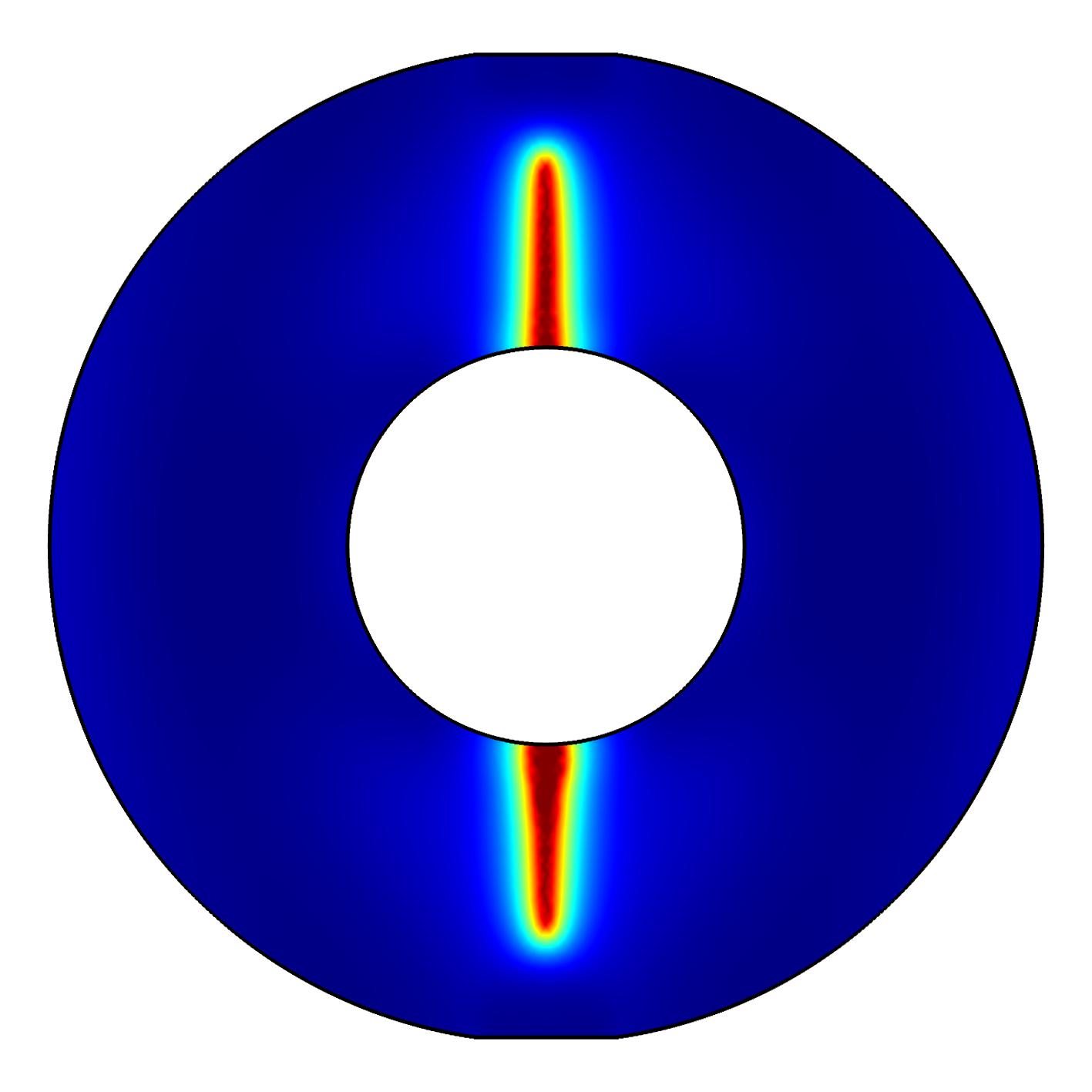}}
	\subfigure[D = 30 mm]{\includegraphics[height = 5cm]{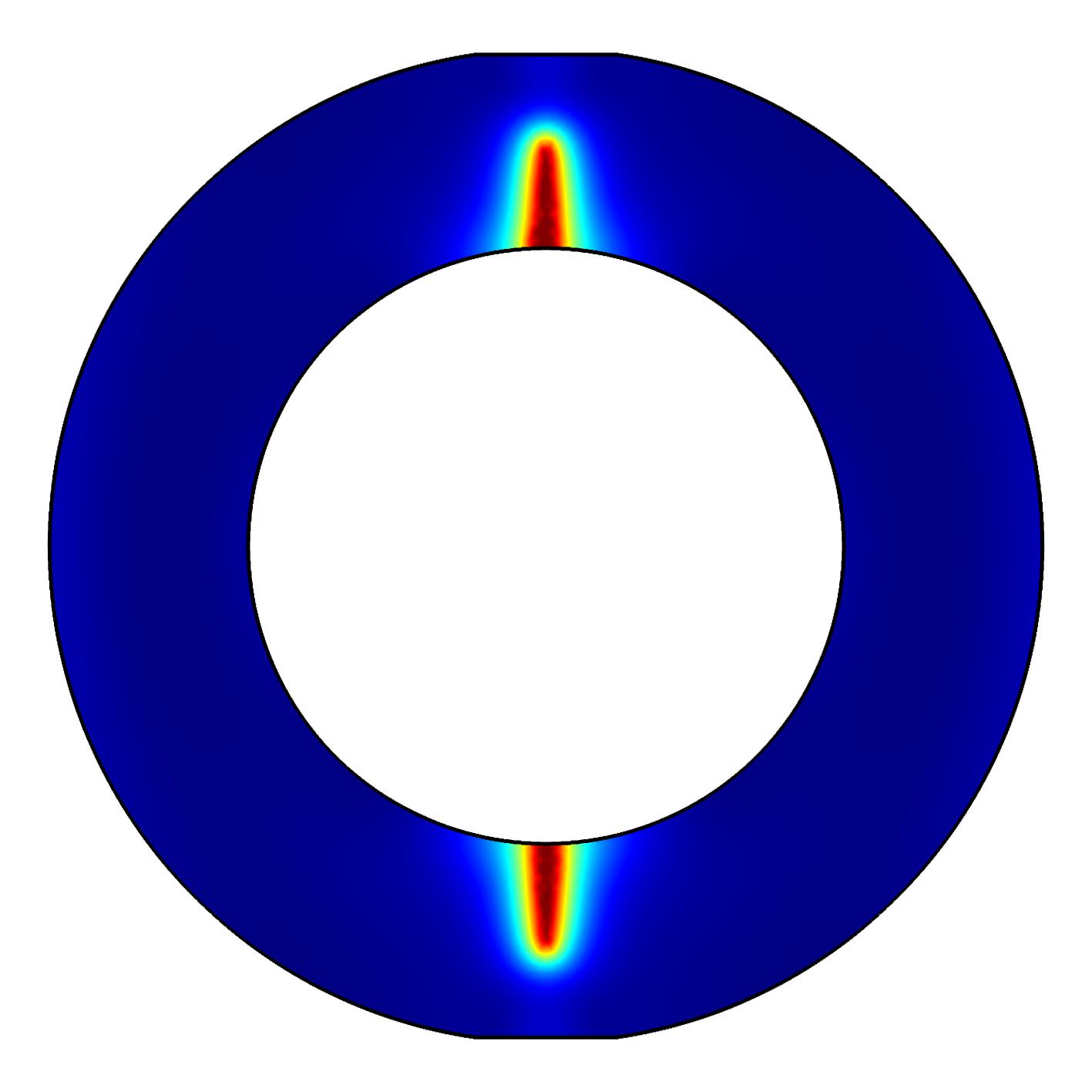}}
	\subfigure{\includegraphics[height = 4cm]{Figure2a_legend.jpg}}
	\caption{Final fracture patterns of the Brazilian discs with a centered void}
	\label{Final fracture patterns of the Brazilian discs with a centered void}
	\end{figure}

We first investigate the fracture patterns in Brazilian discs with a centered void. The diameter of the void affects the crack initiation and propagation, and thereby we choose the diameters of the void as $D=0$, 10, 20, and 30 mm, respectively. The diameter of the disc is still 50 mm. Figure \ref{Final fracture patterns of the Brazilian discs with a centered void} shows the final fracture patterns of the Brazilian discs with a centered void. For the four different diameters, the fractures are similar. A vertical fracture initiates from the top and bottom of the void and propagates towards the two ends of the Brazilian disc. In the Brazilian splitting test, the maximum stress and elastic energy always occur close to the center of the specimen and in the vertical axis. Conversely, the upper and bottom boundaries of the specimen are mainly subjected to compression and have the lowest positive elastic energy, which causes a lower fracture priority. Therefore, the fracture is initiated at the position having the smallest distance from the specimen center and then propagate towards to the specimen boundaries. In addition, when the diameter $D$ increases, it is much easier to split the specimen into two parts.

\begin{figure}[htbp]
	\centering
	\includegraphics[width = 8cm]{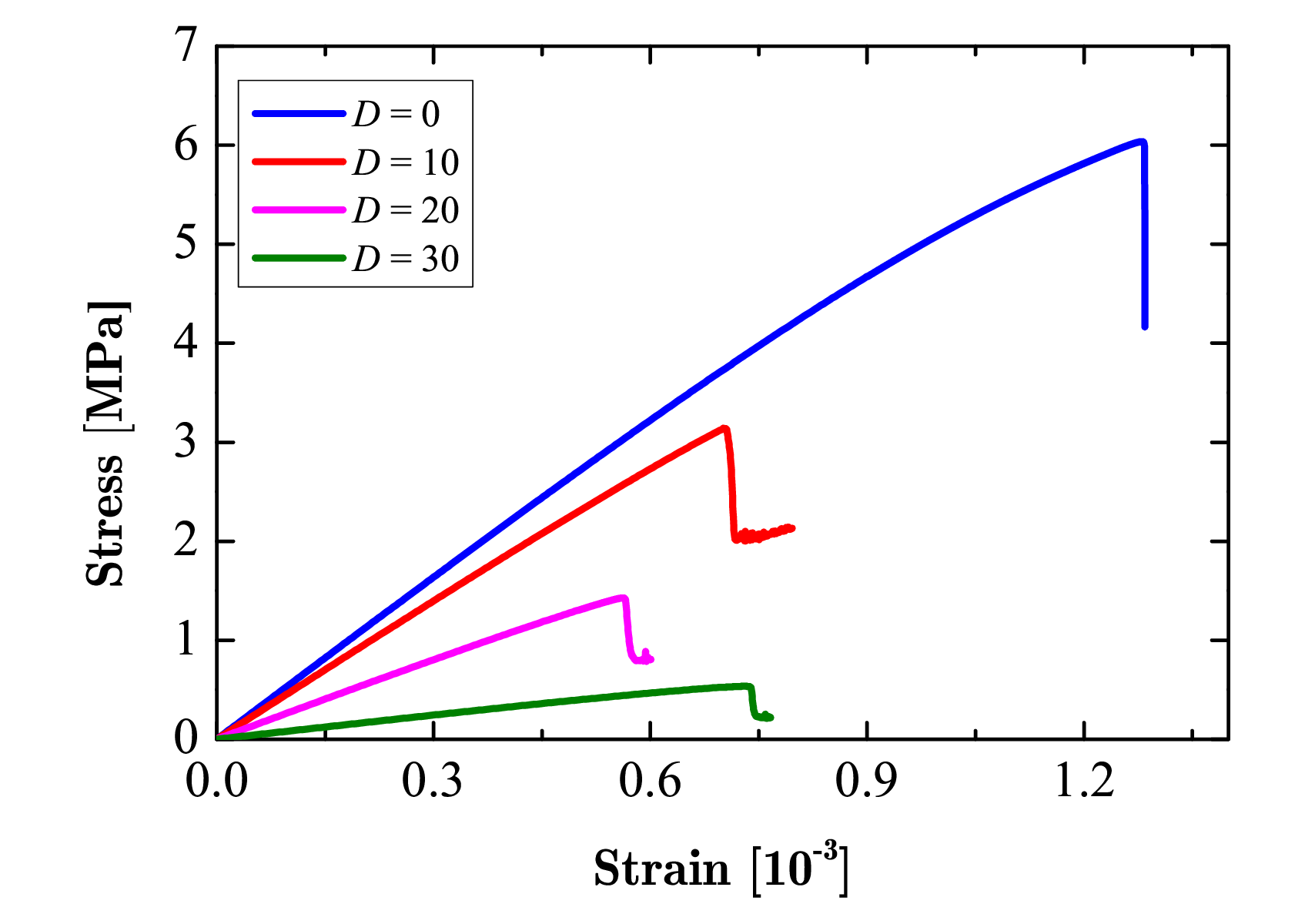}
	\caption{Stress-strain curves of the Brazilian discs with a centered void}
	\label{Stress-strain curves of the Brazilian discs with a centered void}
	\end{figure}

Figure \ref{Stress-strain curves of the Brazilian discs with a centered void} shows the stress-strain curves of the Brazilian discs with a centered void of different diameters. As observed, the void significantly reduces the strength and overall stiffness of the disc. However, the strain that corresponds to the peak strength first decreases and then increases as the diameter increases. The reason is the difference between the decreasing rate of the peak strength and overall stiffness of the disc. In addition, our simulations indicate that the stress for fracture initiation is quite closer to the peaks stress (the ratio is around 99.8\%) because a brittle fracture model is used.

\begin{figure}[H]
	\centering
	\subfigure[D = 0]{\includegraphics[height = 5cm]{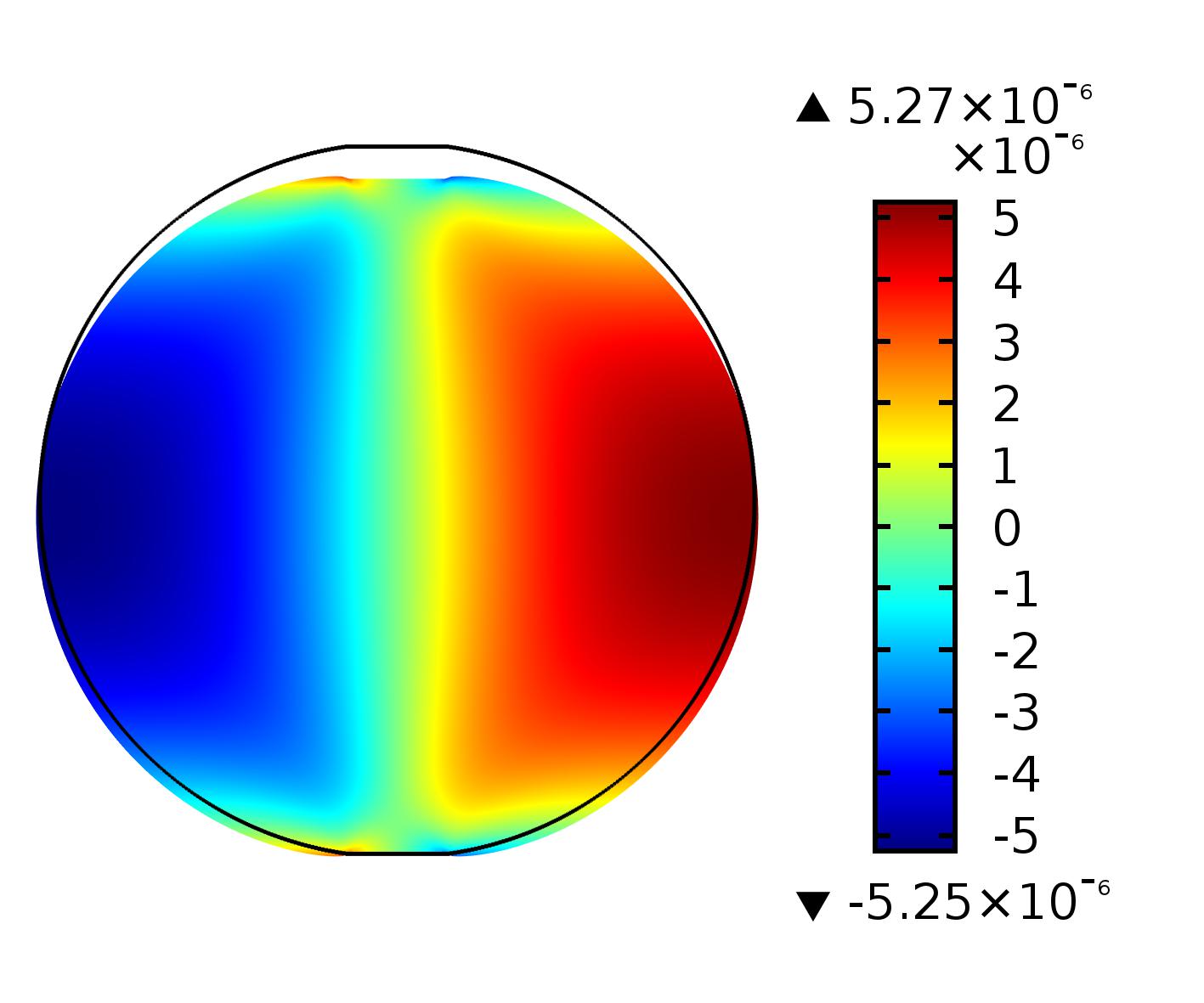}}
	\subfigure[D = 10 mm]{\includegraphics[height = 5cm]{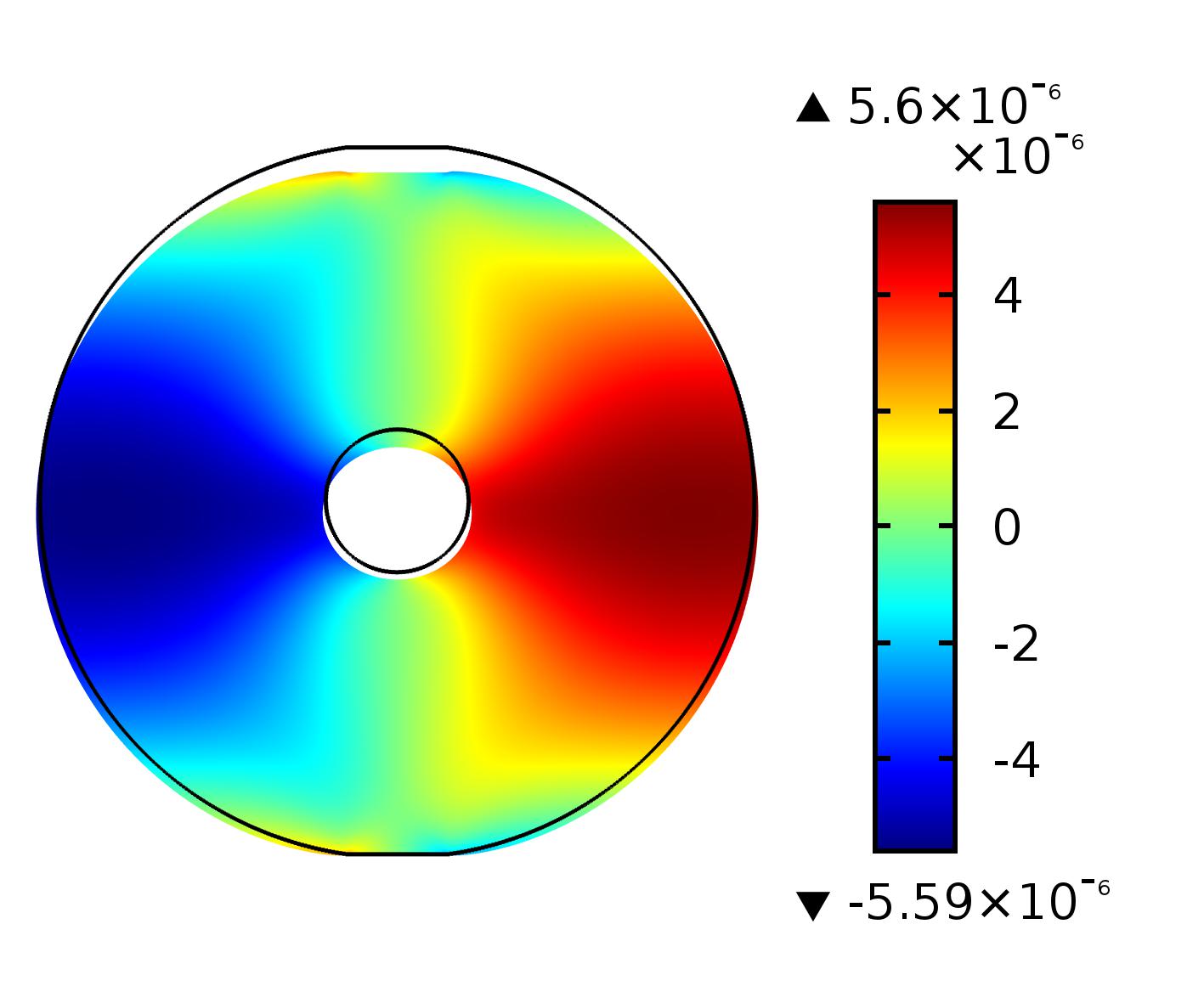}}\\
 	\subfigure[D = 20 mm]{\includegraphics[height = 5cm]{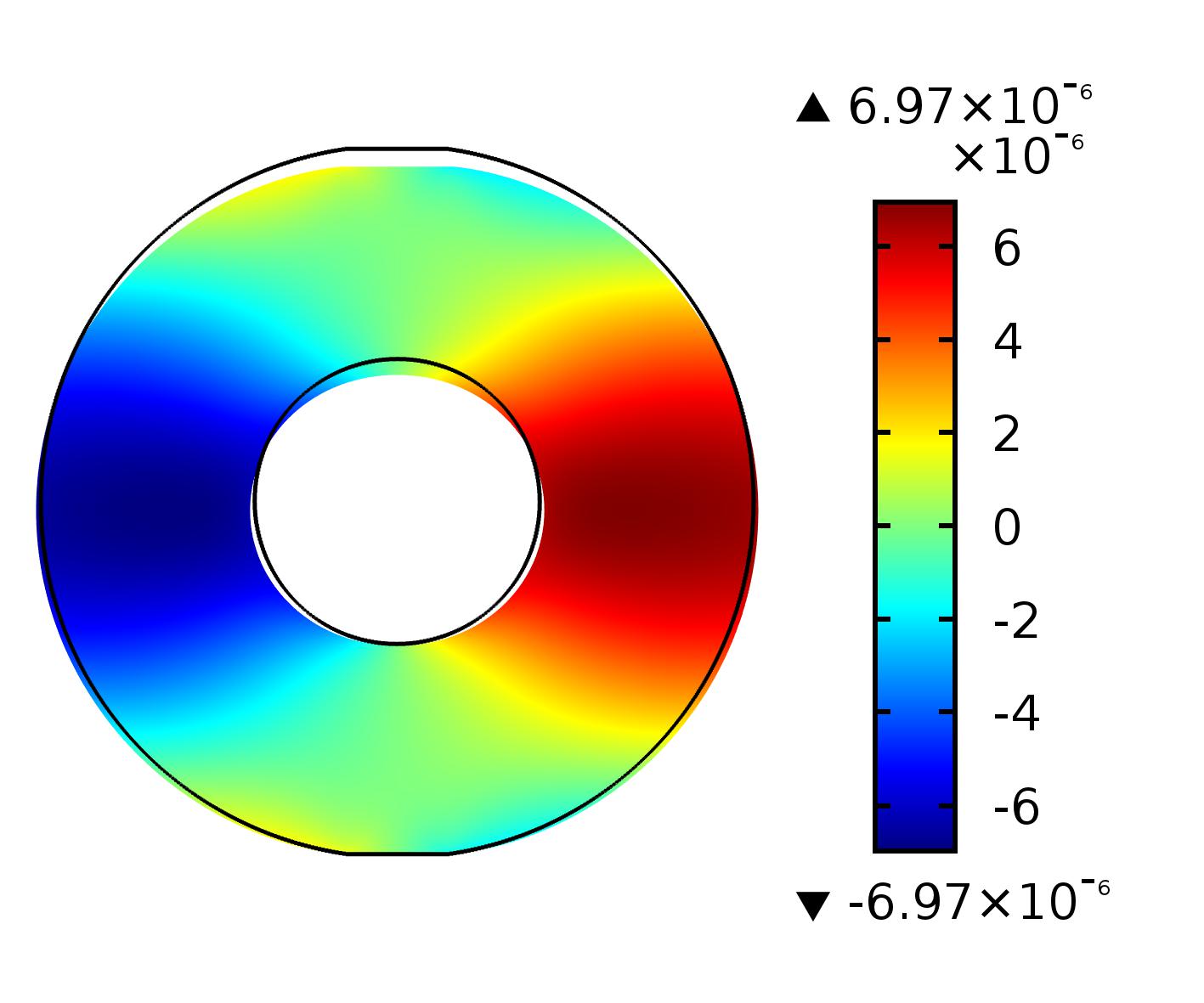}}
	\subfigure[D = 30 mm]{\includegraphics[height = 5cm]{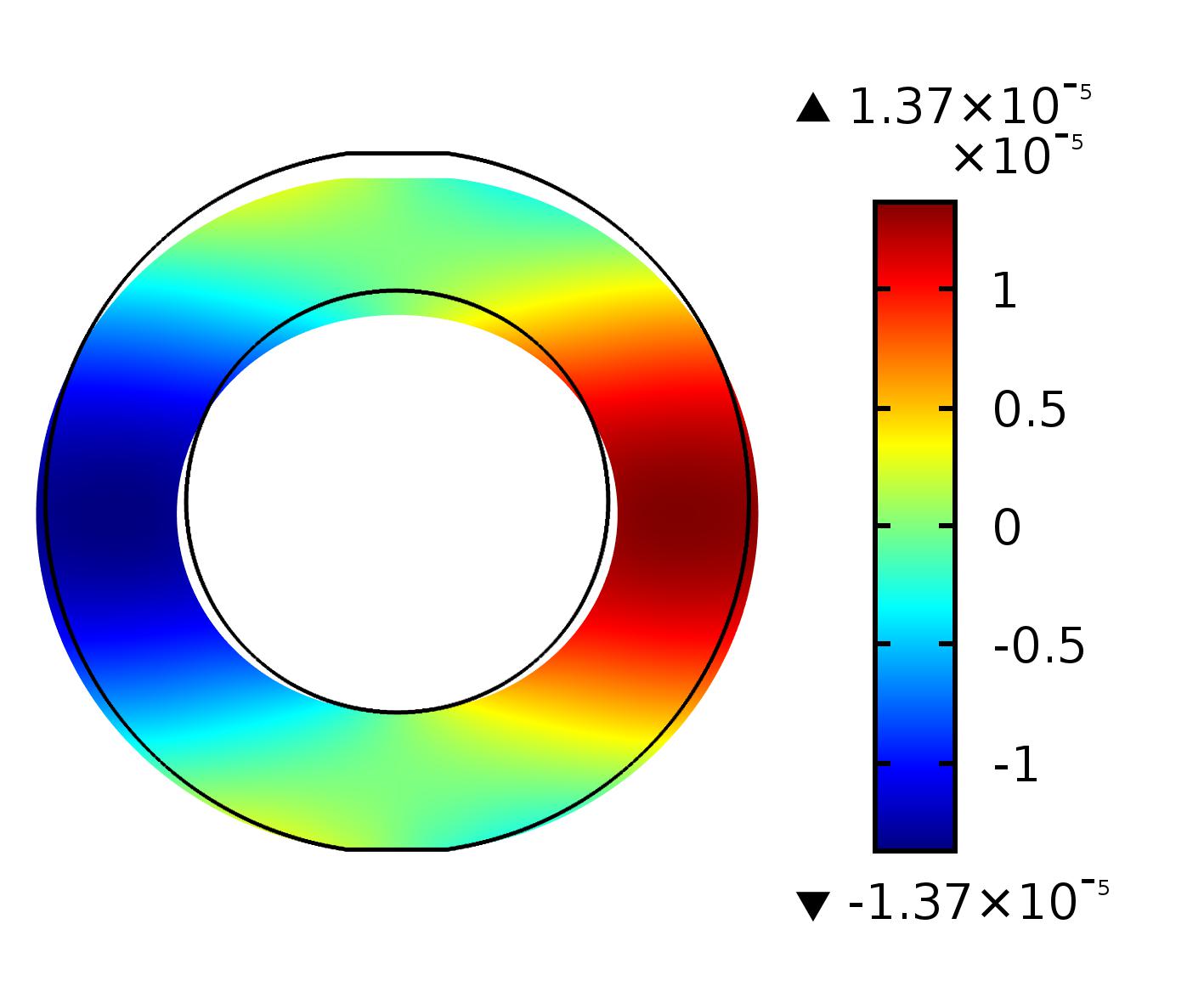}}
	\caption{Horizontal displacement of the Brazilian discs with a centered void}
	\label{Horizontal displacement of the Brazilian discs with a centered void}
	\end{figure}

Figure \ref{Horizontal displacement of the Brazilian discs with a centered void} shows the horizontal displacement of the Brazilian discs with a void. For different diameters of the void, the horizontal displacement fields are similar and symmetry along the vertical axis. Note that in Fig. \ref{Horizontal displacement of the Brazilian discs with a centered void}, the deformation is displayed by an amplification factor of 50.

\subsection{Single void with varying eccentricities along the horizontal axis}

In addition to the diameter of the void, the space distribution of the void also significantly affects the fracture process and stress-strain response. Herein we fix the diameter of the void as 20 mm when changing the position of the void. We use an eccentricity $e=l/R$ in the analysis with $l$ and $R$ being the distance between the centers of the void and Brazilian disc and the radius of the disc, respectively.
 
\begin{figure}[H]
	\centering
	\subfigure[e = 0]{\includegraphics[height = 5cm]{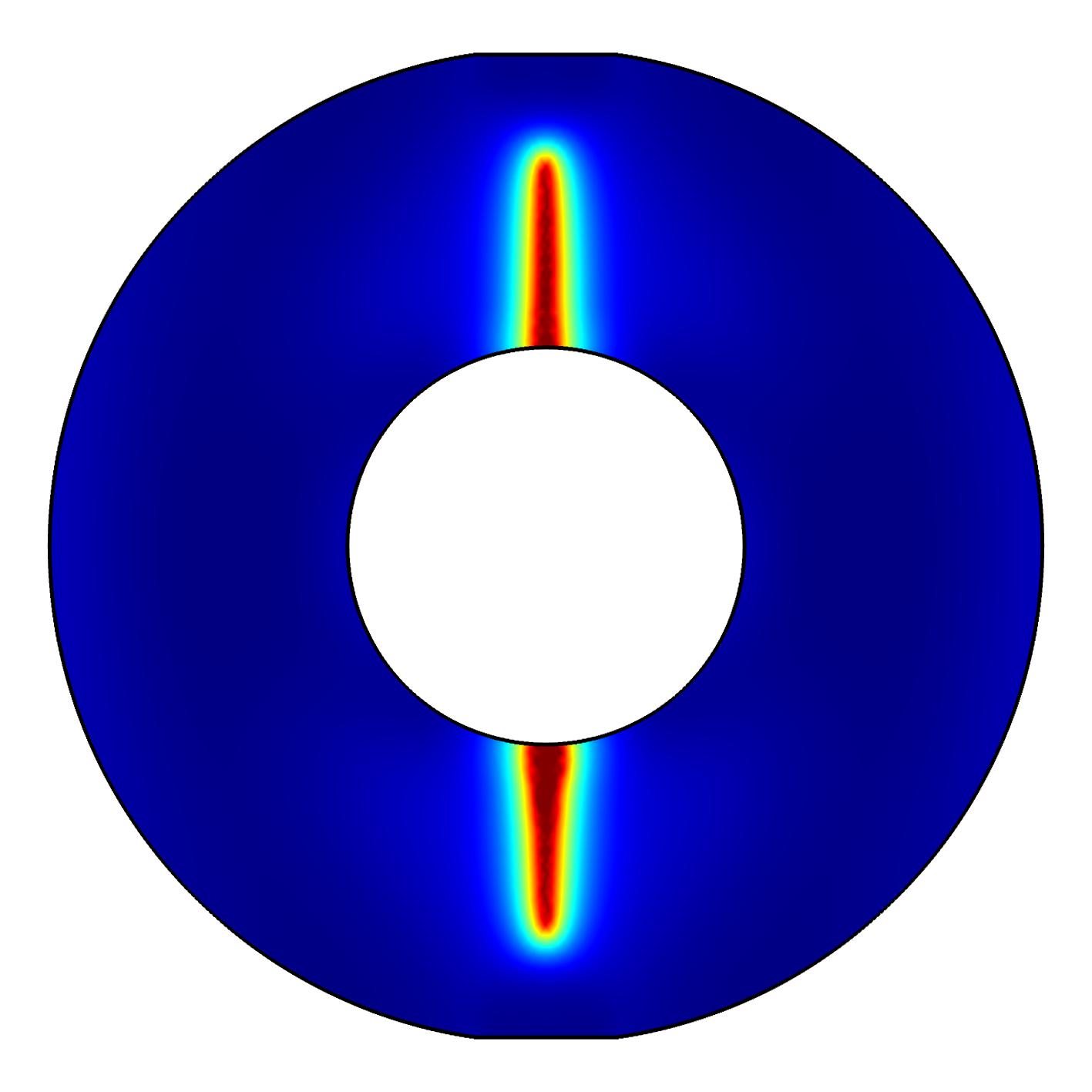}}
	\subfigure[e = 0.2]{\includegraphics[height = 5cm]{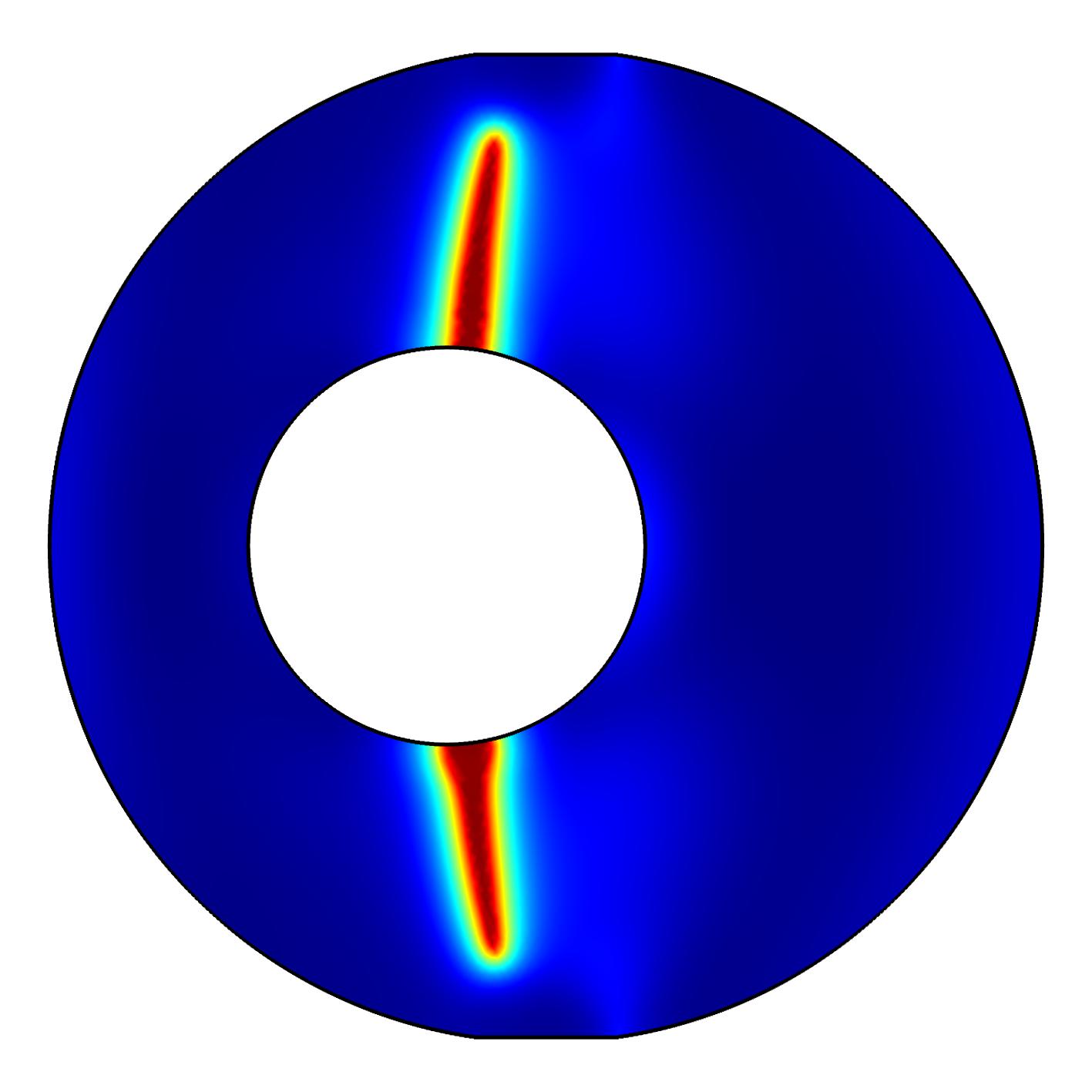}}
 	\subfigure[e = 0.4]{\includegraphics[height = 5cm]{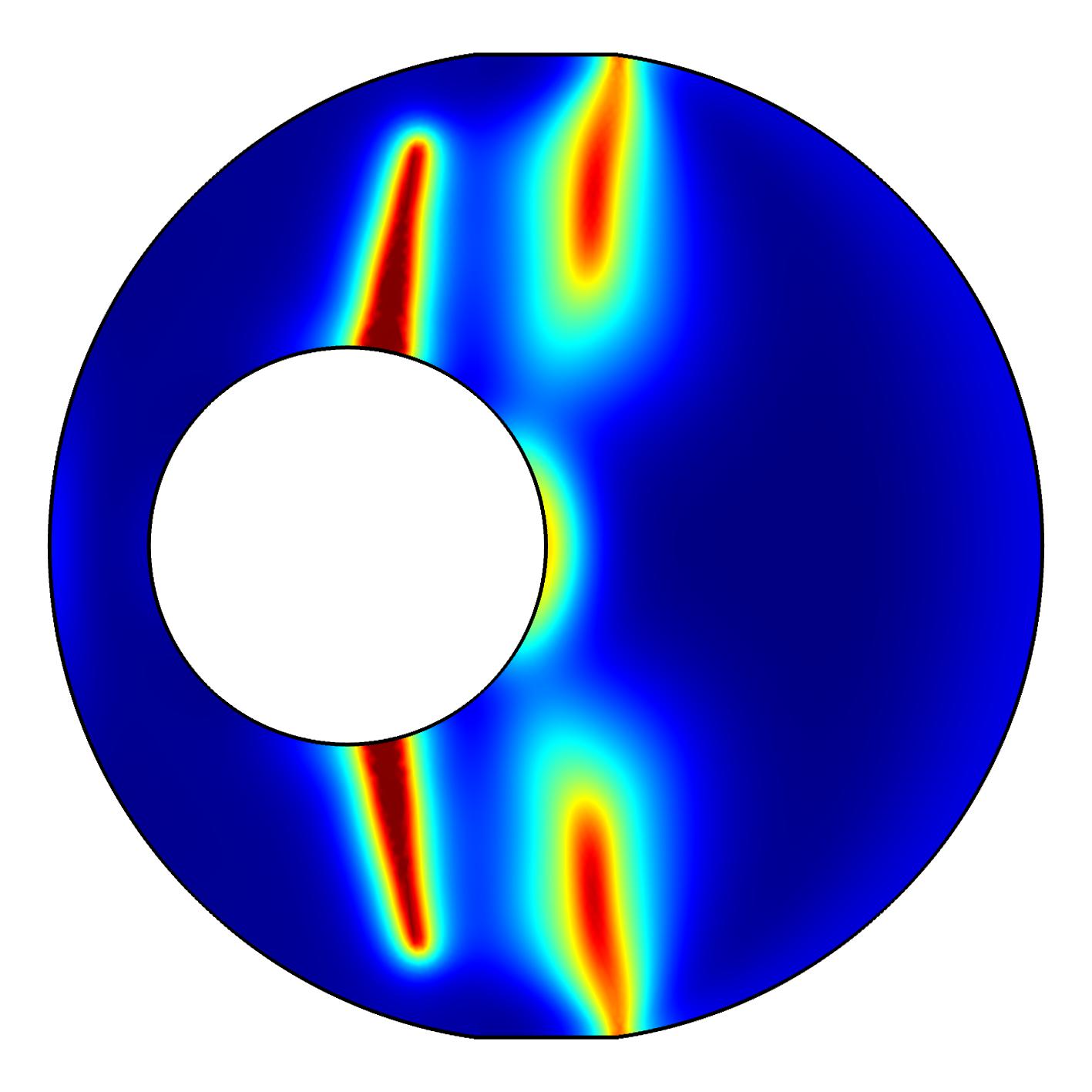}}
 	\subfigure{\includegraphics[height = 4cm]{Figure2a_legend.jpg}}
	\caption{Final fracture patterns of the Brazilian discs with a void along the horizontal axis}
	\label{Final fracture patterns of the Brazilian discs with a void along the horizontal axis}
	\end{figure}

Figure \ref{Final fracture patterns of the Brazilian discs with a void along the horizontal axis} presents the final fractures in the Brazilian discs with a horizontally-arranged void under $e=0$, 0.2, and 0.4. As observed, the fractures initiate from the boundaries of the void and propagate towards the two ends of the disc. When the eccentricity is 0.4, more fractures occur. Two secondary fractures initiate rapidly and propagate along with the propagating dominated fracture. Further analysis indicates that the secondary fracture finally stops at the boundaries of the void, which drives the Brazilian disc into three parts when the eccentricity is large enough.

\begin{figure}[H]
	\centering
	\includegraphics[width = 8cm]{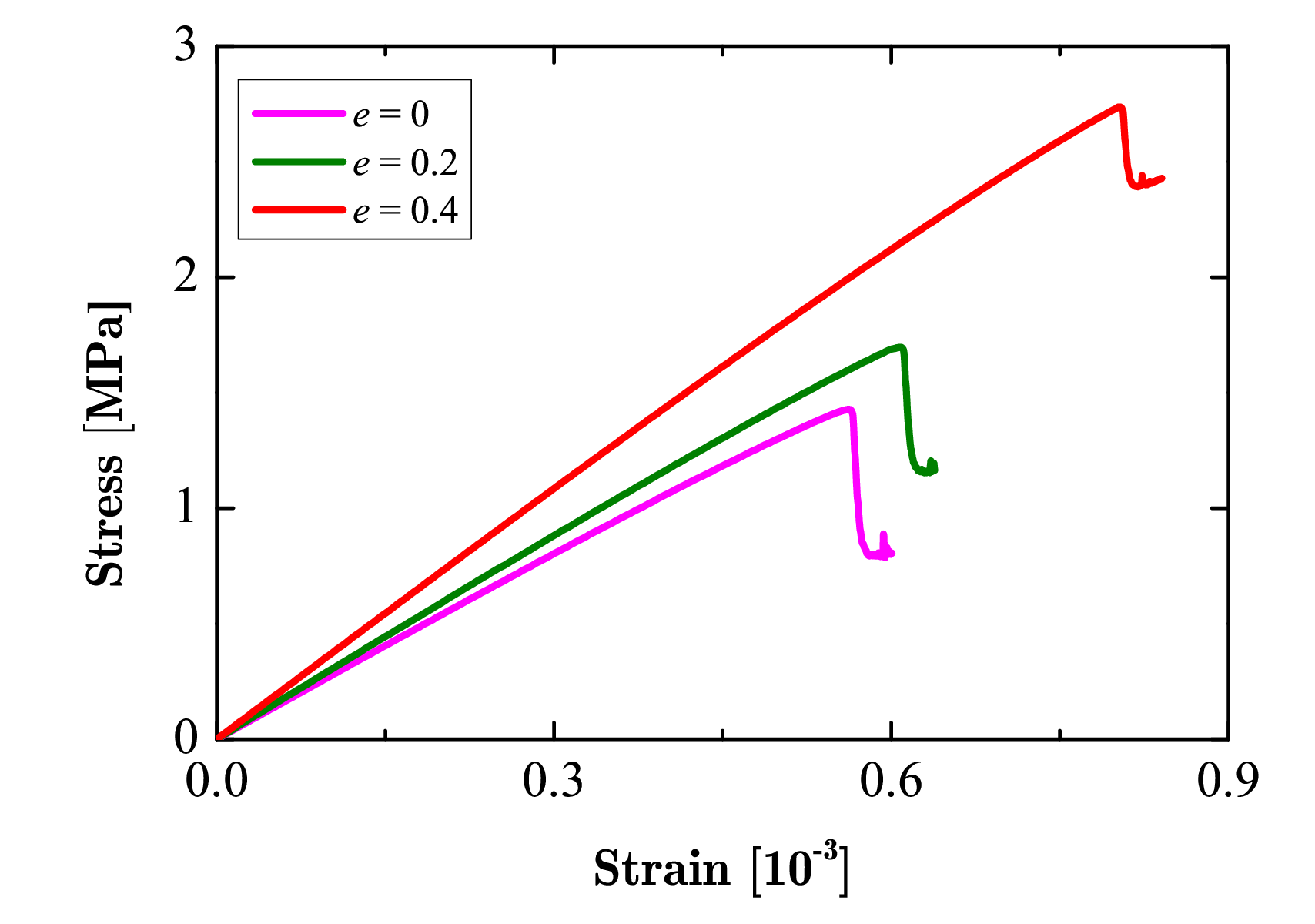}
	\caption{Stress-strain curves of the Brazilian discs with a void along the horizontal axis}
	\label{Stress-strain curves of the Brazilian discs with a void along the horizontal axis}
	\end{figure}

\begin{figure}[H]
	\centering
	\subfigure[e = 0]{\includegraphics[height = 5cm]{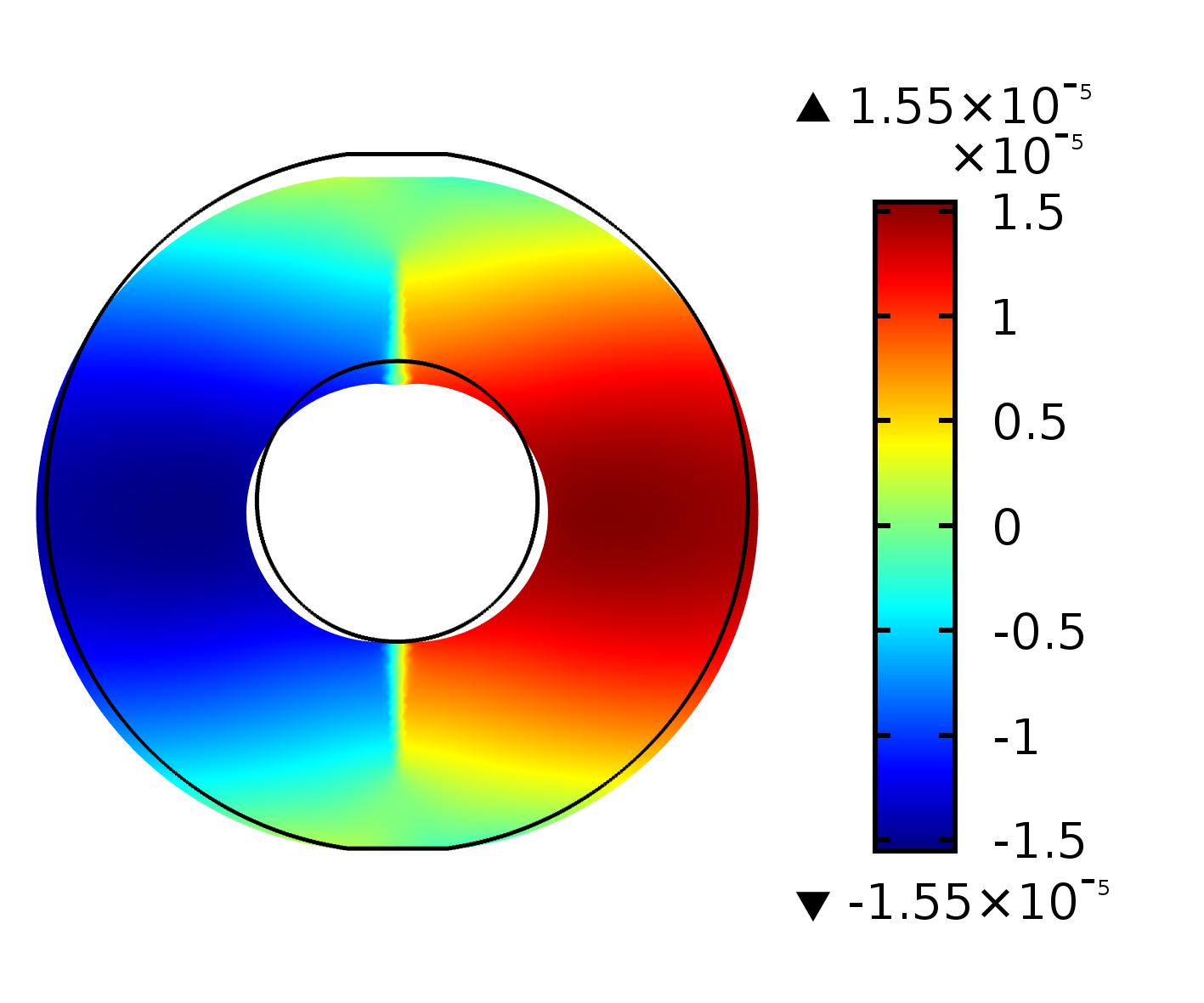}}
	\subfigure[e = 0.2]{\includegraphics[height = 5cm]{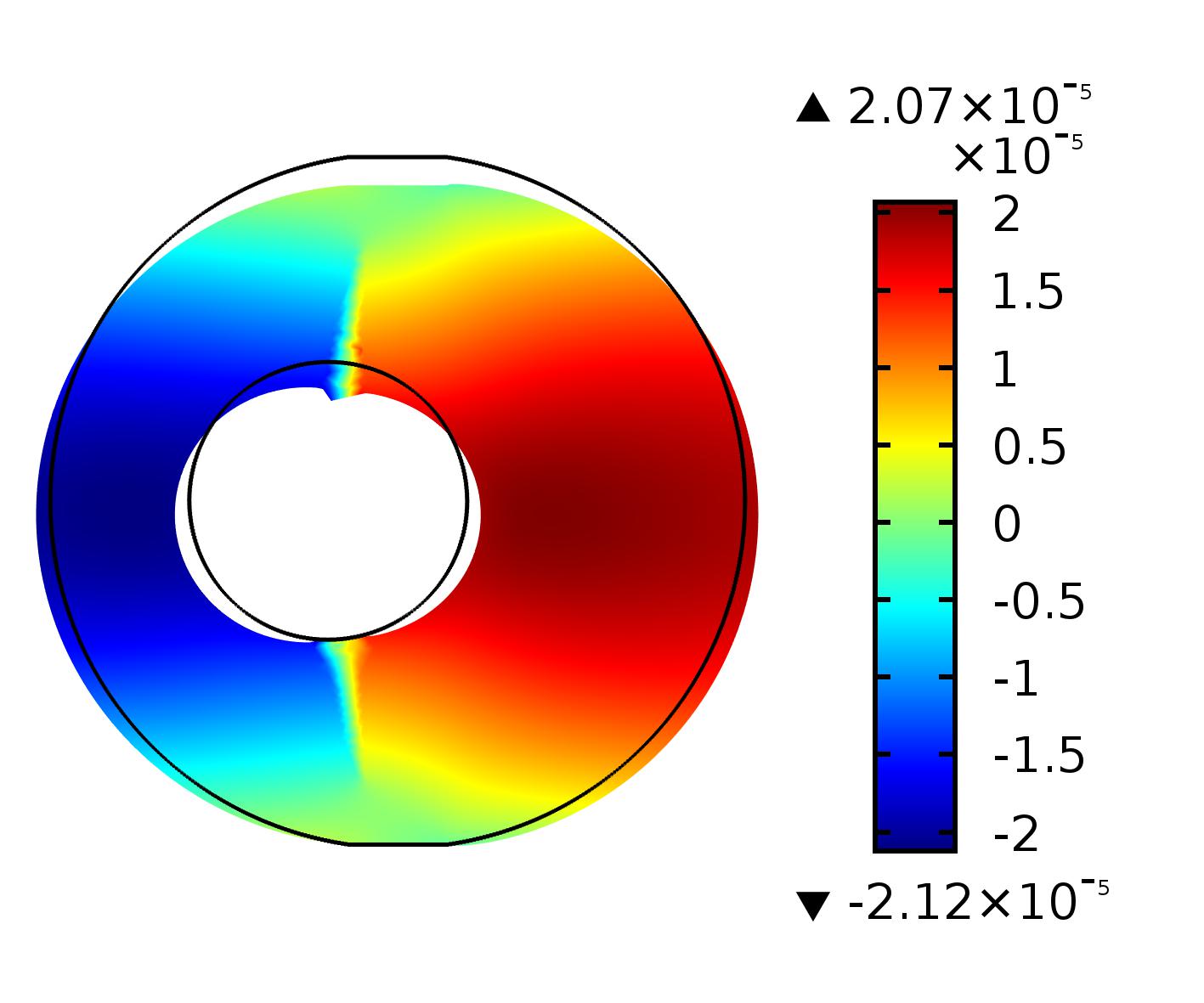}}\\
 	\subfigure[e = 0.4]{\includegraphics[height = 5cm]{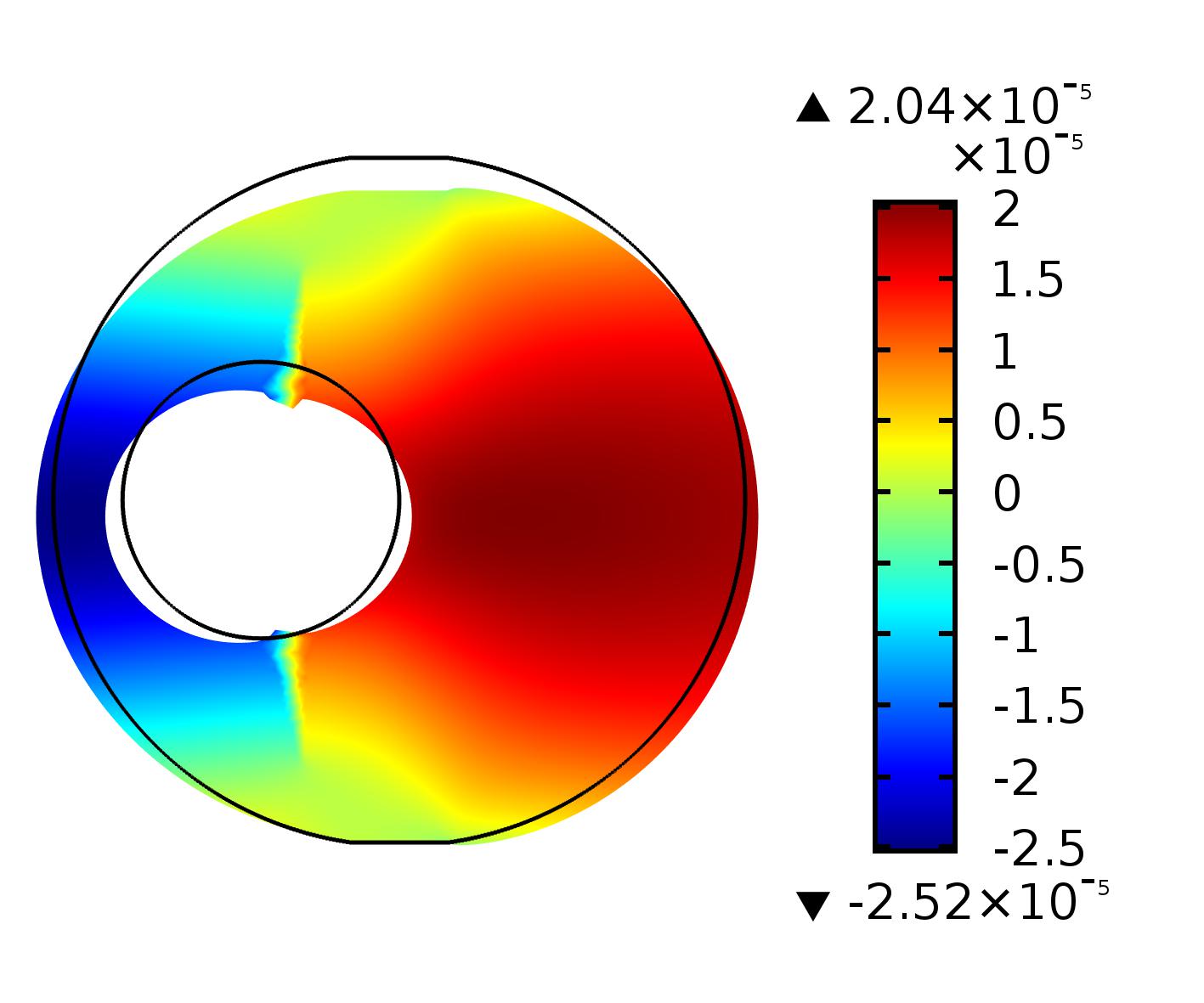}}
	\caption{Horizontal displacement of the Brazilian discs with a void along the horizontal axis}
	\label{Horizontal displacement of the Brazilian discs with a void along the horizontal axis}
	\end{figure}

Figure \ref{Stress-strain curves of the Brazilian discs with a void along the horizontal axis} shows the stress-strain curves of the Brazilian disc with a void in the horizontal axis under different eccentricities. The strength of the disc increases with an increasing eccentricity. Additionally, the overall stiffness of the disc increases with the increase in the eccentricity $e$. Figure \ref{Horizontal displacement of the Brazilian discs with a void along the horizontal axis} shows the horizontal displacement with different void eccentricities under the same vertical displacement increment. The maximum horizontal displacement increases as the eccentricity increases.

\subsection{Single void with varying eccentricities along the vertical axis}

\begin{figure}[H]
	\centering
	\subfigure[e = 0]{\includegraphics[height = 5cm]{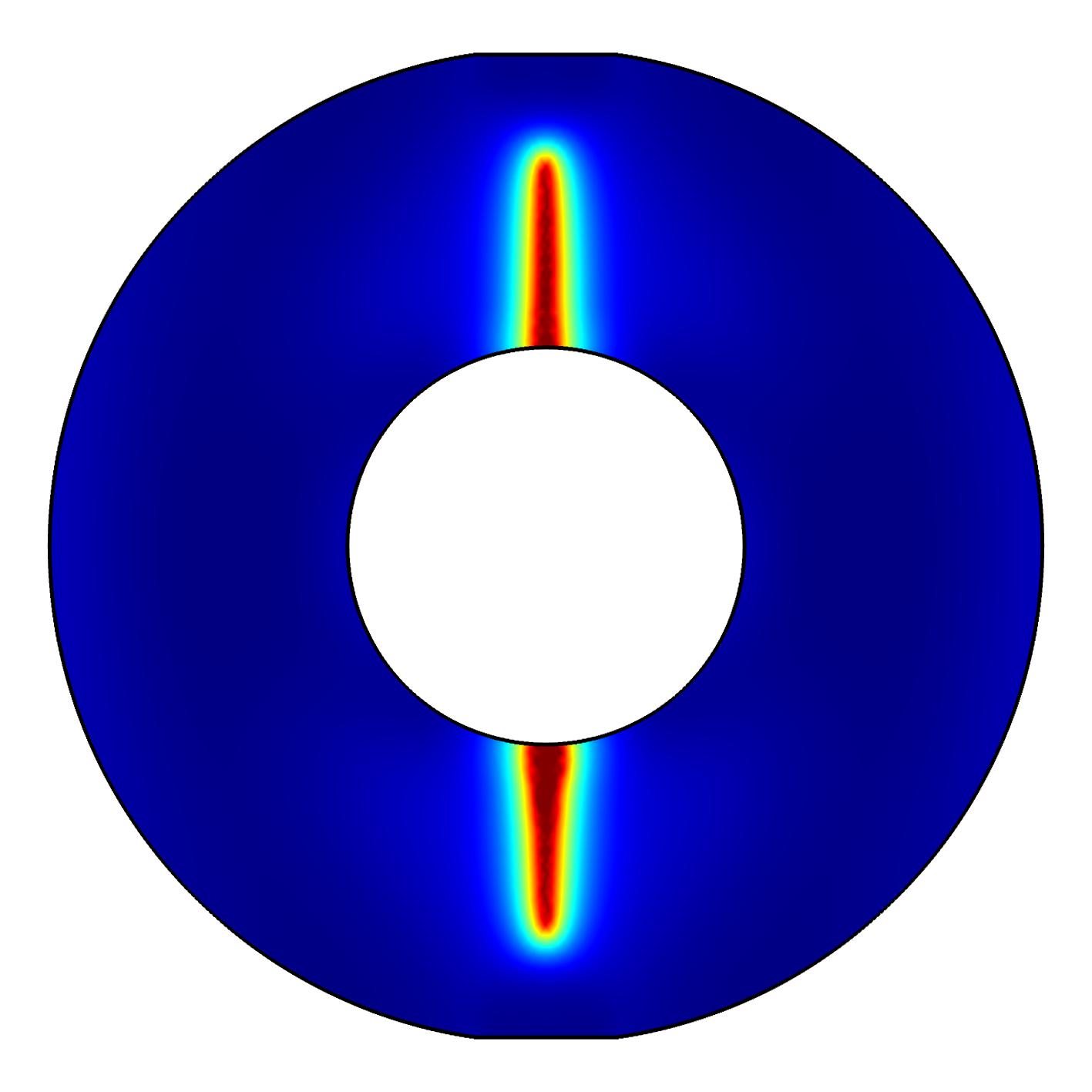}}
	\subfigure[e = 0.2]{\includegraphics[height = 5cm]{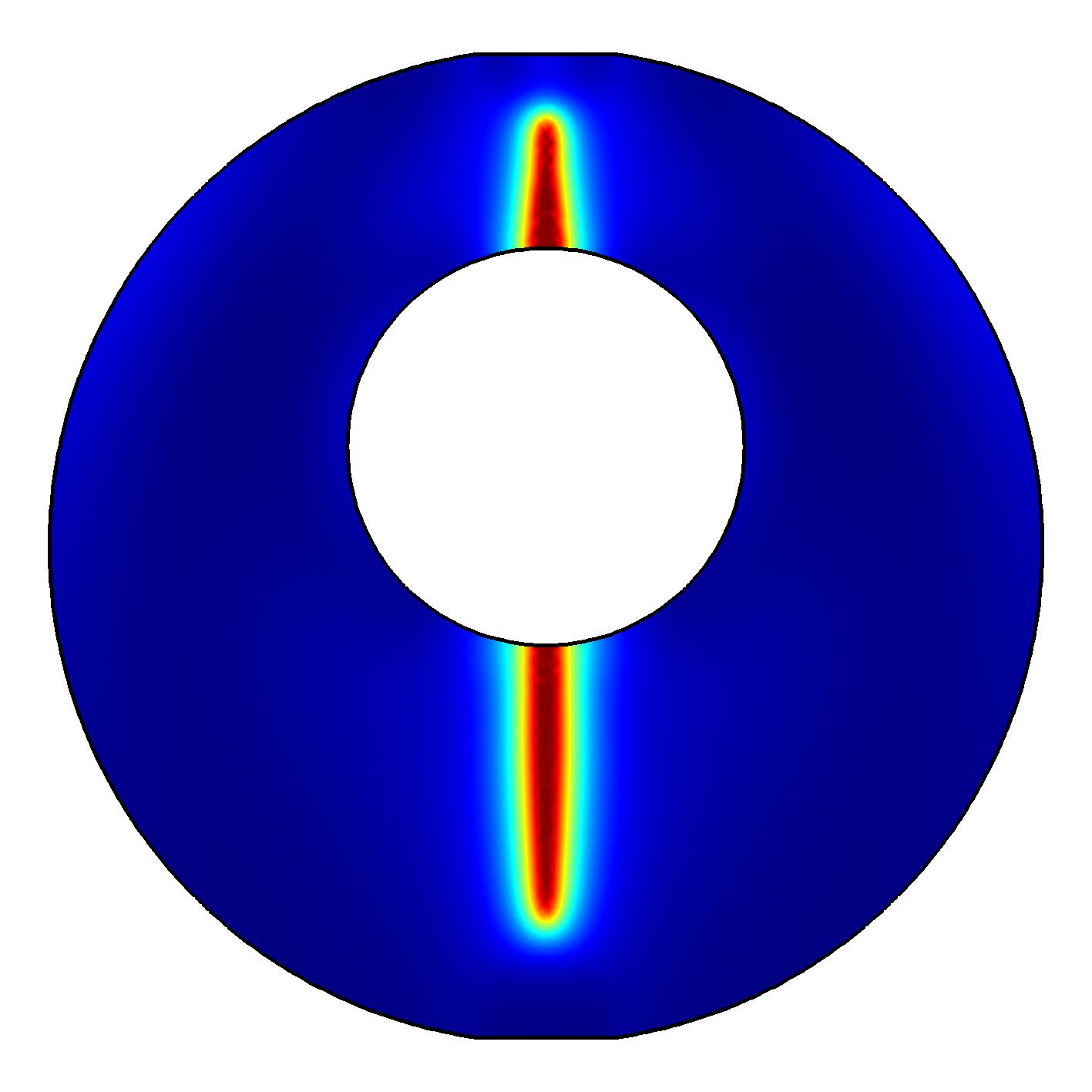}}
	\subfigure[e = 0.4]{\includegraphics[height = 5cm]{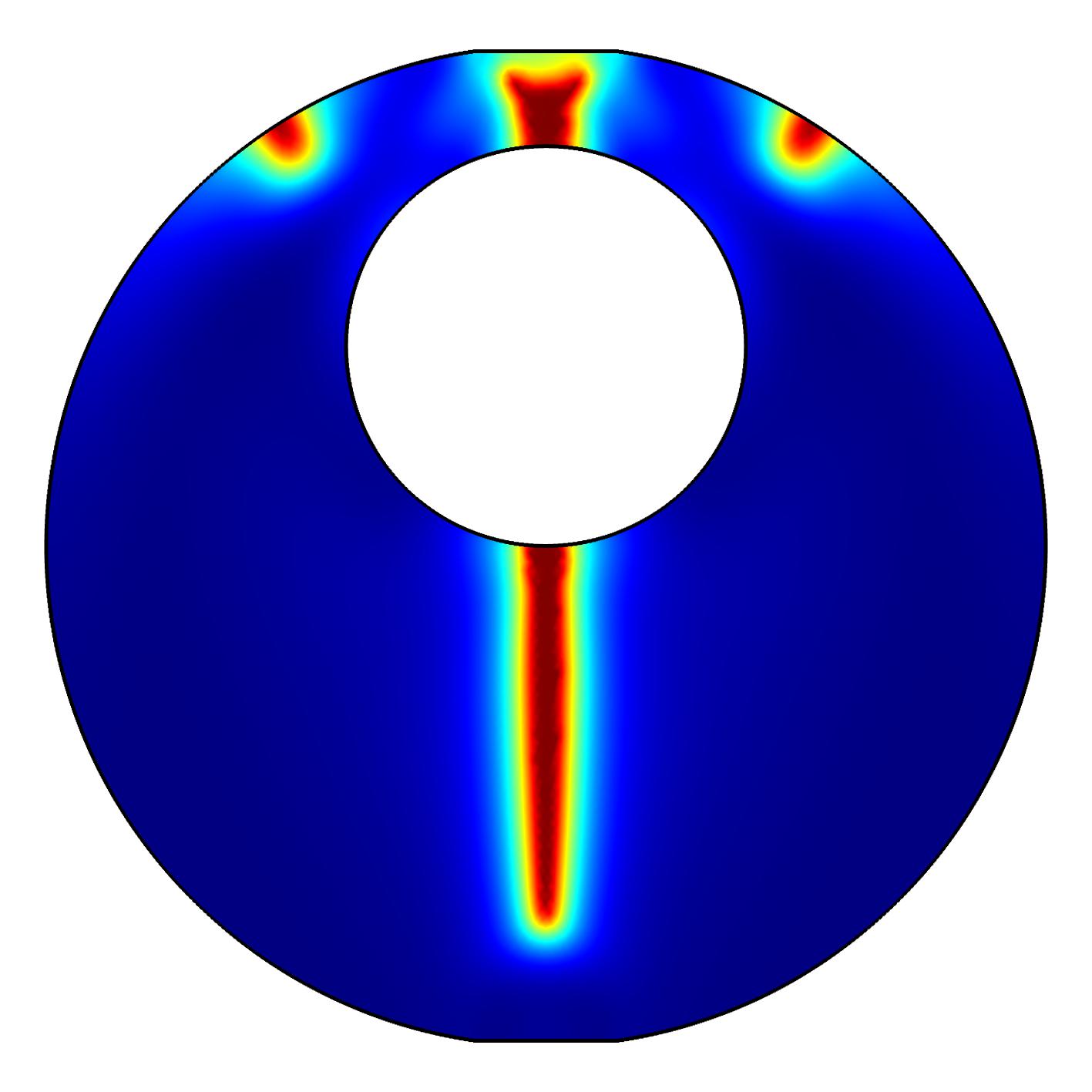}}
	\subfigure{\includegraphics[height = 4cm]{Figure2a_legend.jpg}}
	\caption{Final fracture patterns of the Brazilian discs with a void along the vertical axis}
	\label{Final fracture patterns of the Brazilian discs with a void along the vertical axis}
	\end{figure}

Figure \ref{Final fracture patterns of the Brazilian discs with a void along the vertical axis} presents the final fractures in the Brazilian discs with a vertically-arranged void under $e=0$, 0.2, and 0.4. Different from the case of a void along the horizontal axis, bilateral symmetric results are presented for a void along the vertical axis. The dominated fractures initiate from the top and bottom of the void and propagate vertically towards the two ends of the disc. As the eccentricity $e$ increases, some secondary fractures occur.

\begin{figure}[htbp]
	\centering
	\includegraphics[width = 8cm]{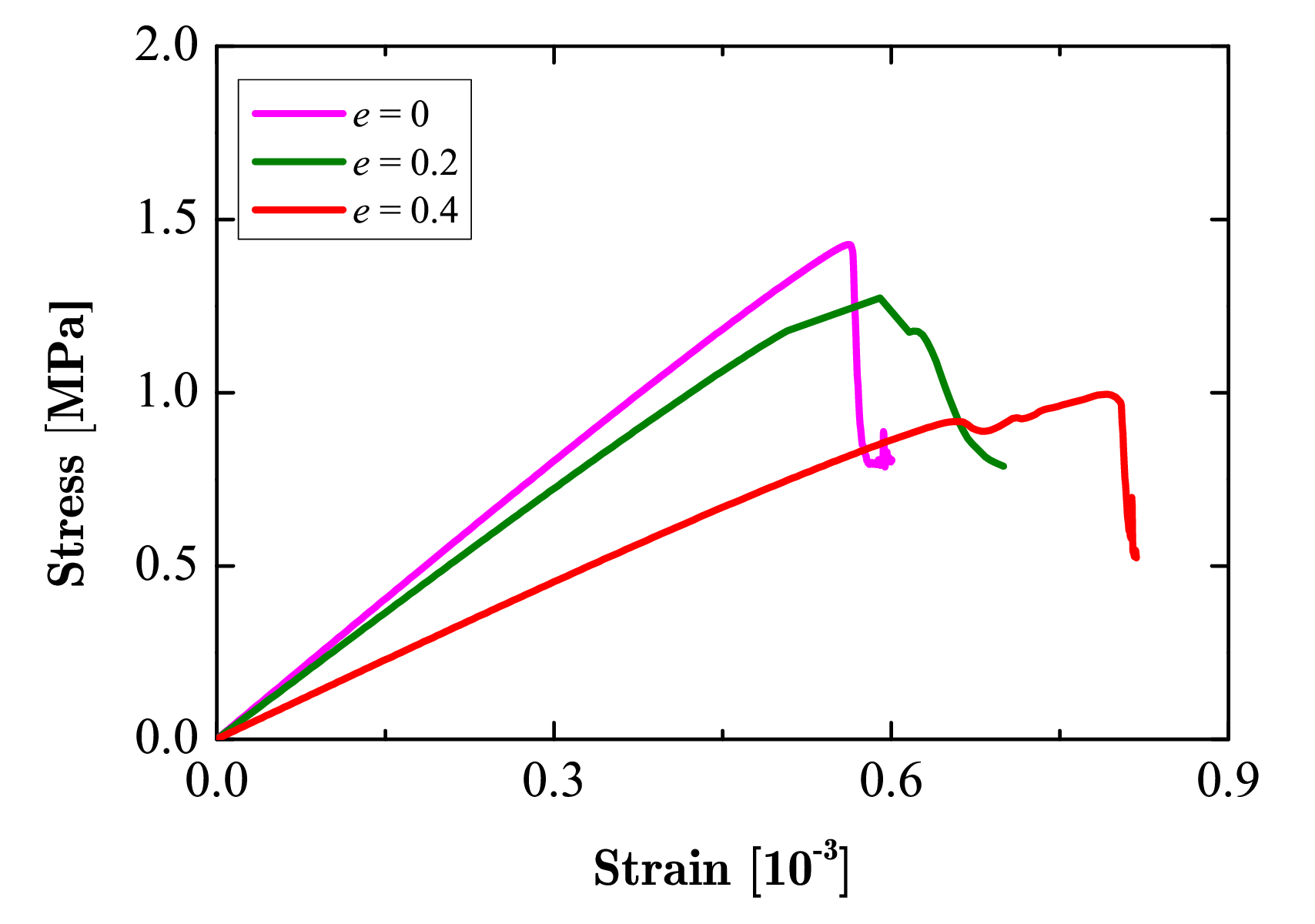}
	\caption{Stress-strain curves of the Brazilian discs with a void along the vertical axis}
	\label{Stress-strain curves of the Brazilian discs with a void along the vertical axis}
	\end{figure}

\begin{figure}[htbp]
	\centering
	\subfigure[e = 0]{\includegraphics[height = 5cm]{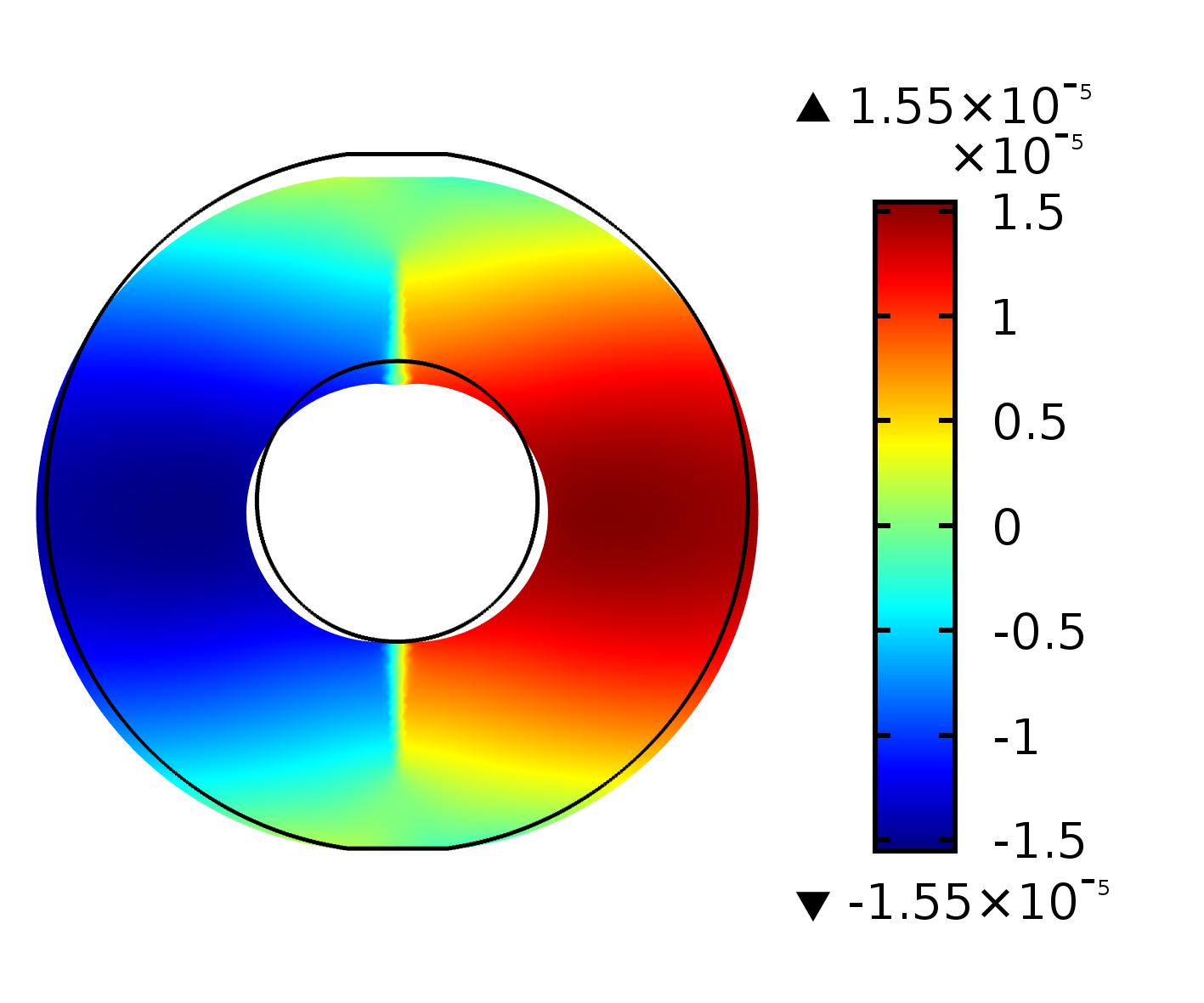}}
	\subfigure[e = 0.2]{\includegraphics[height = 5cm]{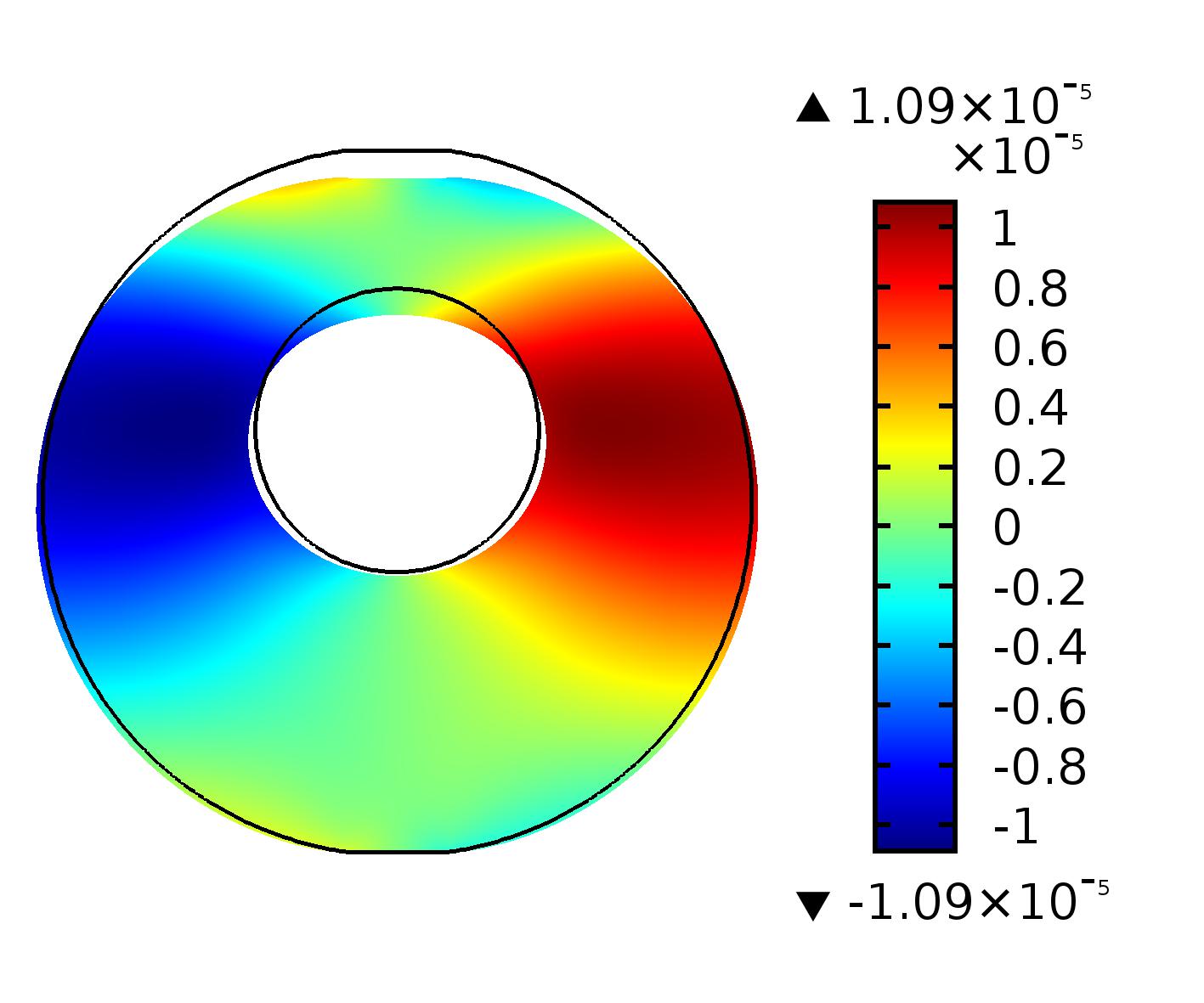}}\\
 	\subfigure[e = 0.4]{\includegraphics[height = 5cm]{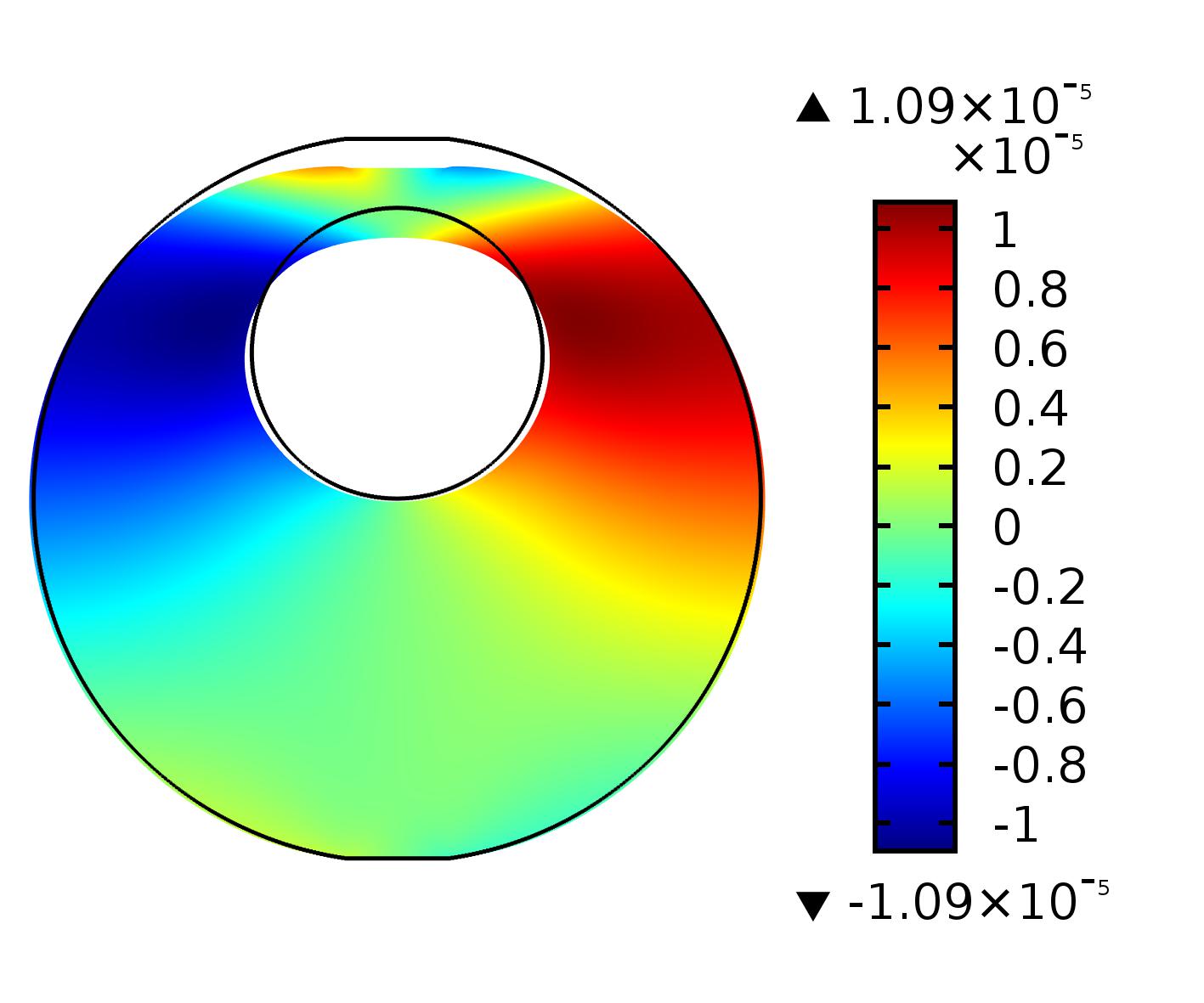}}
	\caption{Horizontal displacement of the Brazilian discs with a void along the vertical axis}
	\label{Horizontal displacement of the Brazilian discs with a void along the vertical axis}
	\end{figure}

Figure \ref{Stress-strain curves of the Brazilian discs with a void along the vertical axis} shows the stress-strain curves of the Brazilian disc with a void along the vertical axis under different eccentricities. The curves vary significantly with an increasing eccentricity. The reason is that a large eccentricity along $y$ axis makes the void closer to the upper loading point, further affecting stability and strength of the disc. The strength of the disc decreases with an increasing eccentricity. Additionally, the overall stiffness of the disc increases with the increase in the eccentricity $e$. Figure \ref{Horizontal displacement of the Brazilian discs with a void along the vertical axis} shows the horizontal displacement with different void eccentricities along the vertical axis. The maximum horizontal displacement decreases as the eccentricity increases because the fracture becomes closer to the disc end.

\subsection{Effects of multiple voids on the fracture patterns}

\begin{figure}[htbp]
	\centering
	\subfigure[$n = 1$]{\includegraphics[height = 5cm]{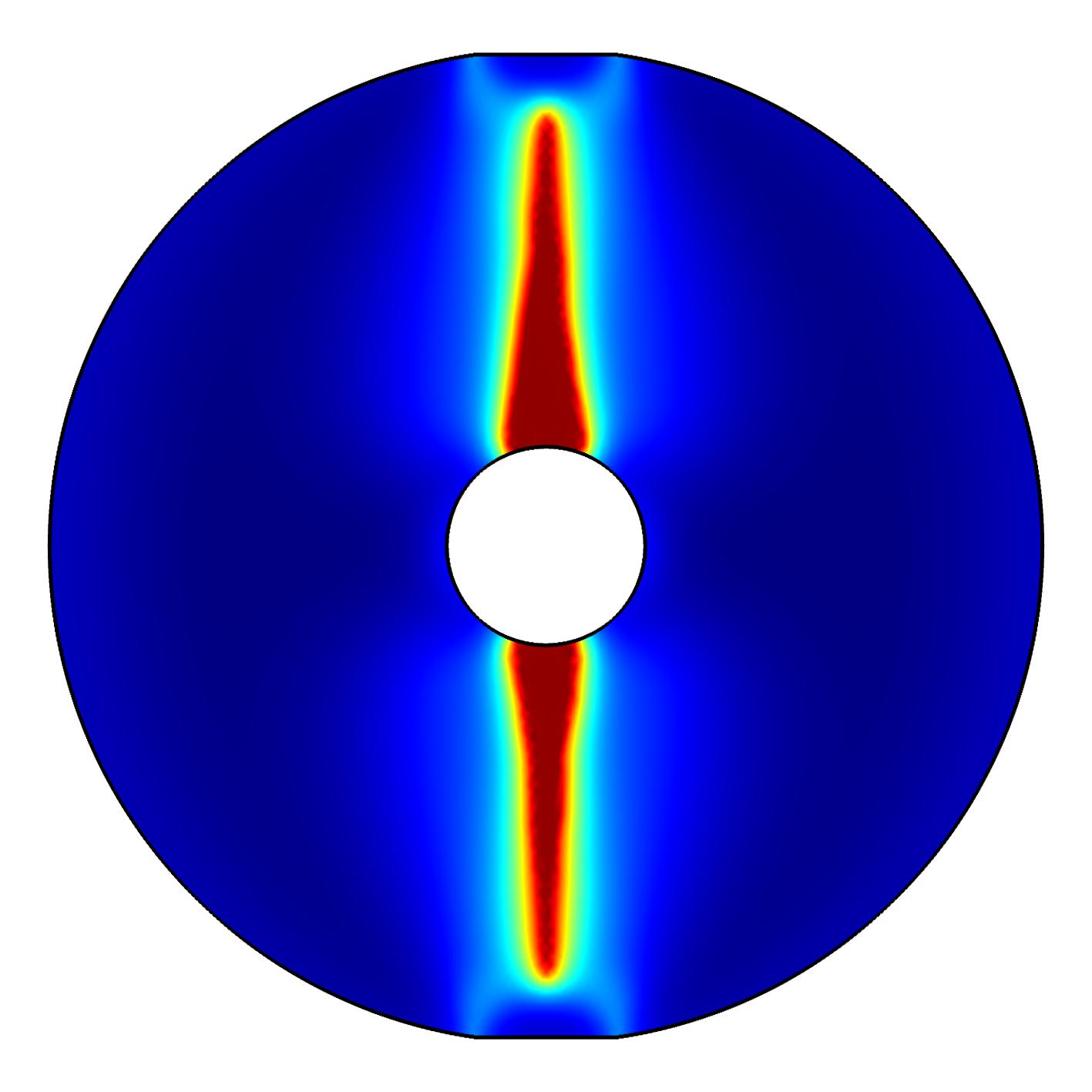}}
	\subfigure[$n = 2$, horizontal]{\includegraphics[height = 5cm]{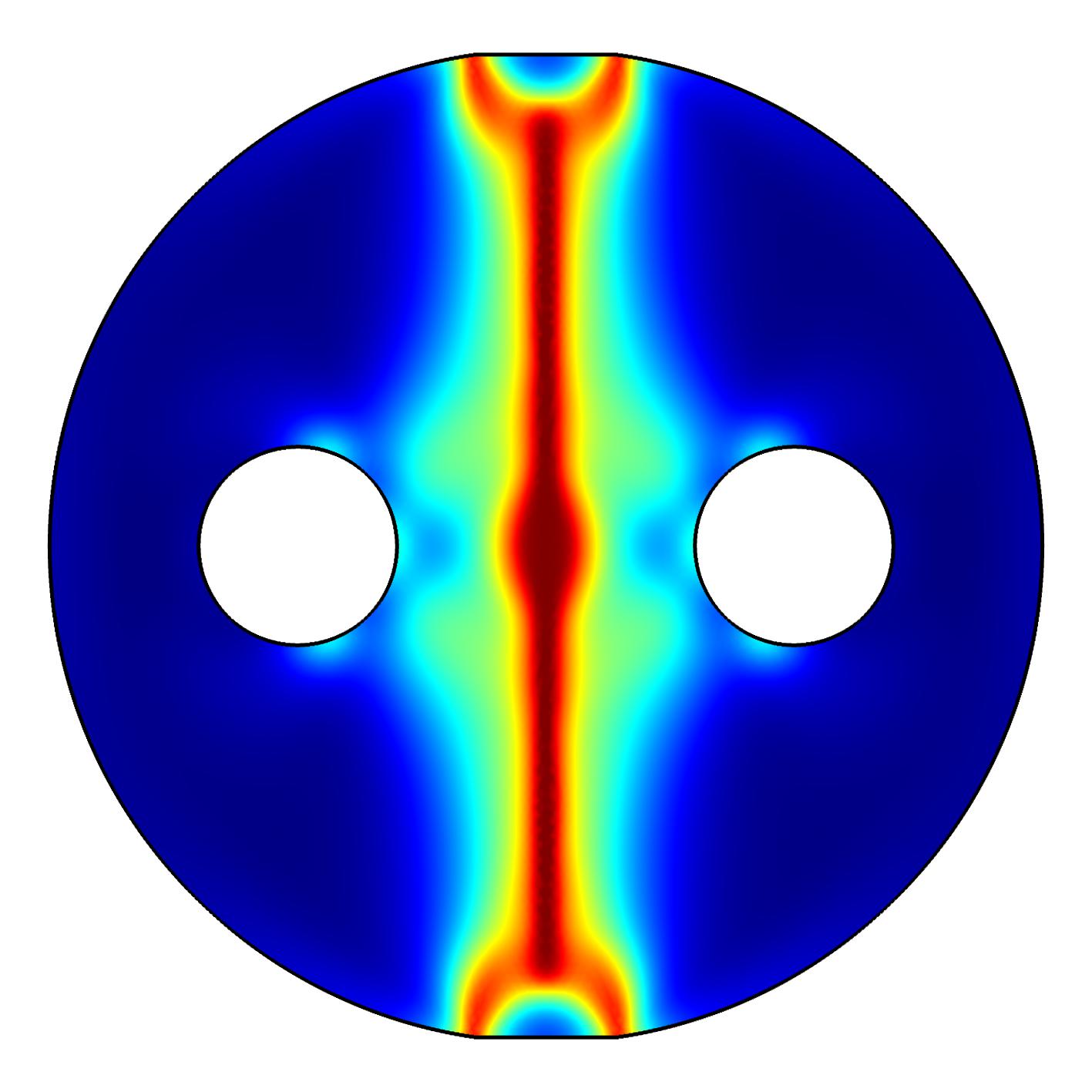}}
	\subfigure[$n = 2$, vertical]{\includegraphics[height = 5cm]{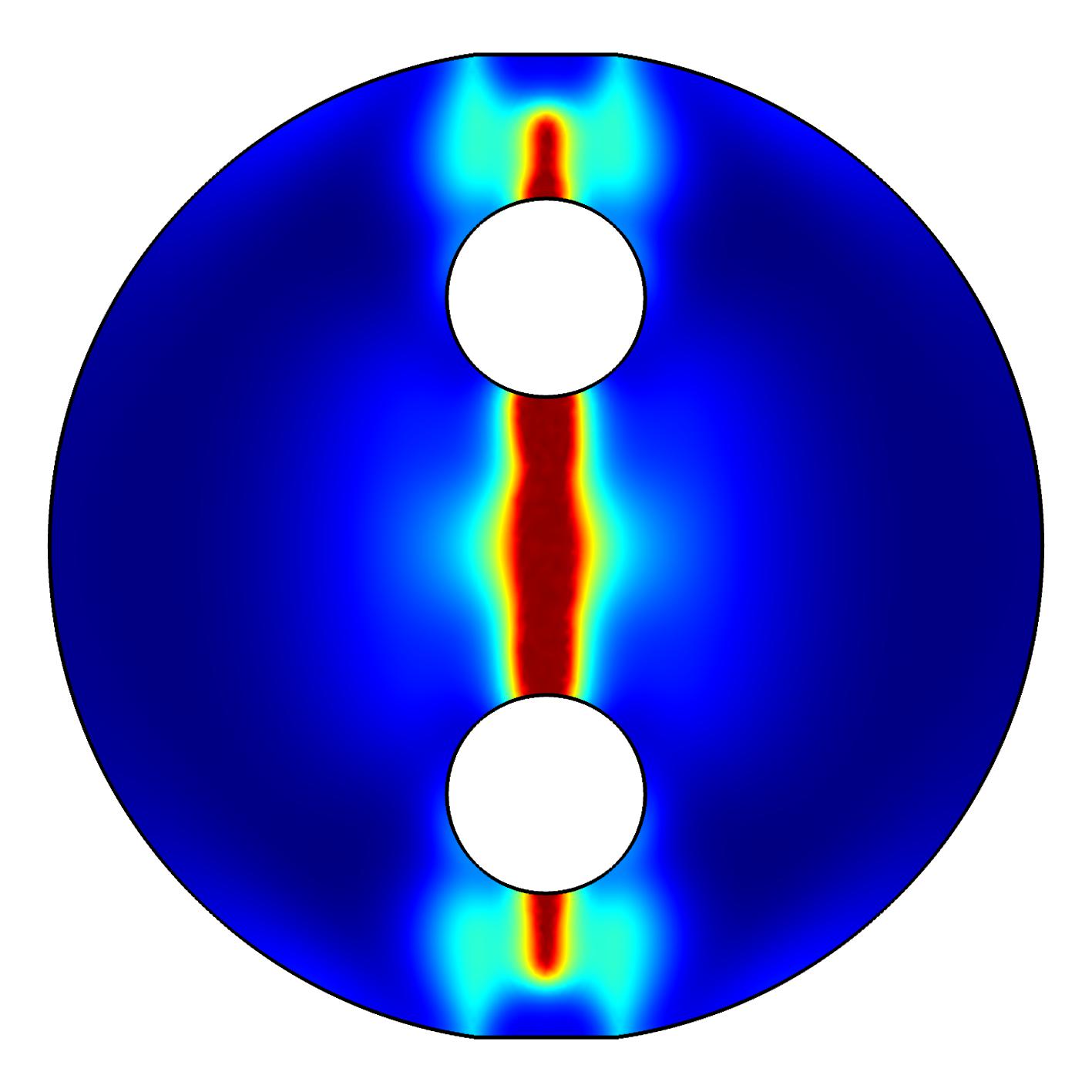}}\\
	\subfigure[$n = 3$]{\includegraphics[height = 5cm]{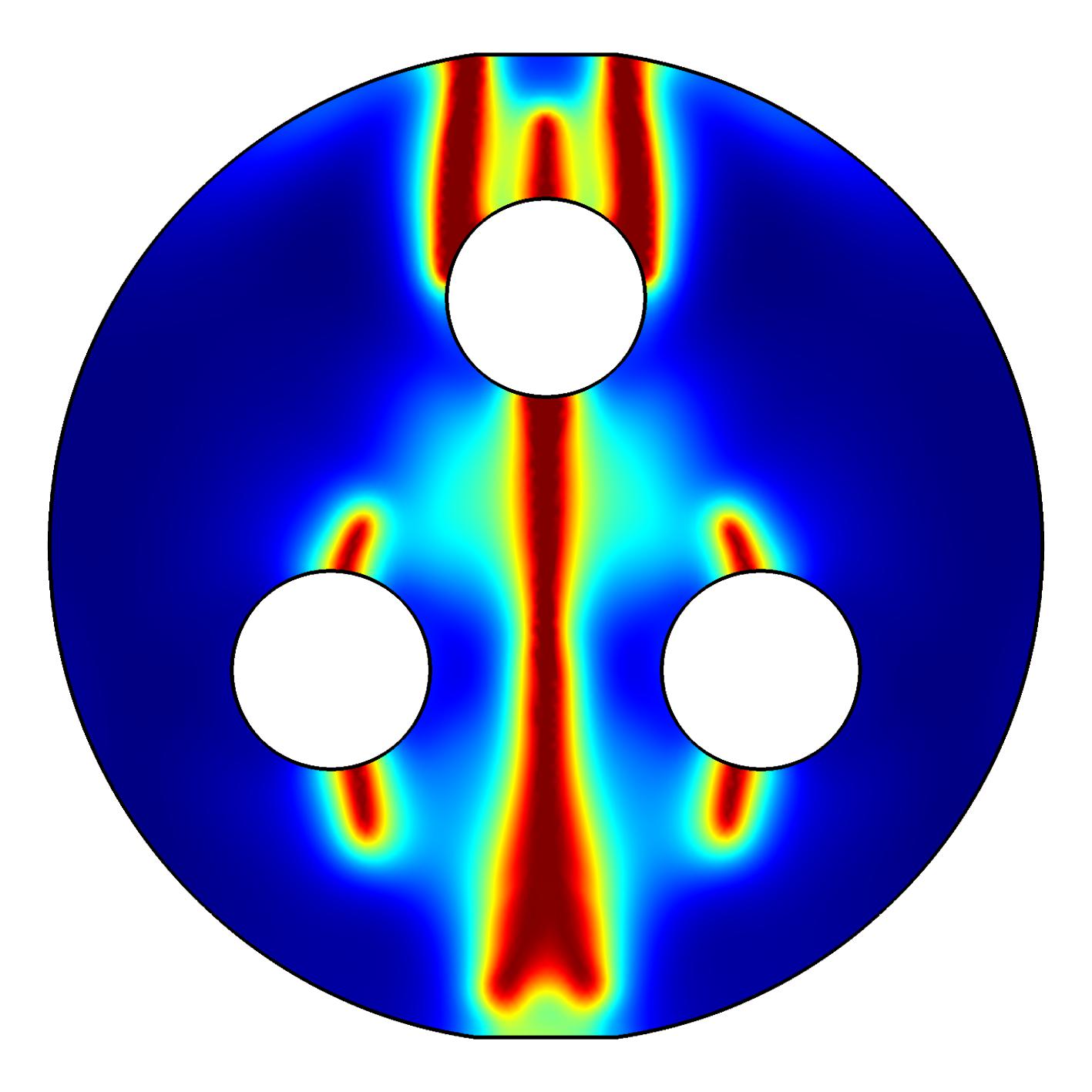}}
	\subfigure[$n = 4$, square]{\includegraphics[height = 5cm]{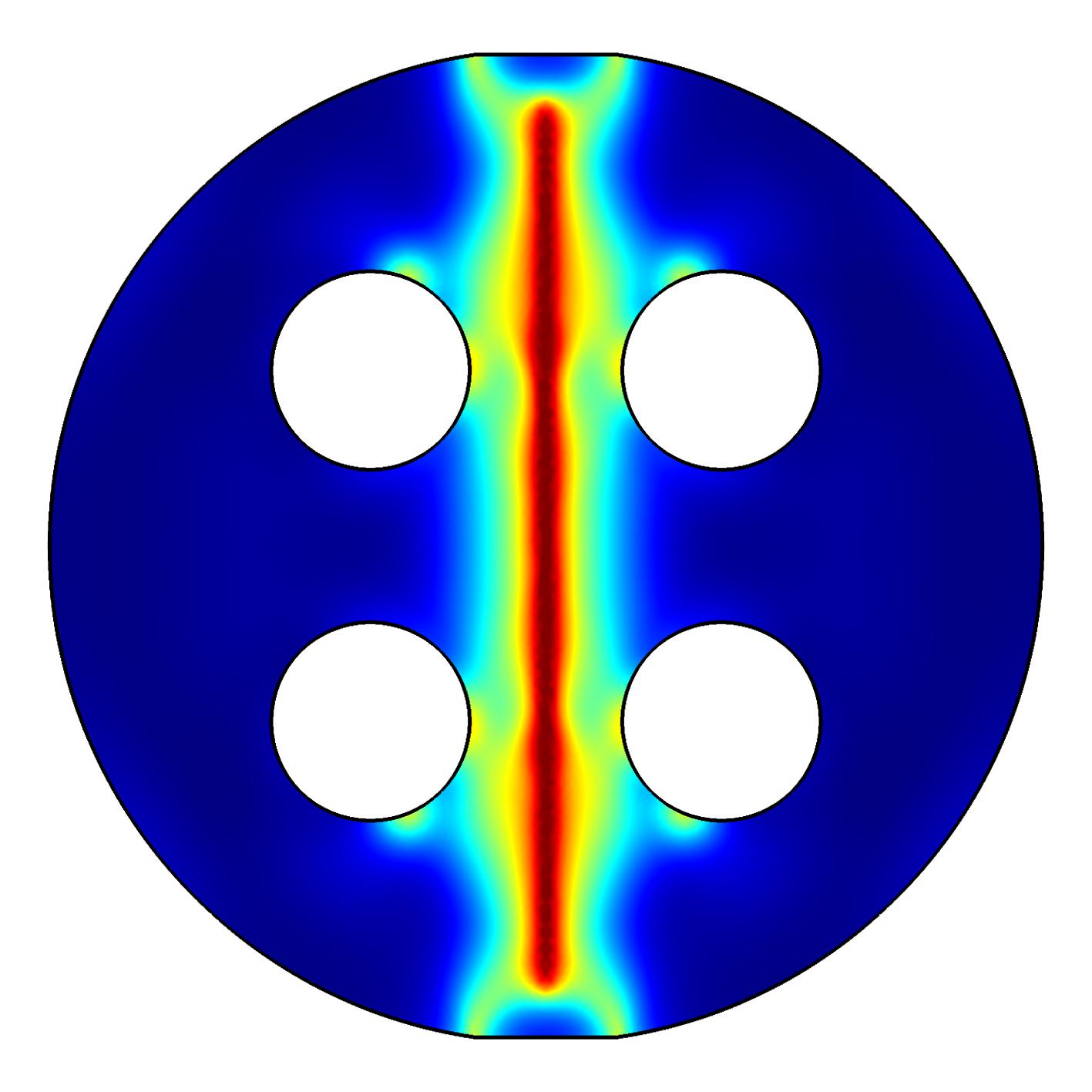}}
	\subfigure[$n = 4$, diamond]{\includegraphics[height = 5cm]{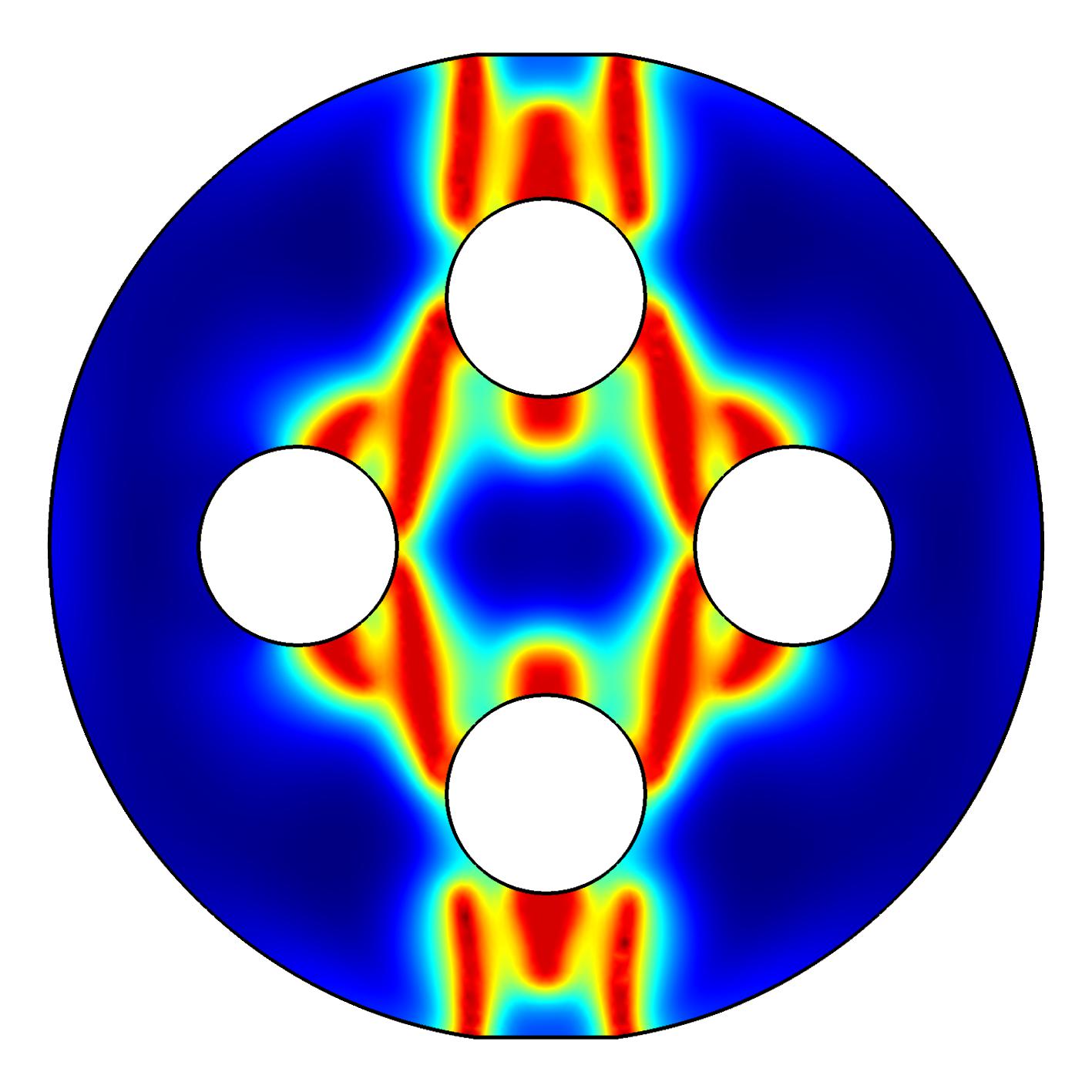}}
	\subfigure{\includegraphics[height = 4cm]{Figure2a_legend.jpg}}
	\caption{Final fracture patterns of the Brazilian discs with multiple voids}
	\label{Final fracture patterns of the Brazilian discs with multiple voids}
	\end{figure}

\begin{figure}[htbp]
	\centering
	\includegraphics[width = 8cm]{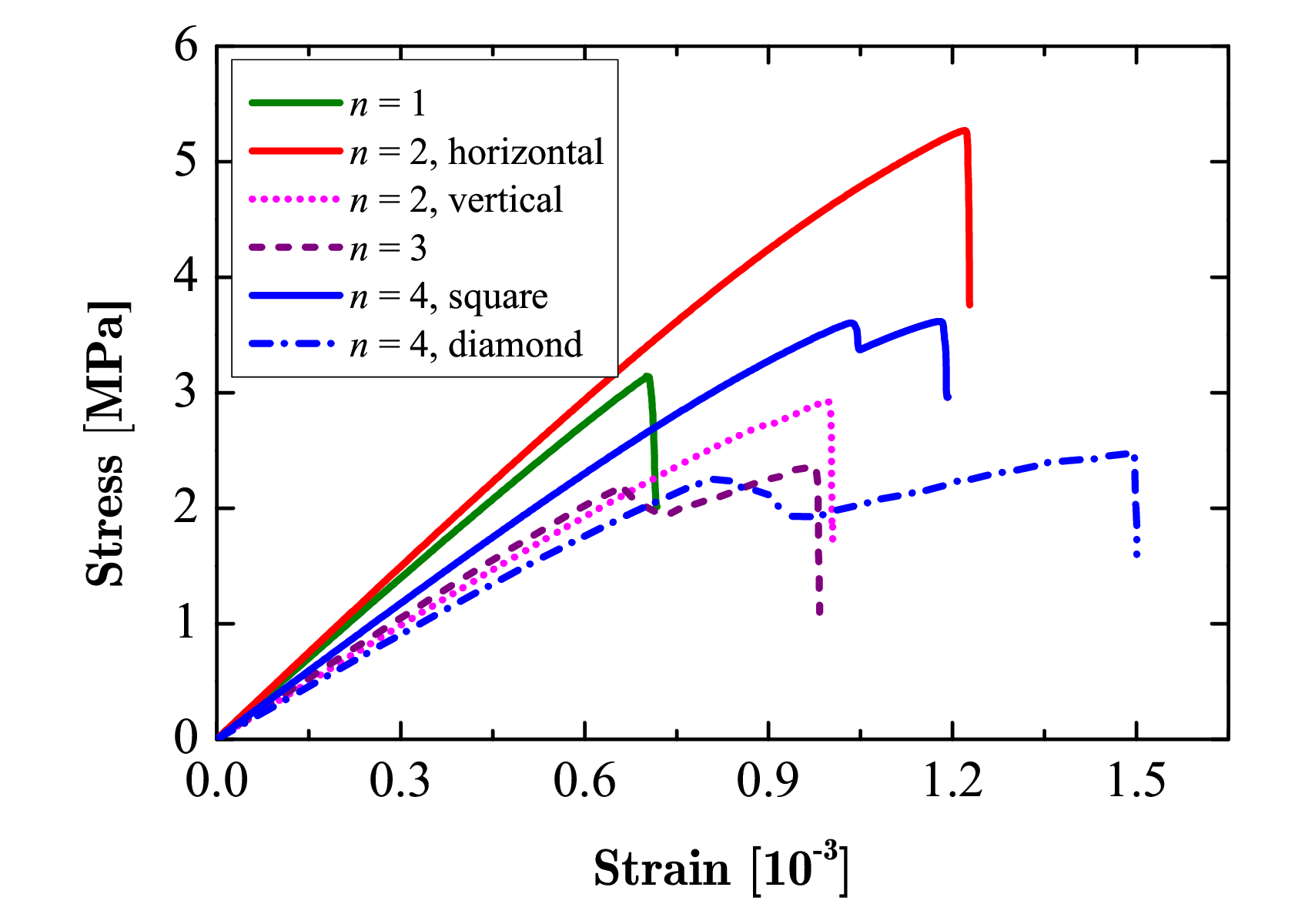}
	\caption{Stress-strain curves of the Brazilian discs with multiple voids}
	\label{Stress-strain curves of the Brazilian discs with multiple voids}
	\end{figure}

\begin{figure}[htbp]
	\centering
	\subfigure[$n = 1$]{\includegraphics[height = 5cm]{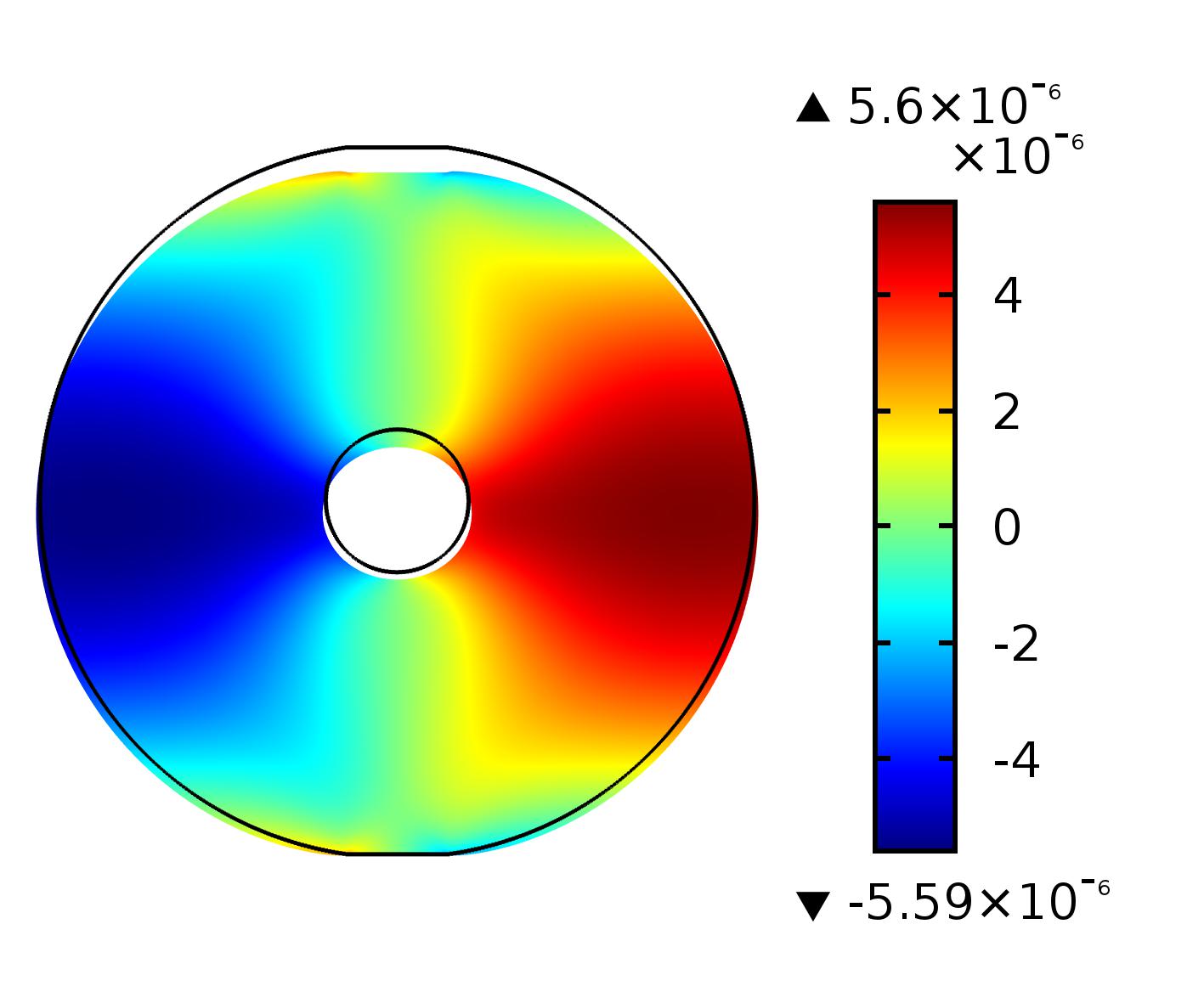}}
	\subfigure[$n = 2$, horizontal]{\includegraphics[height = 5cm]{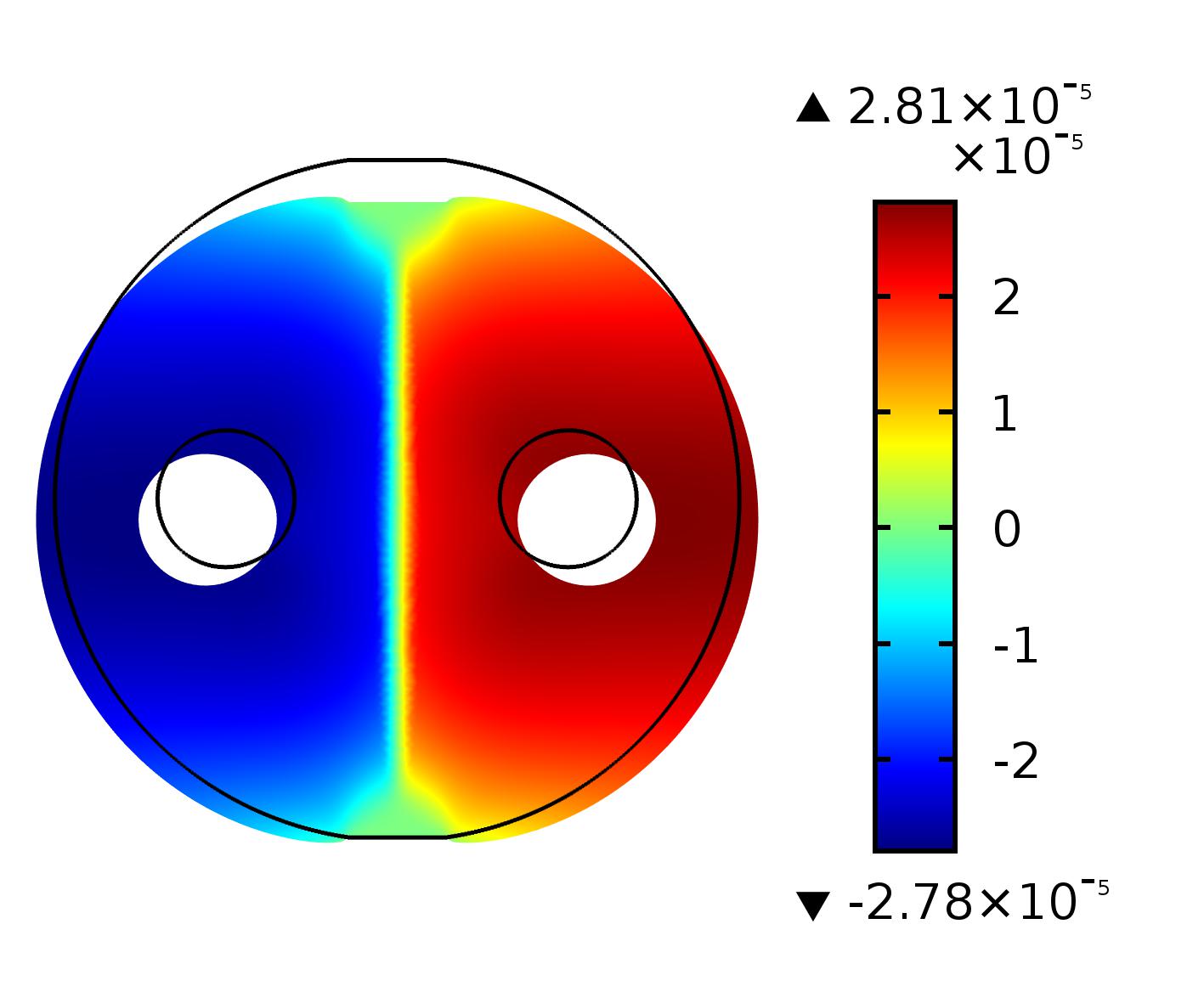}}\\
	\subfigure[$n = 2$, vertical]{\includegraphics[height = 5cm]{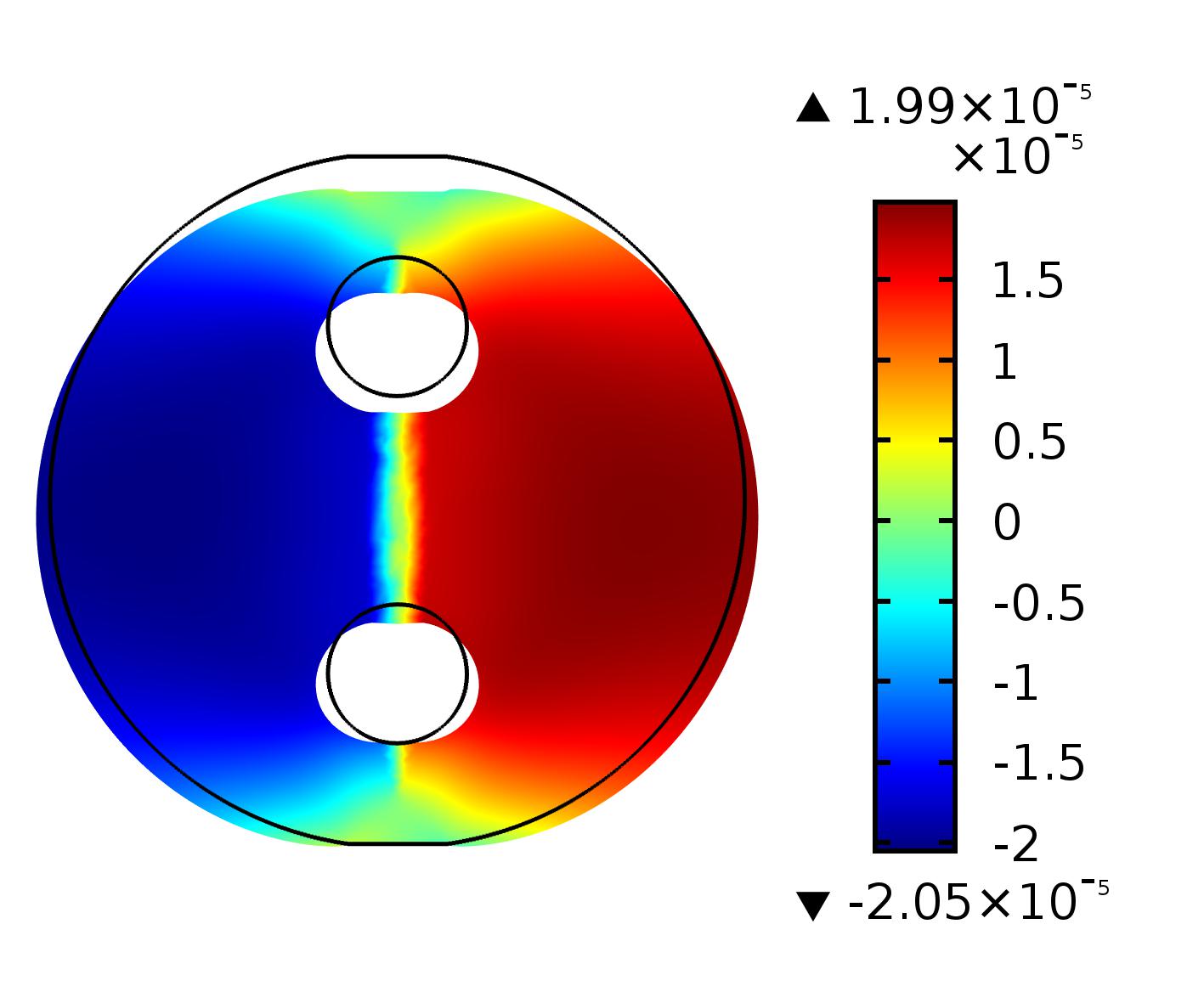}}
	\subfigure[$n = 3$]{\includegraphics[height = 5cm]{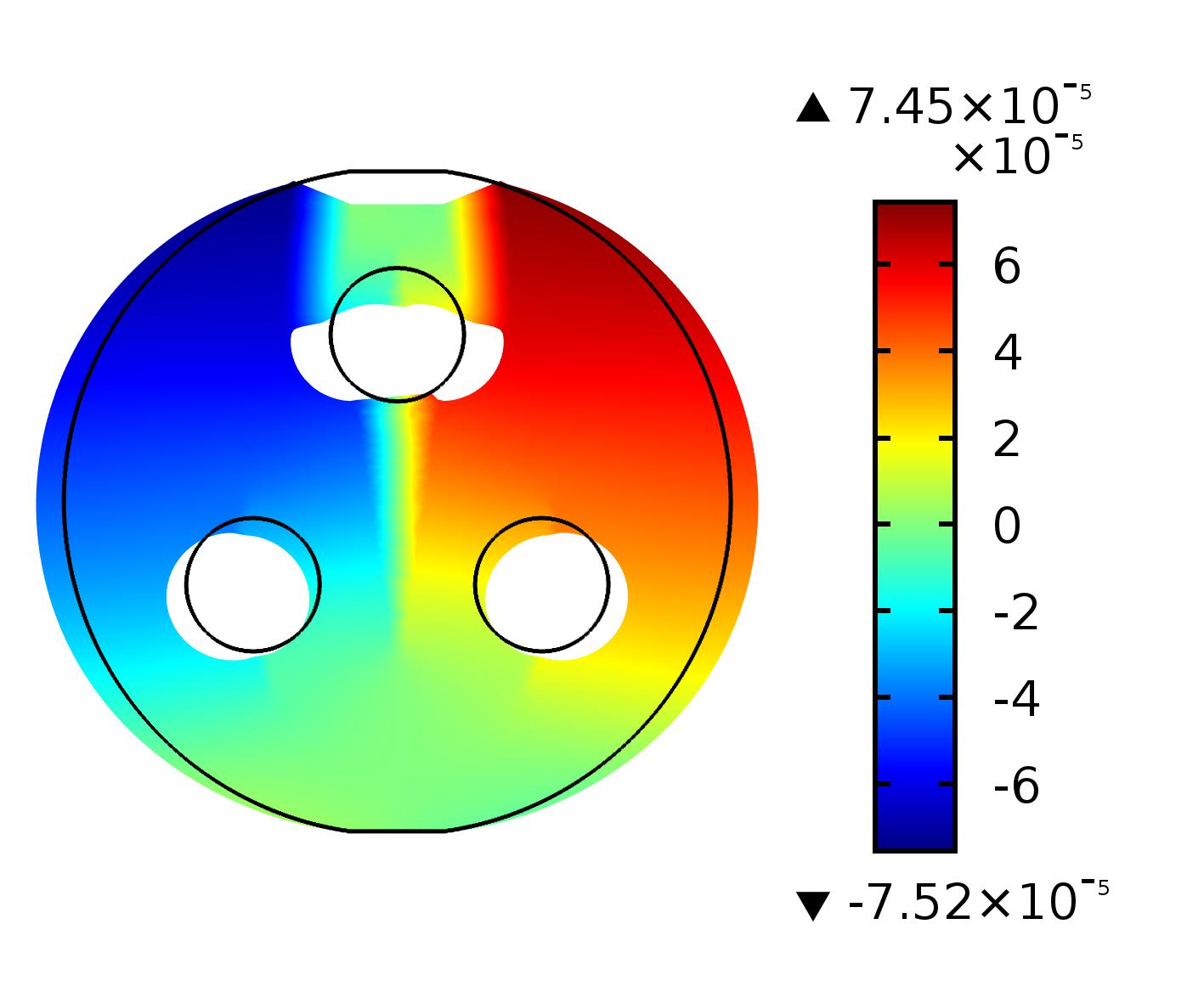}}\\
	\subfigure[$n = 4$, square]{\includegraphics[height = 5cm]{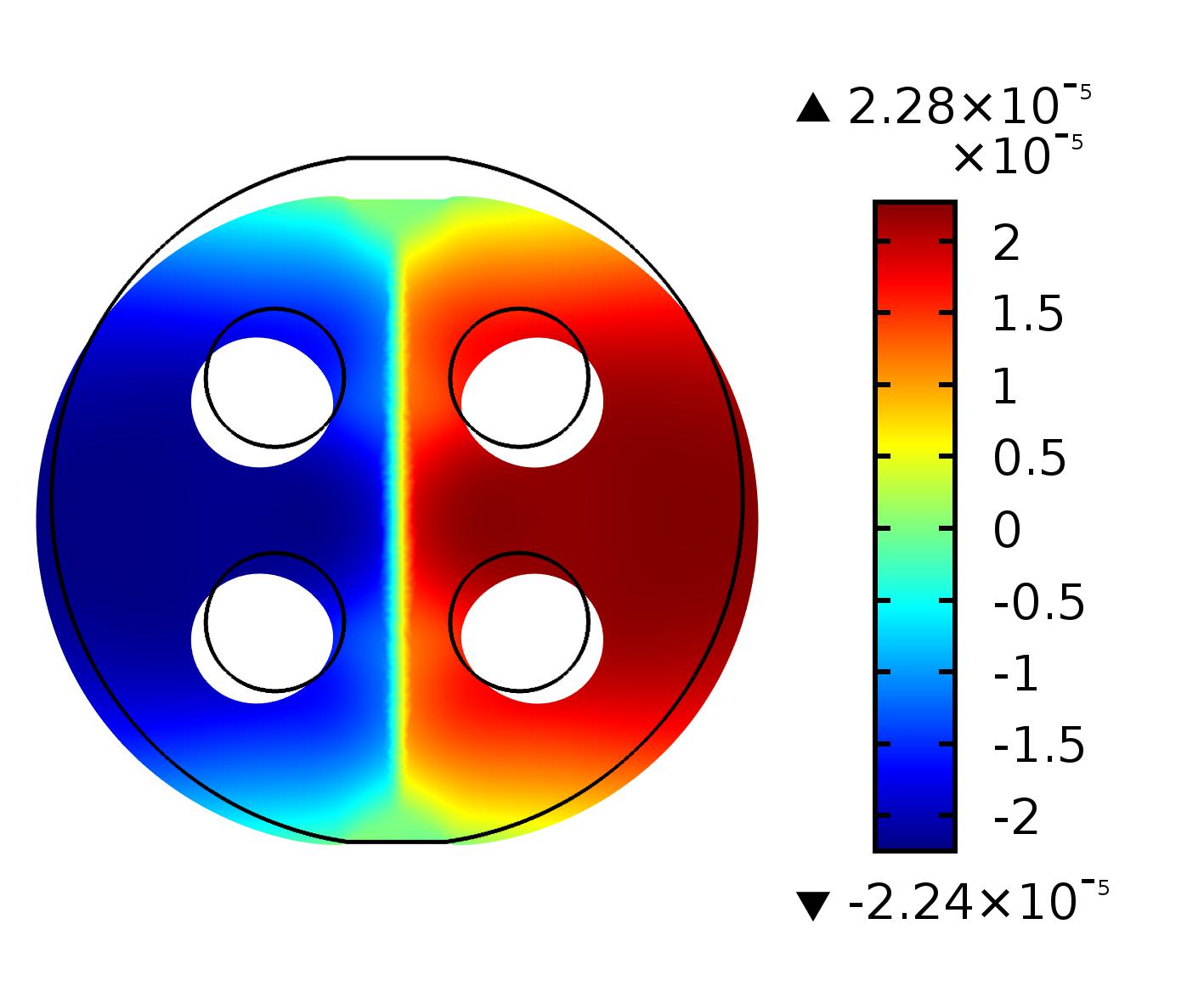}}
	\subfigure[$n = 4$, diamond]{\includegraphics[height = 5cm]{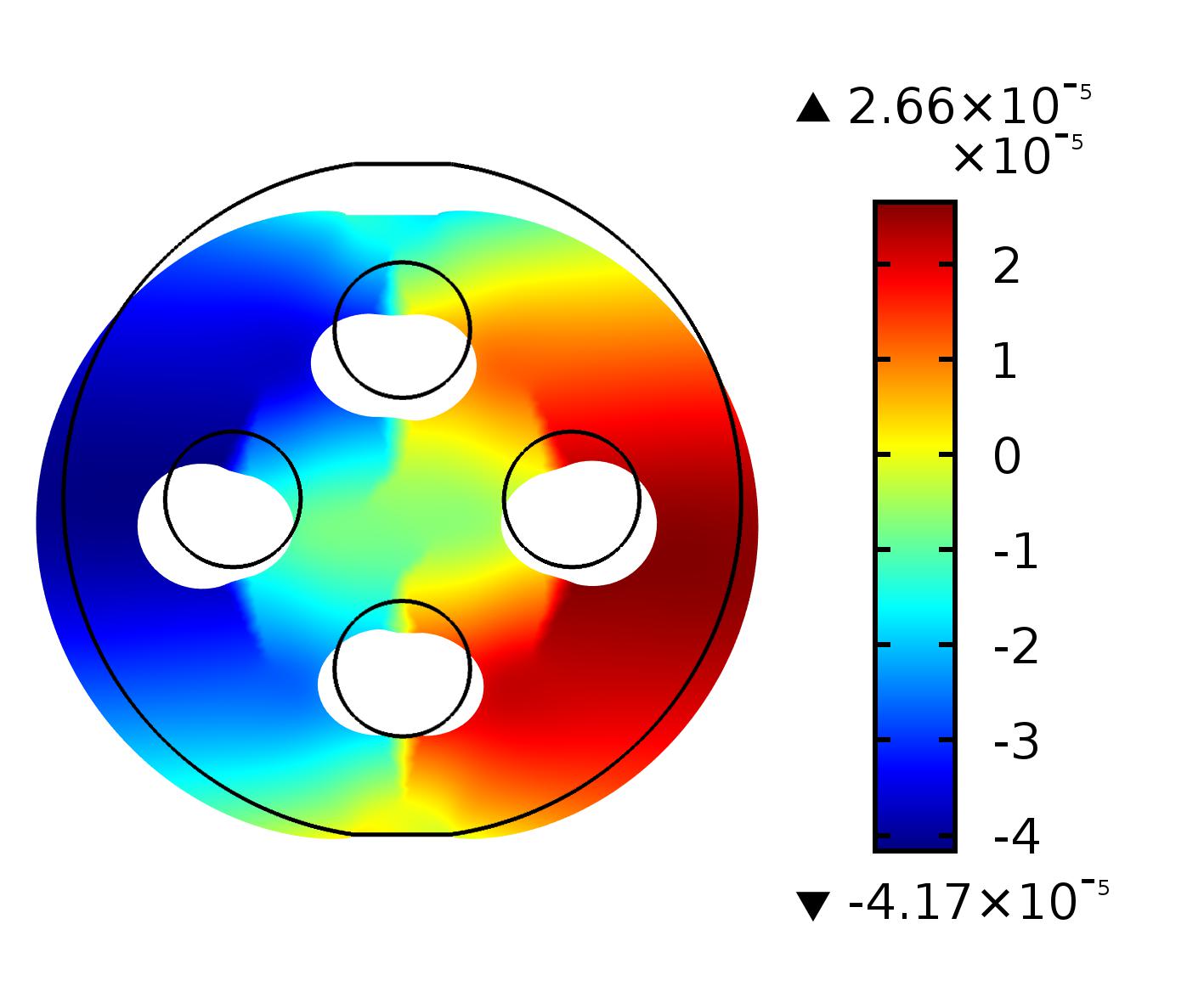}}
	\caption{Horizontal displacement of the Brazilian discs with multiple voids}
	\label{Horizontal displacement of the Brazilian discs with multiple voids}
	\end{figure}

In this subsection, we examine the effect of void number on the fracture patterns in the Brazilian disc. Note that the diameter of the void is fixed to 10 mm in the following analyses. Figure \ref{Final fracture patterns of the Brazilian discs with multiple voids} shows the final fracture patterns with different number of voids in the Brazilian disc. Complex fracture patterns are observed. Fractures initiate from the boundaries of the voids and connect to the disc ends. In addition to dominated fractures, some secondary fractures also occur when the number of voids increases. Figure \ref{Stress-strain curves of the Brazilian discs with multiple voids} shows the stress-strain curves with different number of voids in the Brazilian disc. Both the number and relative position of the voids influence the stress-strain responses. In general, if the voids are closer to the disc ends or there are more voids in the vertical direction, the overall strength of the Brazilian disc will decrease. In addition, a more complex horizontal displacement field is shown in Fig. \ref{Horizontal displacement of the Brazilian discs with multiple voids}.

\section {Fracture patterns in Brazilian discs with filled inclusions} \label{Fracture patterns in Brazilian discs with inclusions}

This section describes the fractures pattern in Brazilian discs with a circular inclusion using the phase field method. The parameters for the examples are identical to those in the previous section. The filling materials of the inclusions are assigned with different Young's moduli compared to the rest of the disc. Two sets of  Young's moduli of the inclusion namely $E_i=15$ GPa and 60 GPa, whereas the former is softer and the latter is harder than the matrix material, are used in the testing examples to show the effect of inclusions on the fracture patterns in the Brazilian discs. 

\subsection{Brazilian discs with a center inclusion}

\begin{figure}[H]
	\centering
	\subfigure[$E_i$ = 15 GPa, D = 10 mm]{\includegraphics[height = 5cm]{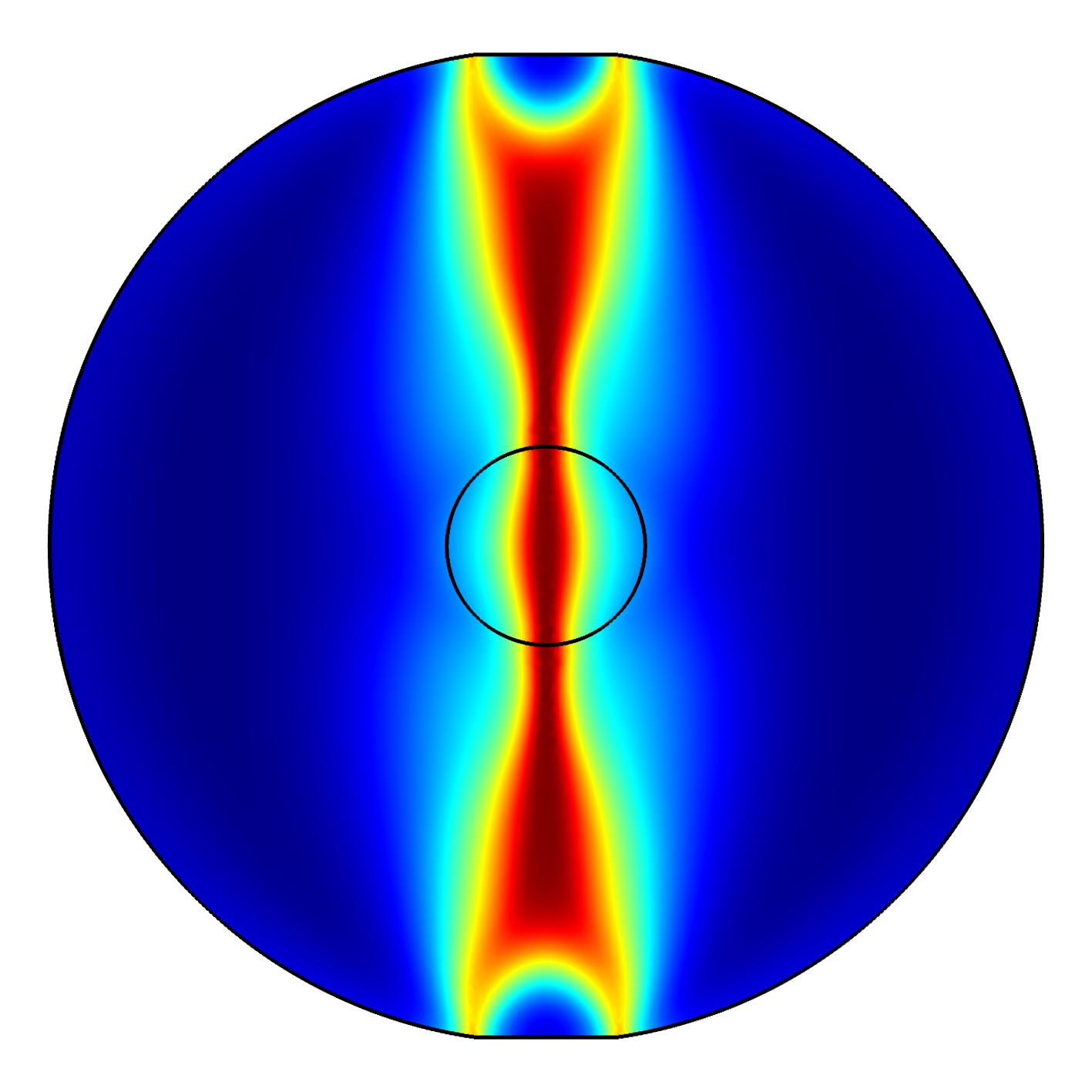}}
 	\subfigure[$E_i$ = 15 GPa, D = 20 mm]{\includegraphics[height = 5cm]{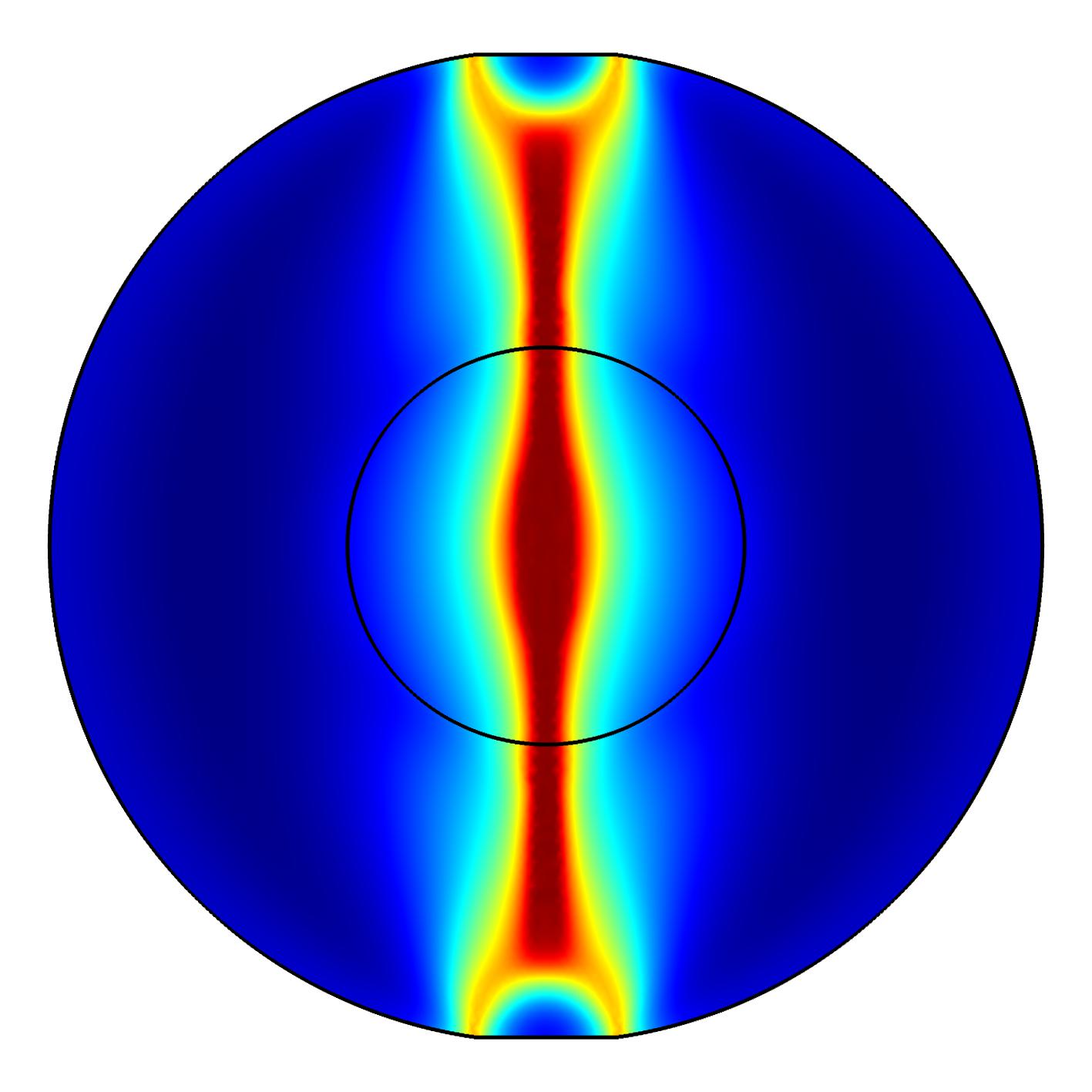}}
	\subfigure[$E_i$ = 15 GPa, D = 30 mm]{\includegraphics[height = 5cm]{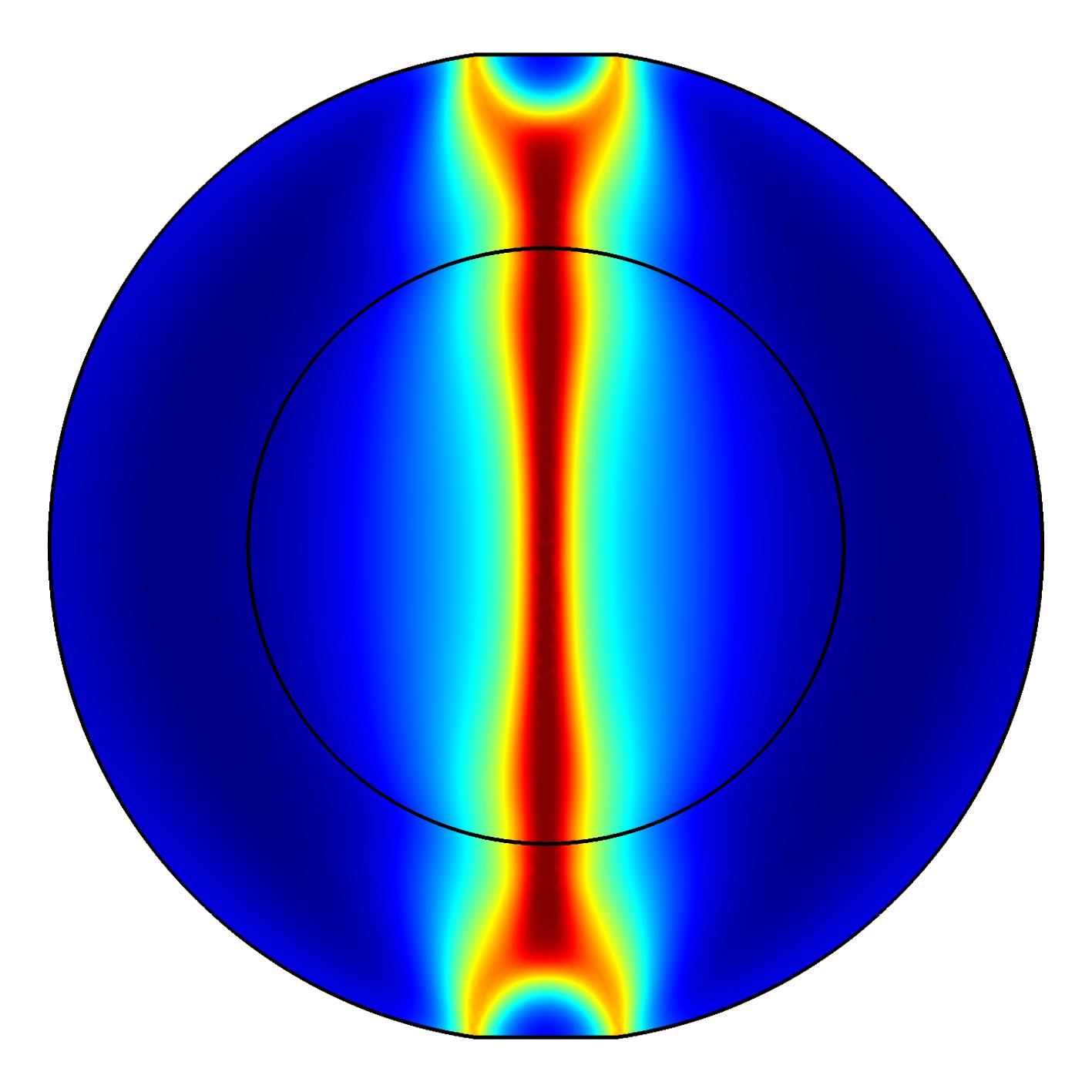}}\\
	\subfigure[$E_i$ = 60 GPa, D = 10 mm]{\includegraphics[height = 5cm]{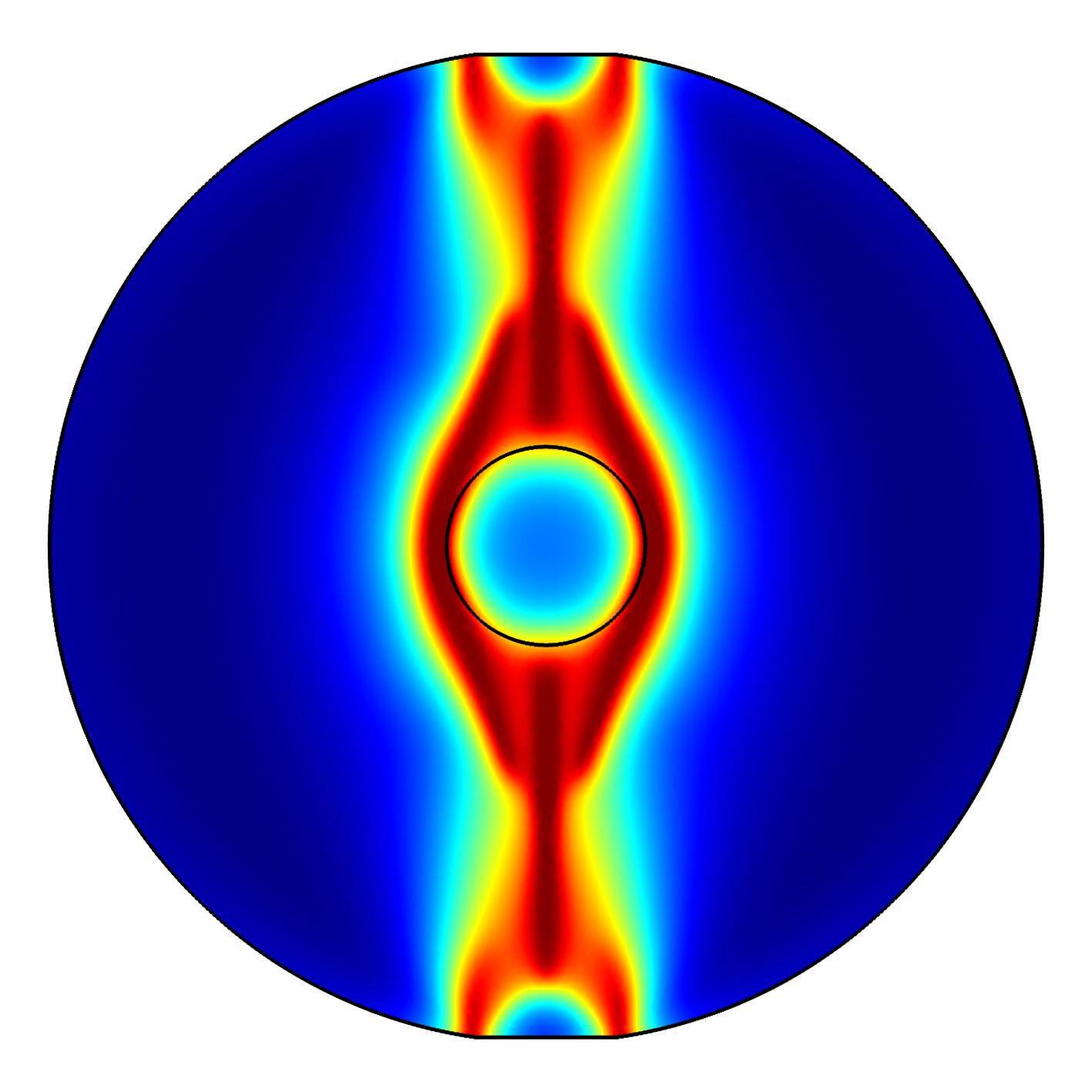}}
 	\subfigure[$E_i$ = 60 GPa, D = 20 mm]{\includegraphics[height = 5cm]{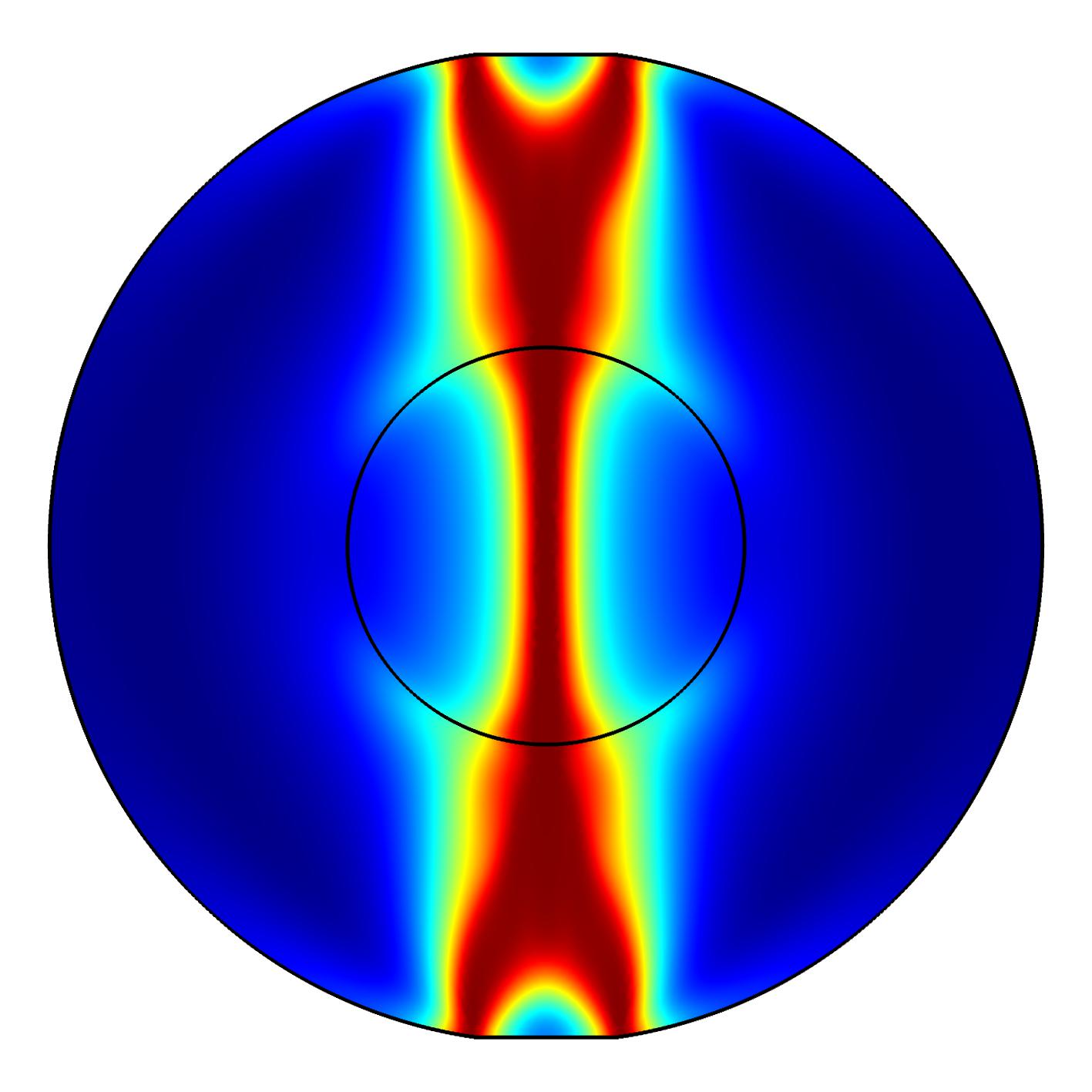}}
	\subfigure[$E_i$ = 60 GPa, D = 30 mm]{\includegraphics[height = 5cm]{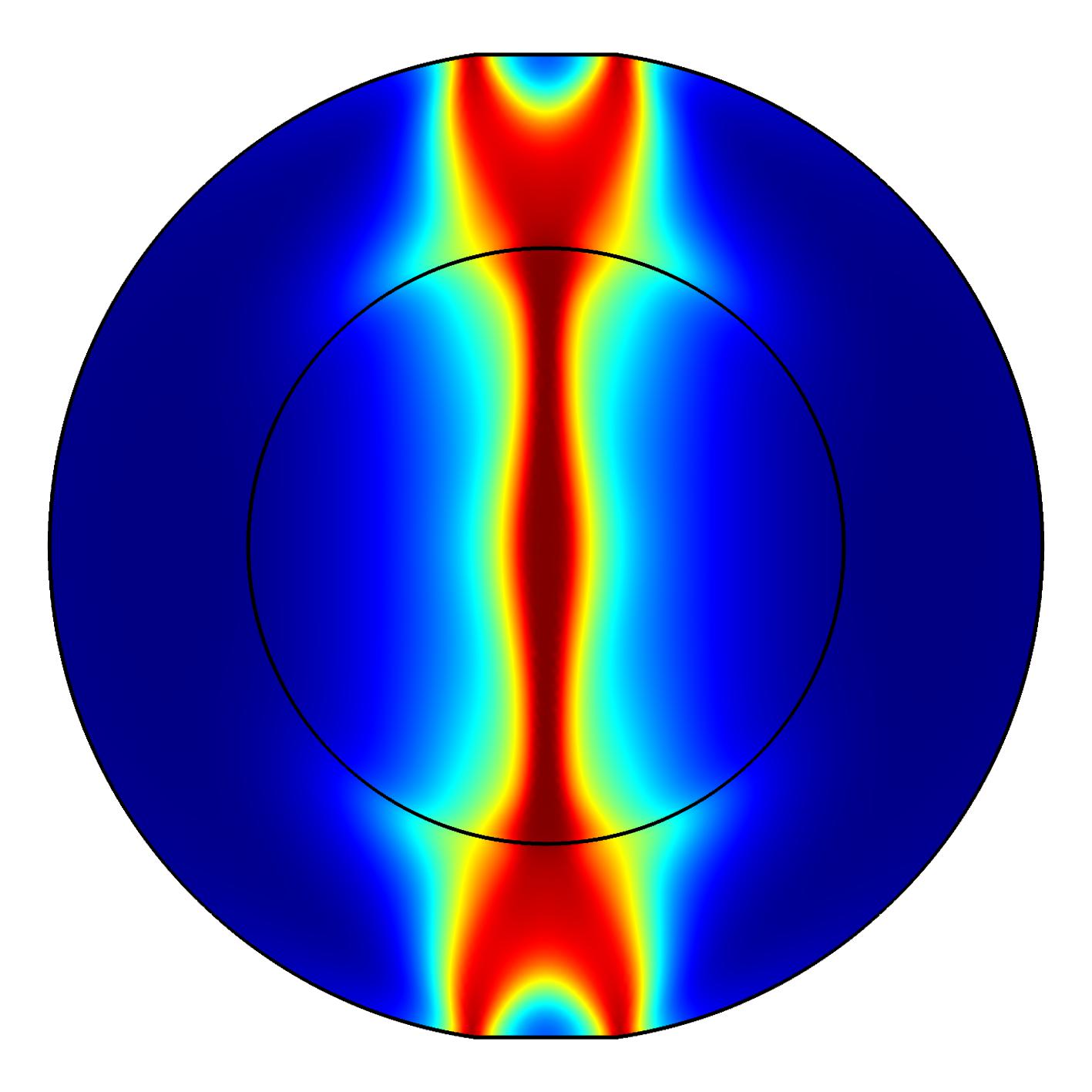}}
	\subfigure{\includegraphics[height = 4cm]{Figure2a_legend.jpg}}
	\caption{Final fracture patterns of the Brazilian discs with a centered inclusion}
	\label{Final fracture patterns of the Brazilian discs with a centered inclusion}
	\end{figure}

We first investigate the fracture patterns in Brazilian discs with a centered inclusion. The diameter of the inclusion affects the crack initiation and propagation, and thereby we choose the diameters of the inclusion as $D=0$, 10, 20, and 30 mm when fixing the diameter of the disc to 50 mm. Figure \ref{Final fracture patterns of the Brazilian discs with a centered inclusion} shows the final fracture patterns of the Brazilian discs with a centered inclusion for $E_i$ = 15 GPa and 60 GPa. Note that the results of $D=0$ (without inclusions) is presented in the previous section. For an inclusion that is softer than the disc matrix, i.e. $E_i$ = 15 GPa, the fracture initiates in the center of inclusion and propagates towards the two ends of the disc. However, for a stiffer inclusion ($E_i$ = 60 GPa), there are two different types of fracture patterns. When the diameter of the inclusion is small, fractures initiate and propagate first around the interface of the inclusion and matrix. Vertical dominated fractures and inclined secondary fractures are observed. In contrast, when the inclusion diameter is large, the fracture patterns are similar with those in the case of a softer inclusion.

\begin{figure}[H]
	\centering
	\subfigure[$E_i$ = 15 GPa]{\includegraphics[height = 5cm]{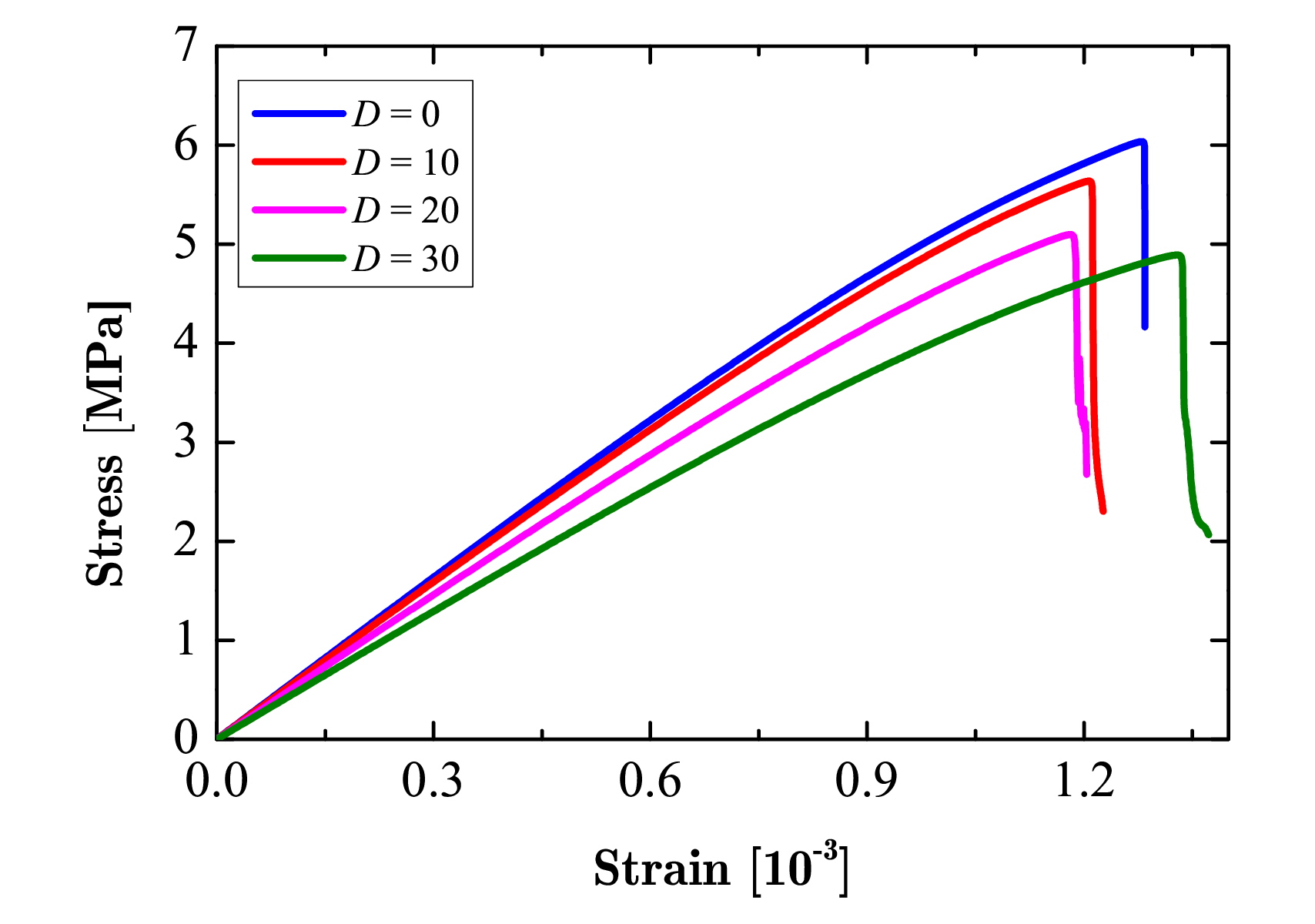}}
 	\subfigure[$E_i$ = 60 GPa]{\includegraphics[height = 5cm]{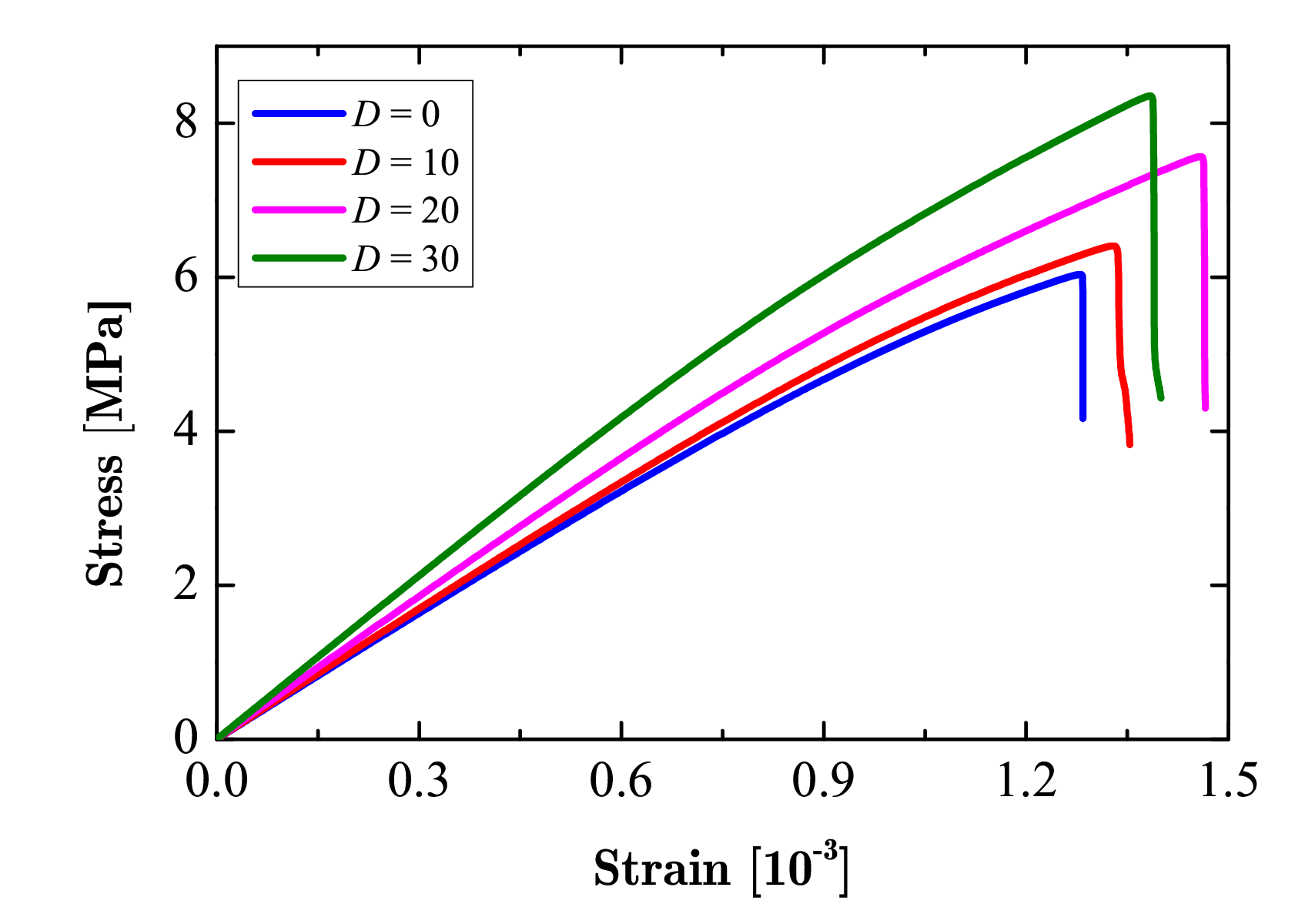}}
	\caption{Stress-strain curves of the Brazilian discs with a centered inclusion}
	\label{Stress-strain curves of the Brazilian discs with a centered inclusion}
	\end{figure}

\begin{figure}[H]
	\centering
	\includegraphics[width = 8cm]{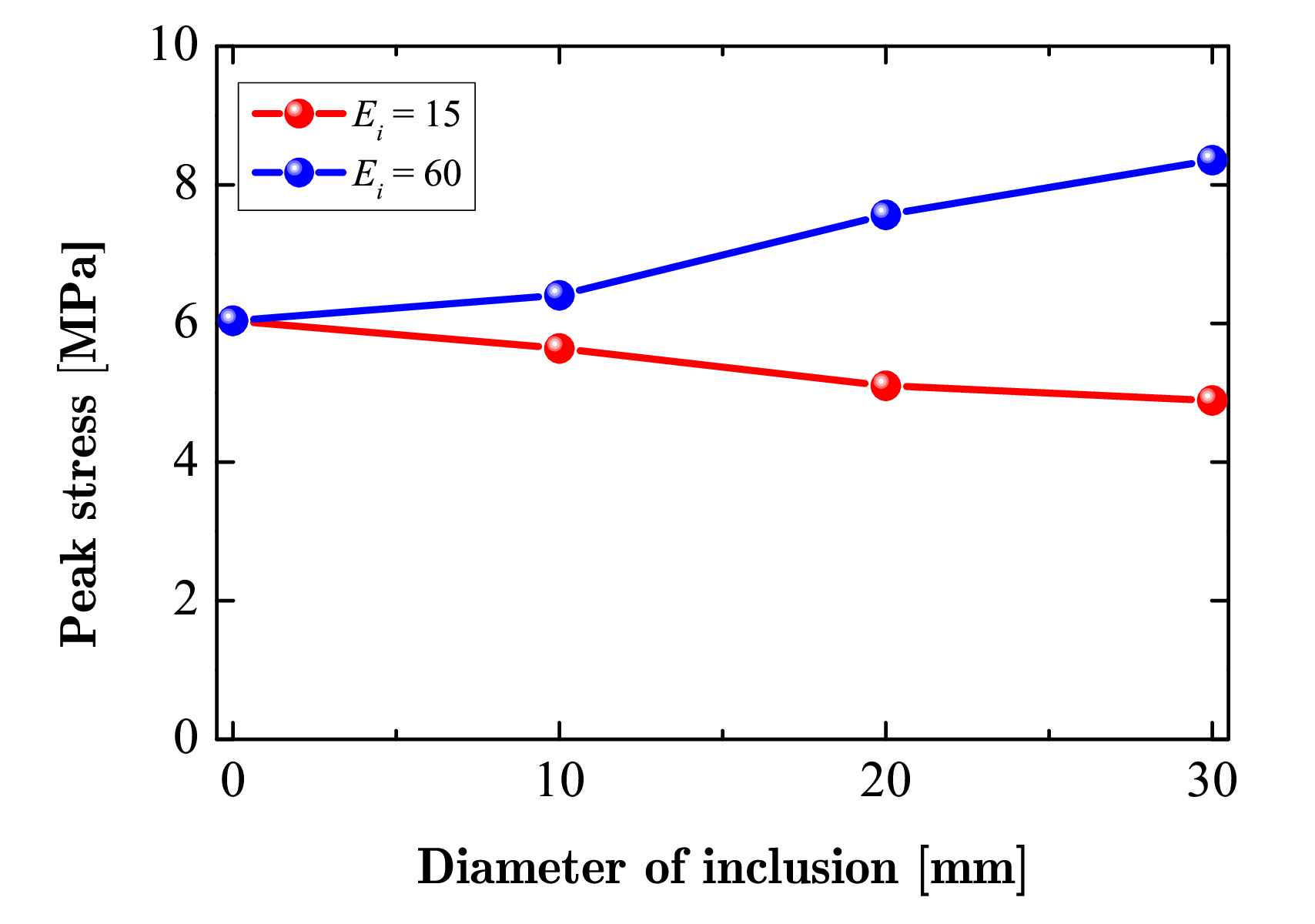}
	\caption{Peak stresses of the Brazilian discs with a centered inclusion}
	\label{Peak stresses of the Brazilian discs with a centered inclusion}
	\end{figure}

Figure \ref{Stress-strain curves of the Brazilian discs with a centered inclusion} shows the stress-strain curves of the Brazilian discs with a center inclusion of different diameters. As it can be observed, an increasing inclusion diameter significantly reduces the strength and overall stiffness of the disc when $E_i=15$ GPa. The slope of the curve gradually decreases as the diameter increases.  And conversely, for a stiffer inclusion, the slope of the curve increases as the diameter increases. Figure \ref{Peak stresses of the Brazilian discs with a centered inclusion} shows the peak stresses of the Brazilian discs with a centered inclusion. As observed, the peak stress shows a decreasing trend and an increasing trend with an increasing inclusion diameter when $E_i$ = 15 GPa and 60 GPa, respectively.

\subsection{Single inclusion with varying eccentricities along the horizontal axis}

\begin{figure}[htbp]
	\centering
	\subfigure[$E_i=15$ GPa, $e = 0$]{\includegraphics[height = 3.8cm]{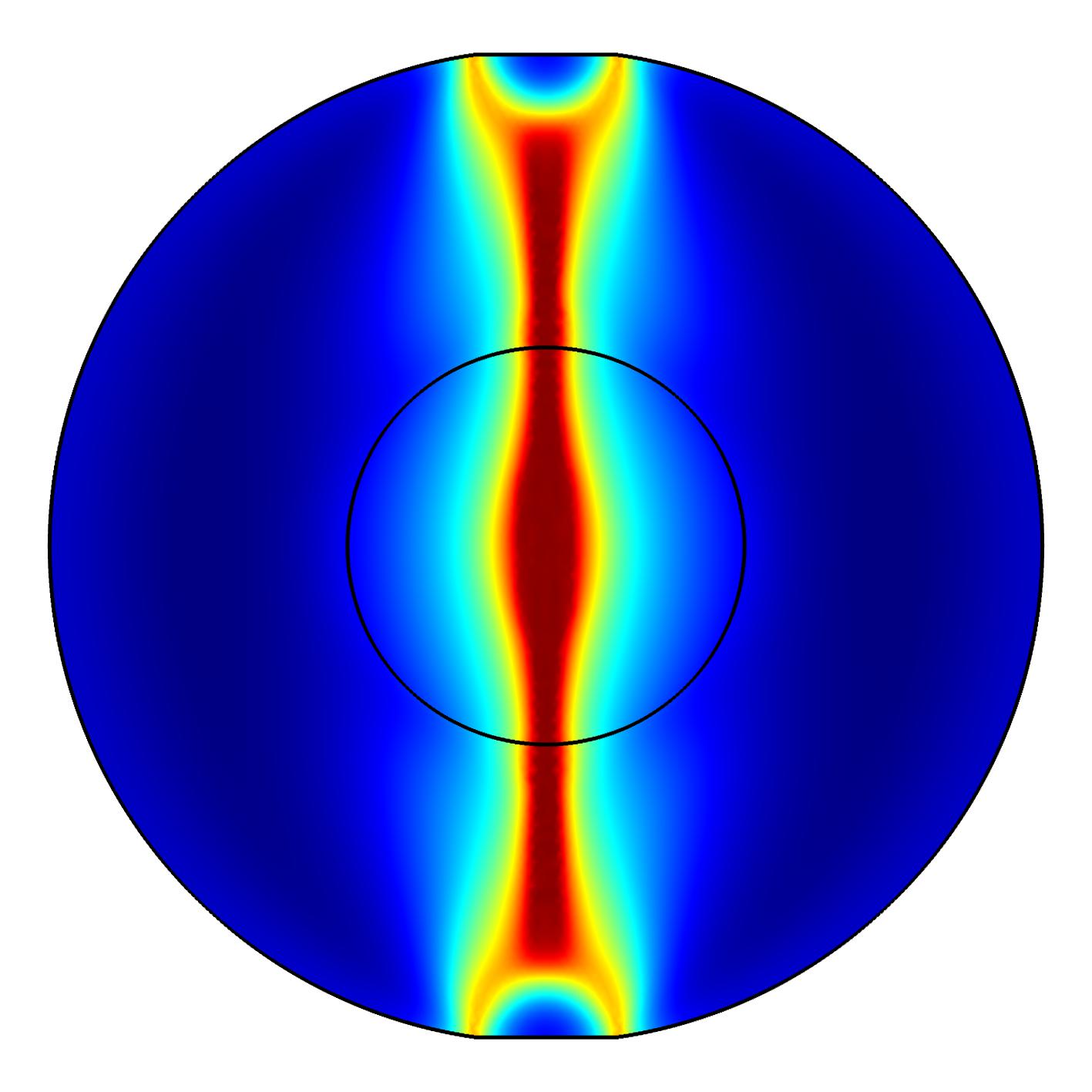}}
	\subfigure[$E_i=15$ GPa, $e = 0.2$]{\includegraphics[height = 3.8cm]{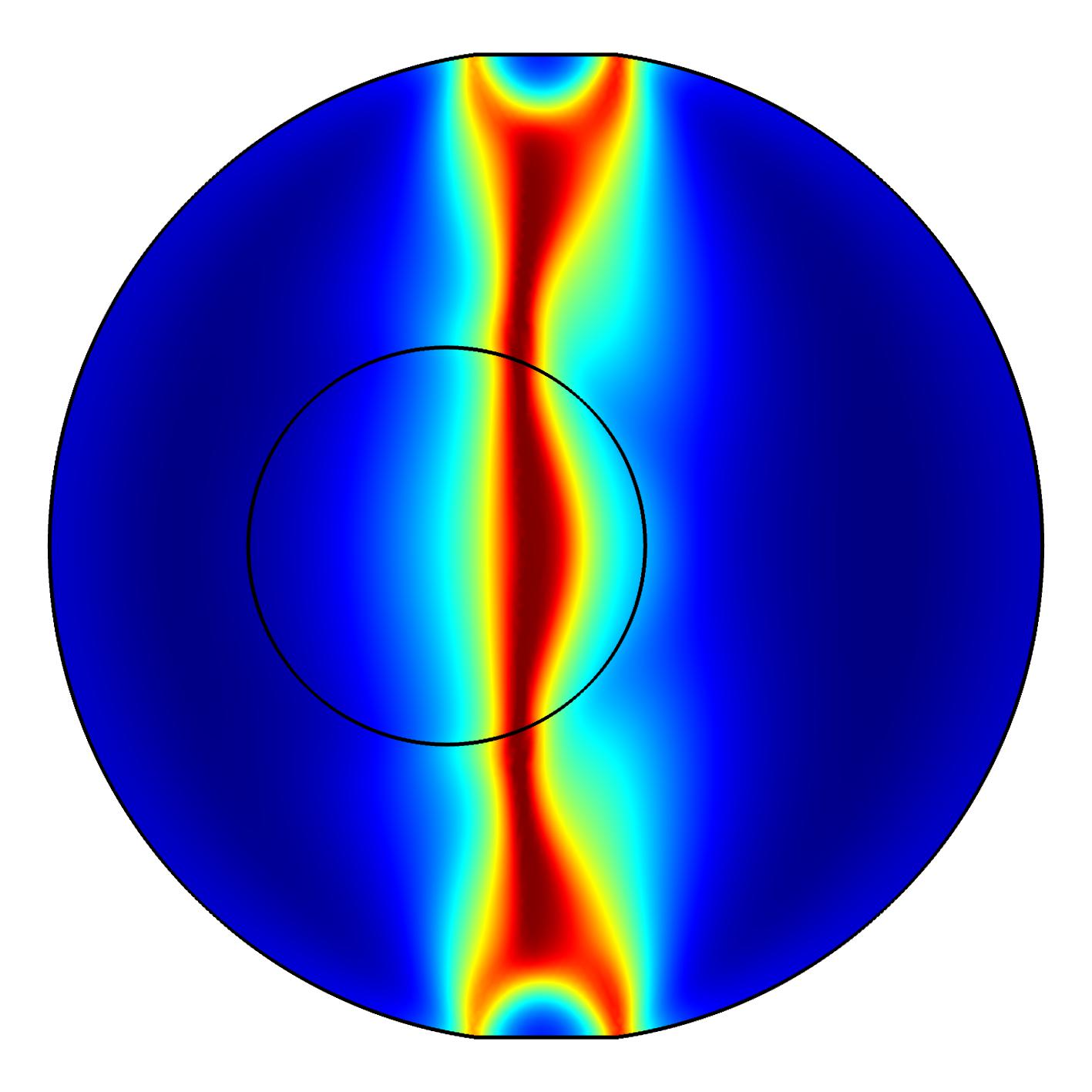}}
	\subfigure[$E_i=15$ GPa, $e = 0.4$]{\includegraphics[height = 3.8cm]{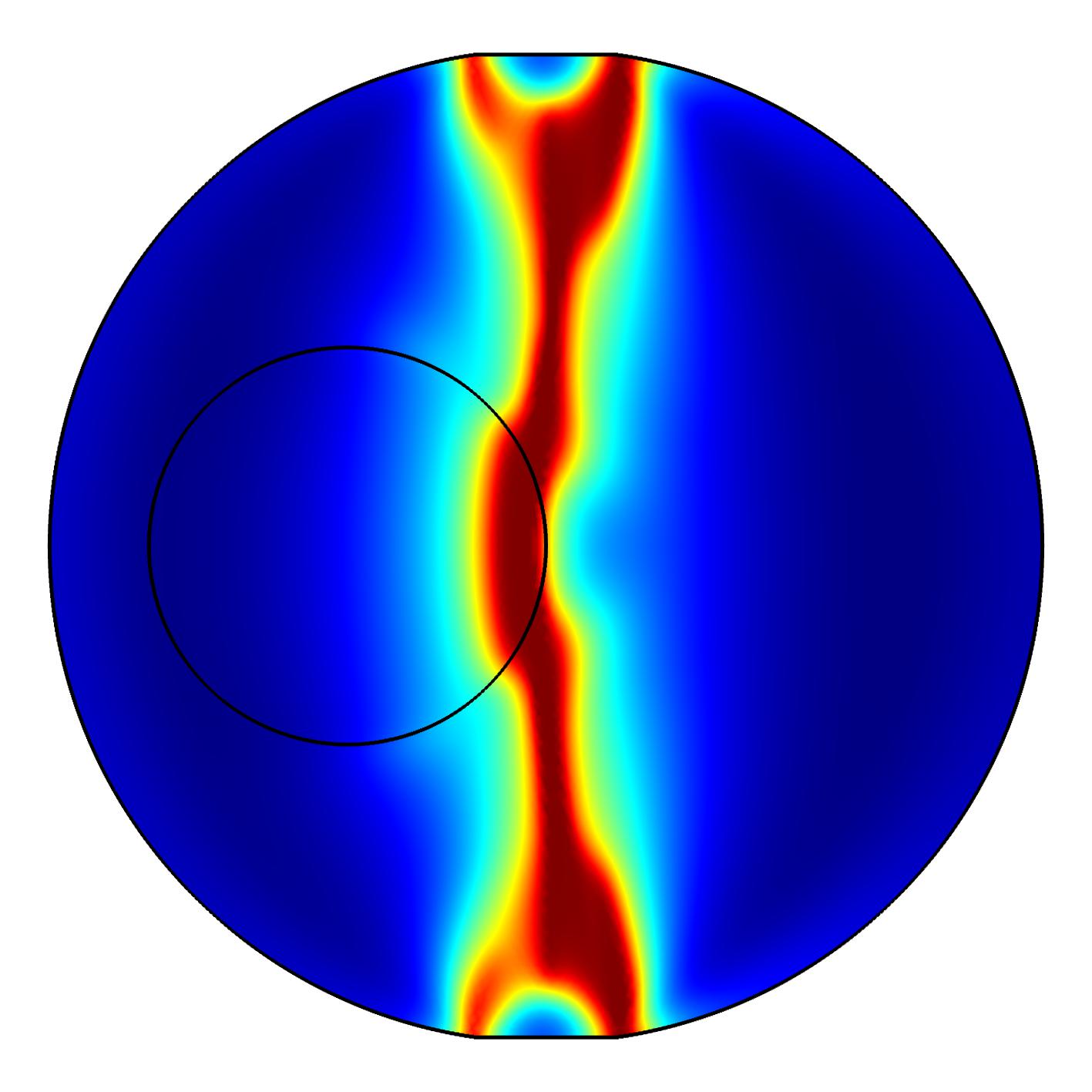}}
	\subfigure[$E_i=15$ GPa, $e = 0.6$]{\includegraphics[height = 3.8cm]{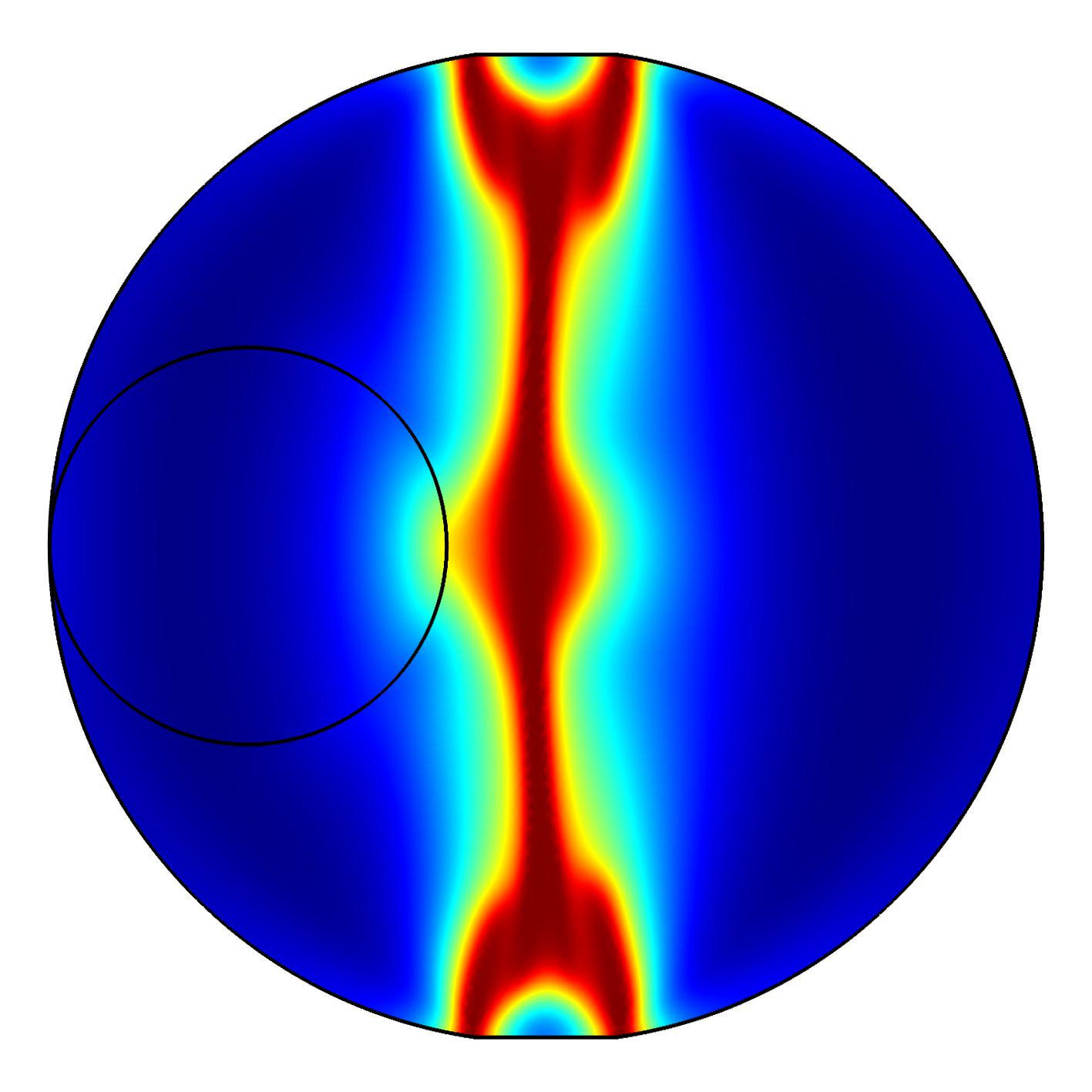}}
	\subfigure[$E_i=60$ GPa, $e = 0$]{\includegraphics[height = 3.8cm]{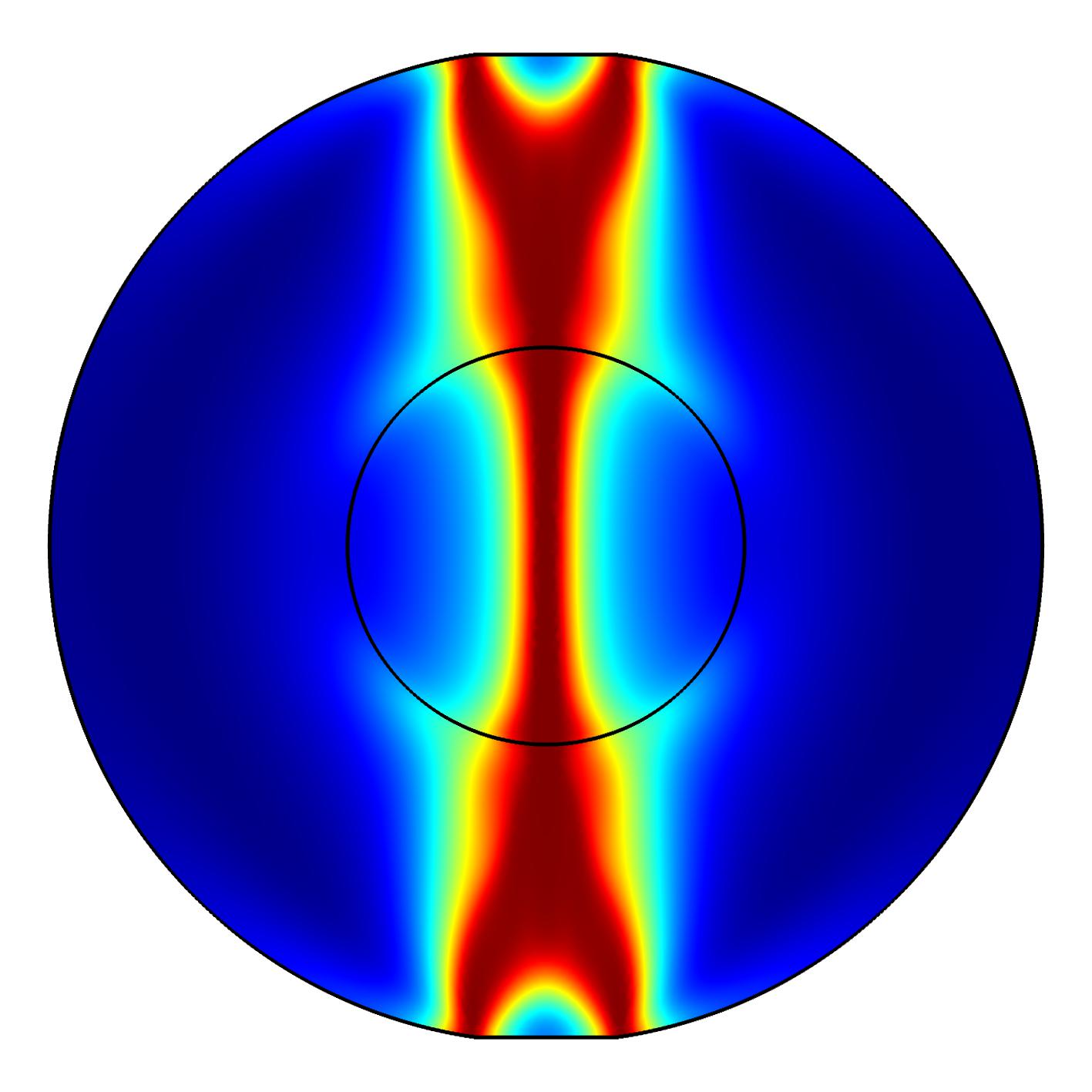}}
	\subfigure[$E_i=60$ GPa, $e = 0.2$]{\includegraphics[height = 3.8cm]{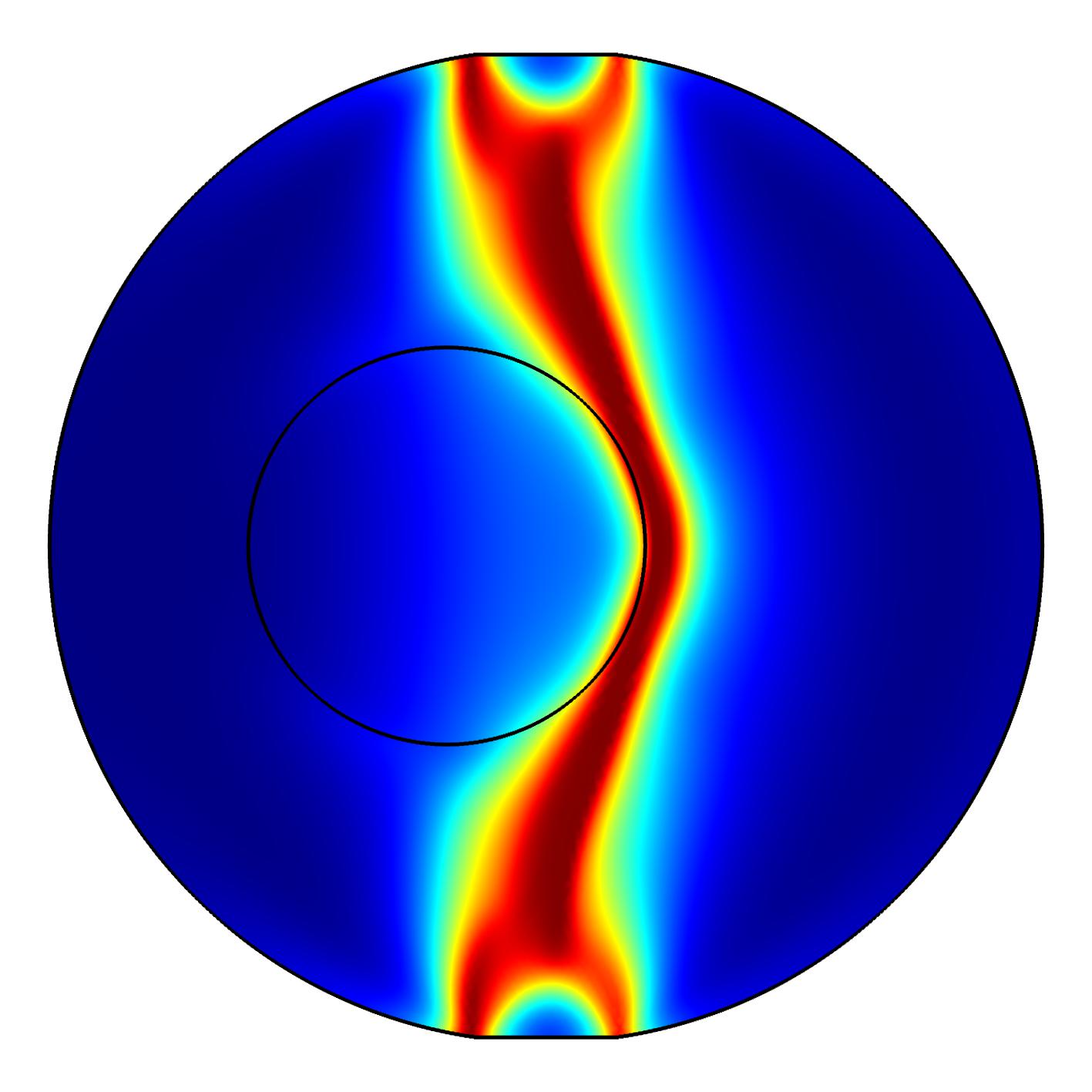}}
	\subfigure[$E_i=60$ GPa, $e = 0.4$]{\includegraphics[height = 3.8cm]{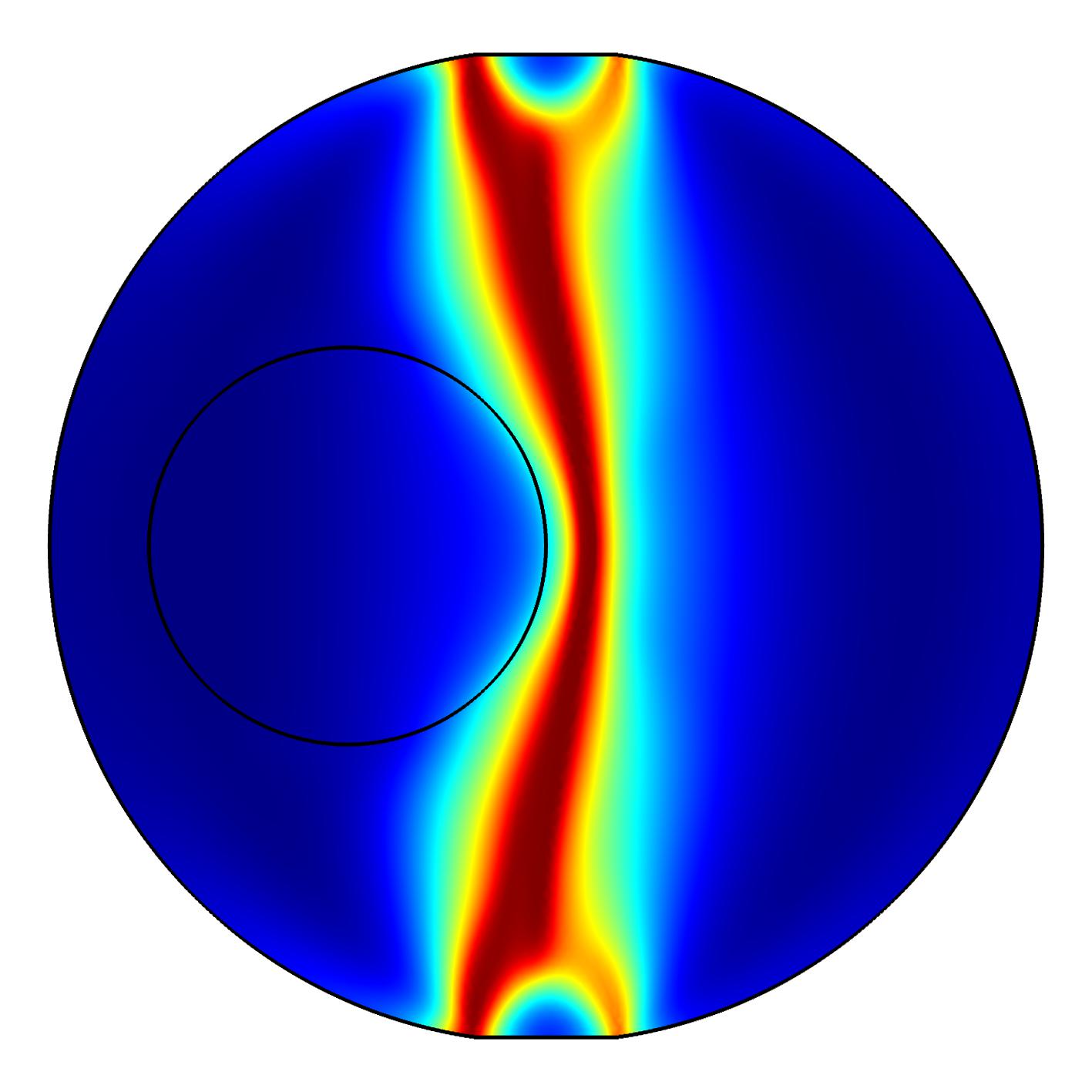}}
	\subfigure[$E_i=60$ GPa, $e = 0.6$]{\includegraphics[height = 3.8cm]{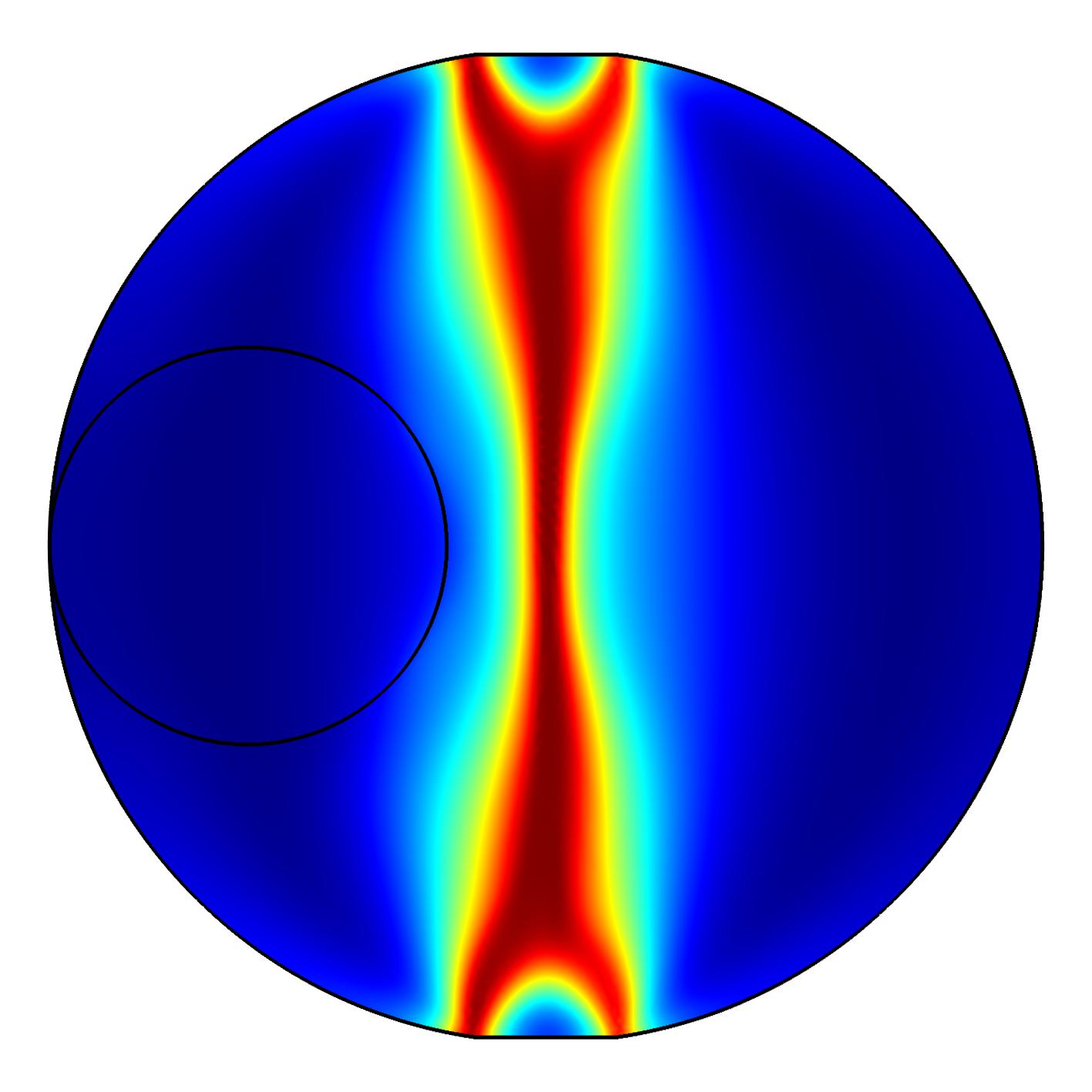}}
	\subfigure{\includegraphics[height = 3cm]{Figure2a_legend.jpg}}
	\caption{Final fracture patterns of the Brazilian discs with an inclusion along the horizontal axis}
	\label{Final fracture patterns of the Brazilian discs with an inclusion along the horizontal axis}
	\end{figure}

Figure \ref{Final fracture patterns of the Brazilian discs with an inclusion along the horizontal axis} shows the final fracture patterns of Brazilian discs with an inclusion along the horizontal direction by using the phase field model. The diameter of the inclusion is fixed to 20 mm. The results are presented for $E_i$ = 15 GPa and 60 GPa. If the stiffness of the inclusion is lower than that of the disc matrix ($E_i$ = 15 GPa), the fracture initiates in the middle of the disc and has a deflection towards the interface between the matrix and the inclusion. On the contrary, if the stiffness of the inclusion is larger than that of the matrix, the fracture initiates in the center of the disc and propagates almost vertically towards the disc ends for $e=0$ or $e=0.6$. However, the fracture propagates along the inclusion interface and towards the disc ends for $e=0.2$ or $e=0.4$. 

\begin{figure}[htbp]
	\centering
	\subfigure[$E_i=15$ GPa]{\includegraphics[width = 8cm]{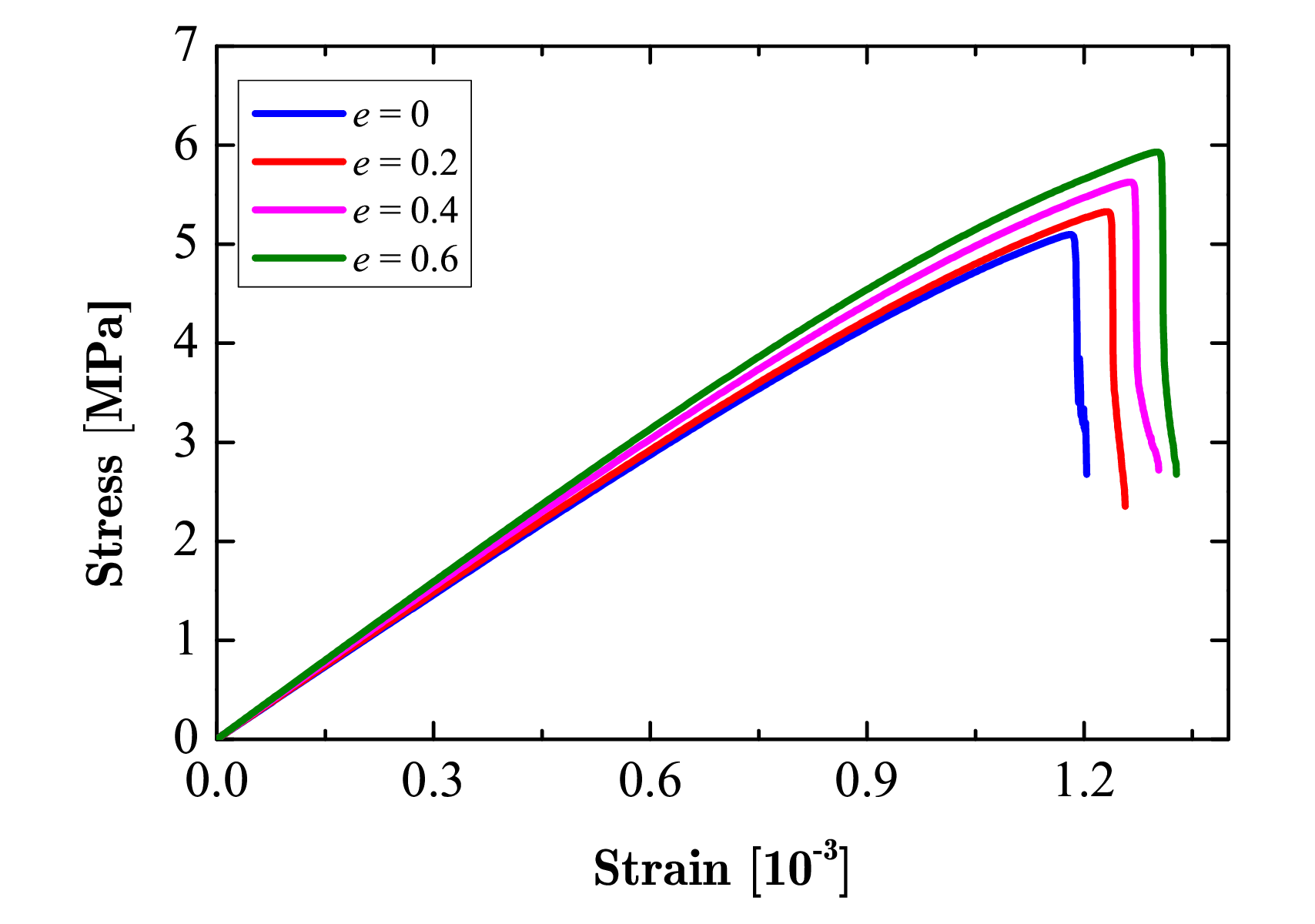}}
	\subfigure[$E_i=60$ GPa]{\includegraphics[width = 8cm]{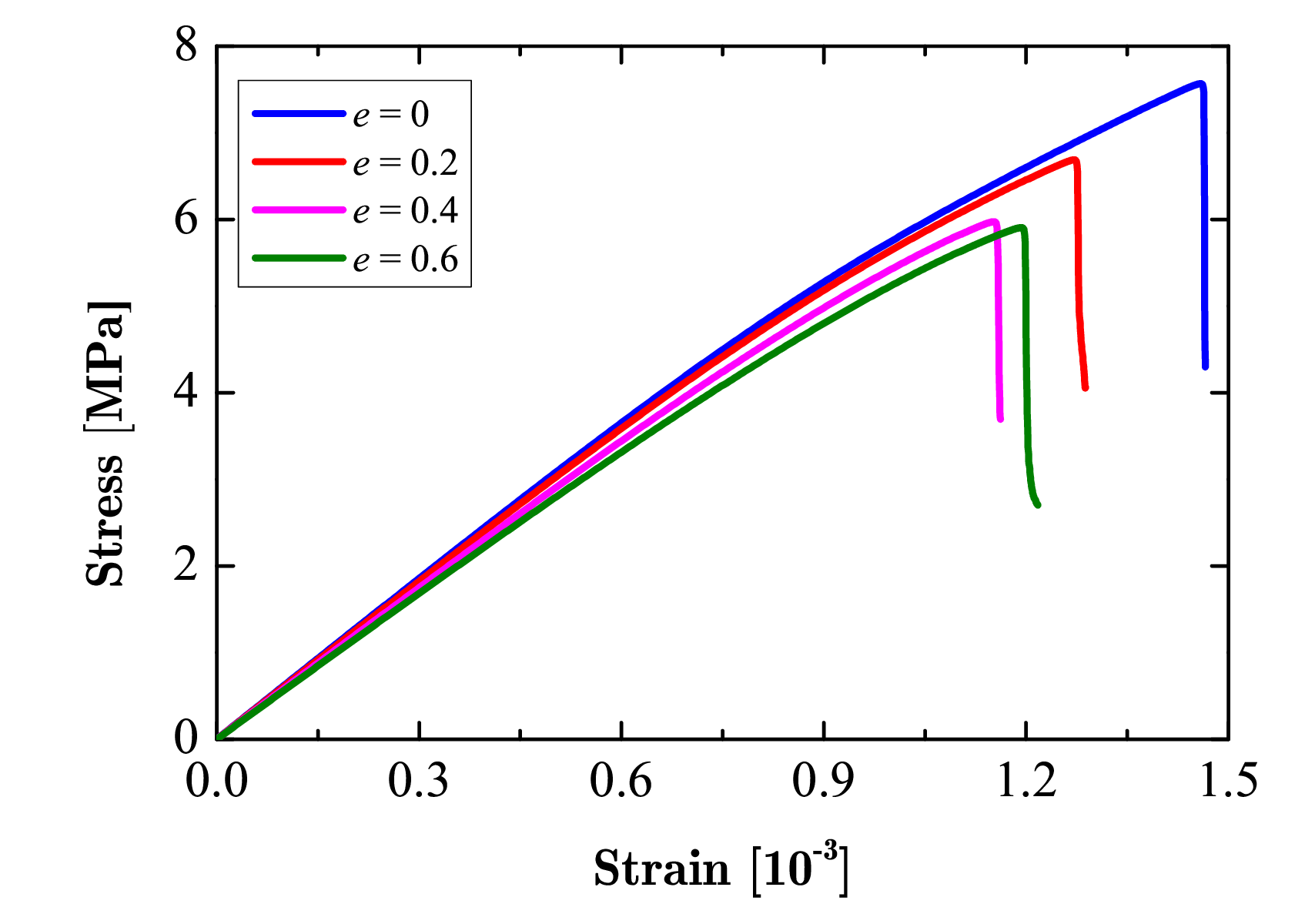}}
	\caption{Stress-strain curves of the Brazilian discs with an inclusion along the horizontal axis}
	\label{Stress-strain curves of the Brazilian discs with an inclusion along the horizontal axis}
	\end{figure}

\begin{figure}[htbp]
	\centering
	\includegraphics[width = 8cm]{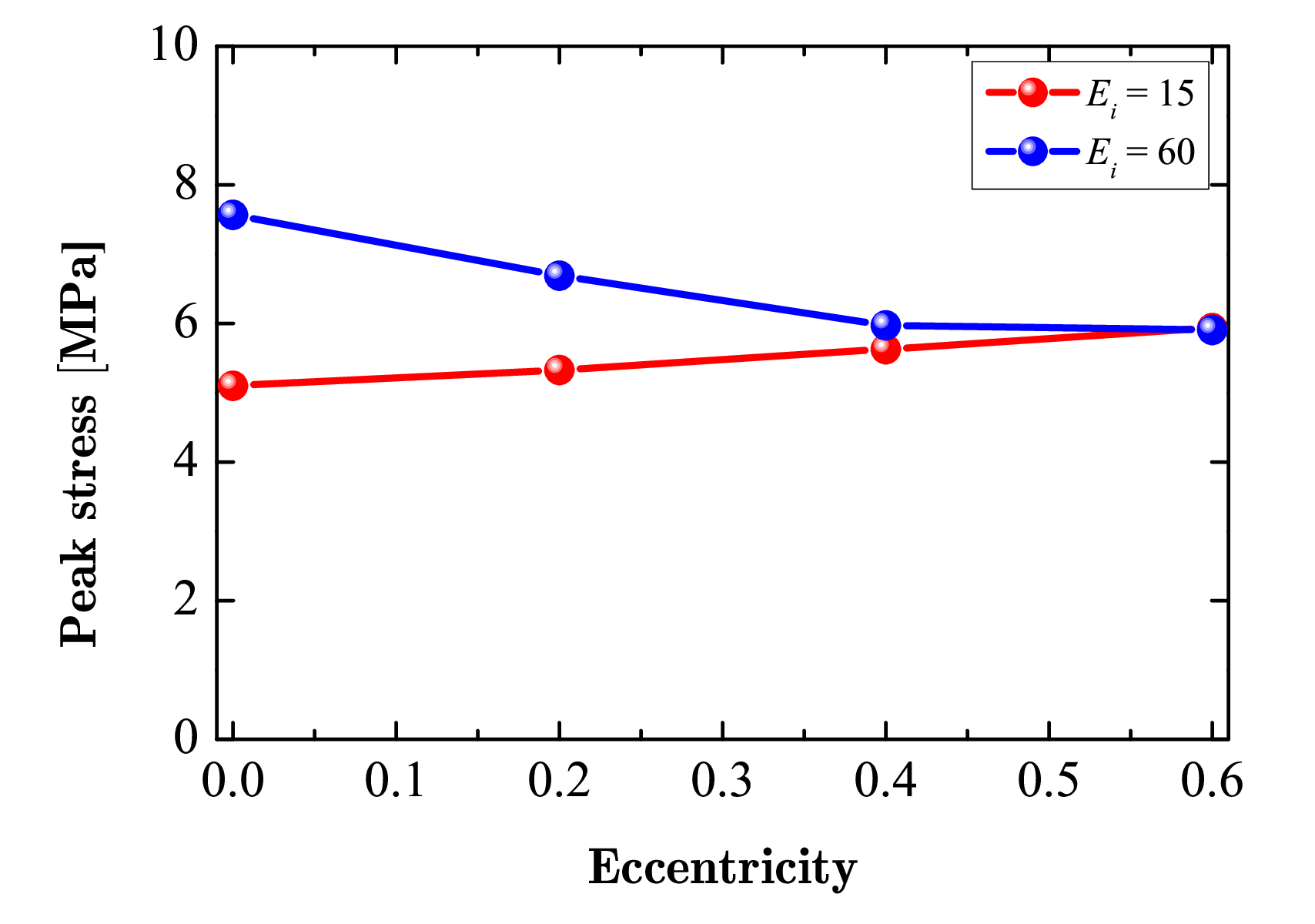}
	\caption{Peak stresses of the Brazilian discs with an inclusion along the horizontal axis}
	\label{Peak stresses of the Brazilian discs with an inclusion along the horizontal axis}
	\end{figure}

Figure \ref{Stress-strain curves of the Brazilian discs with an inclusion along the horizontal axis} shows the stress-strain curves of Brazilian discs with an inclusion along the horizontal axis for different eccentricities. For a softer inclusion, the overall strength of the disc increases with an increasing eccentricity. Additionally, the overall stiffness of the disc increases with the increase in the eccentricity $e$. On the other hand, the overall strength and stiffness of the disc decrease with the increase in the eccentricity for a stiffer inclusion. The peak stresses of the Brazilian discs with an inclusion along the horizontal axis are shown in Fig. \ref{Peak stresses of the Brazilian discs with an inclusion along the horizontal axis}, which supports the observations in Fig. \ref{Stress-strain curves of the Brazilian discs with an inclusion along the horizontal axis}. A gradually increasing trend is observed for the peak stress with an increasing eccentricity under $E_i=15$ GPa. Meanwhile, for $E_i=60$ GPa, the peak stress decreases as the eccentricity increases.

\subsection{Single inclusion with eccentricities along the vertical axis}

\begin{figure}[htbp]
	\centering
	\subfigure[$E_i=15$ GPa, $e = 0$]{\includegraphics[height = 3.8cm]{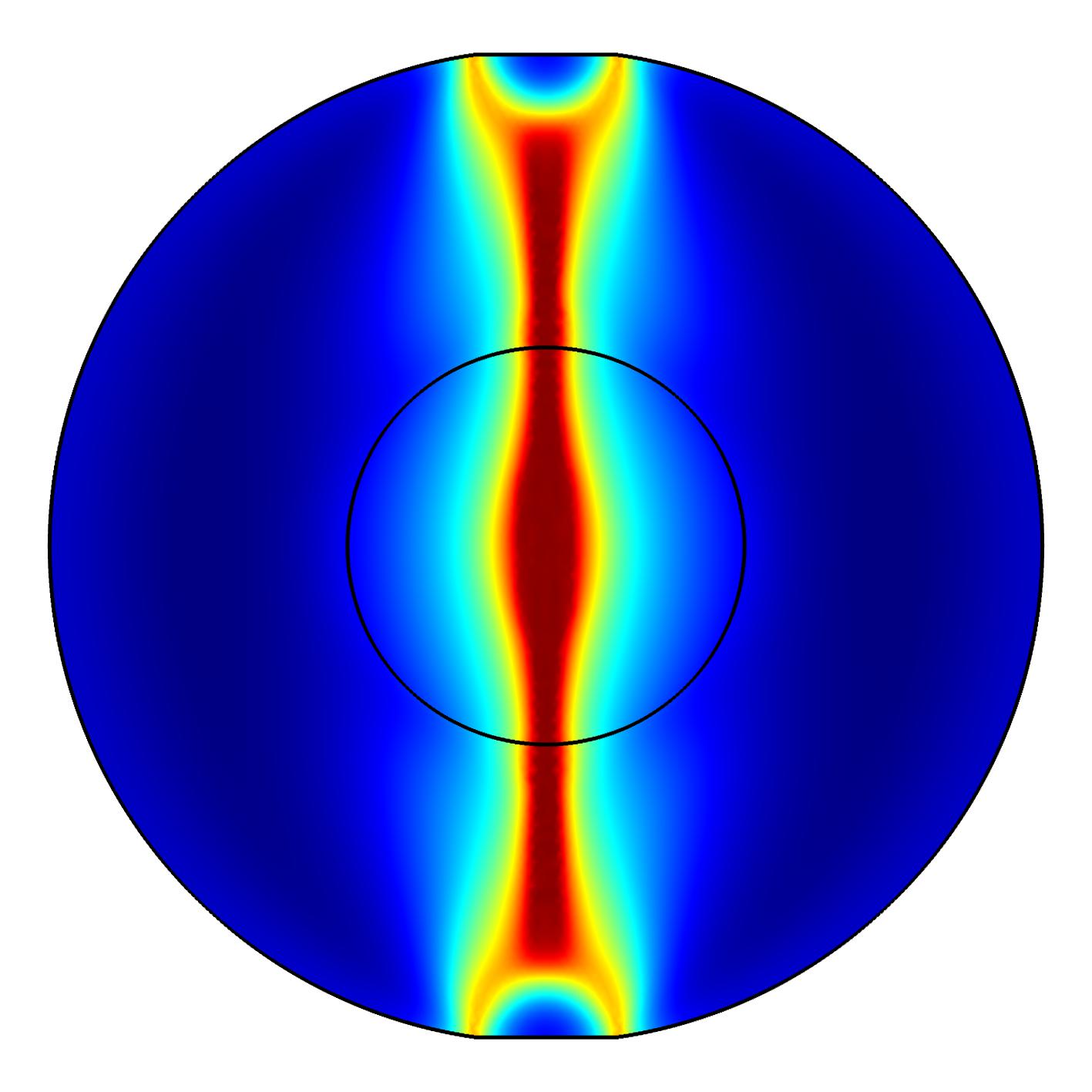}}
	\subfigure[$E_i=15$ GPa, $e = 0.2$]{\includegraphics[height = 3.8cm]{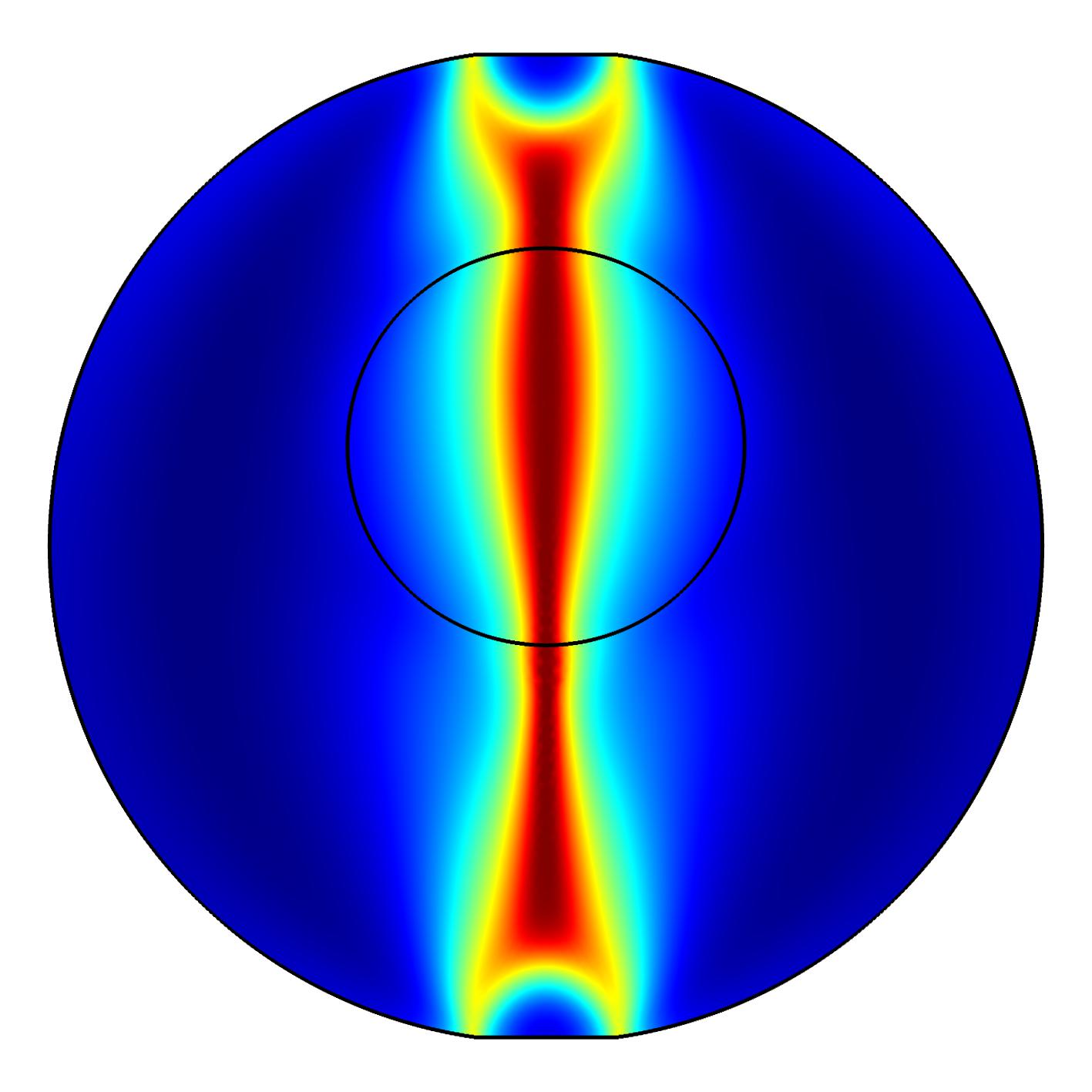}}
	\subfigure[$E_i=15$ GPa, $e = 0.4$]{\includegraphics[height = 3.8cm]{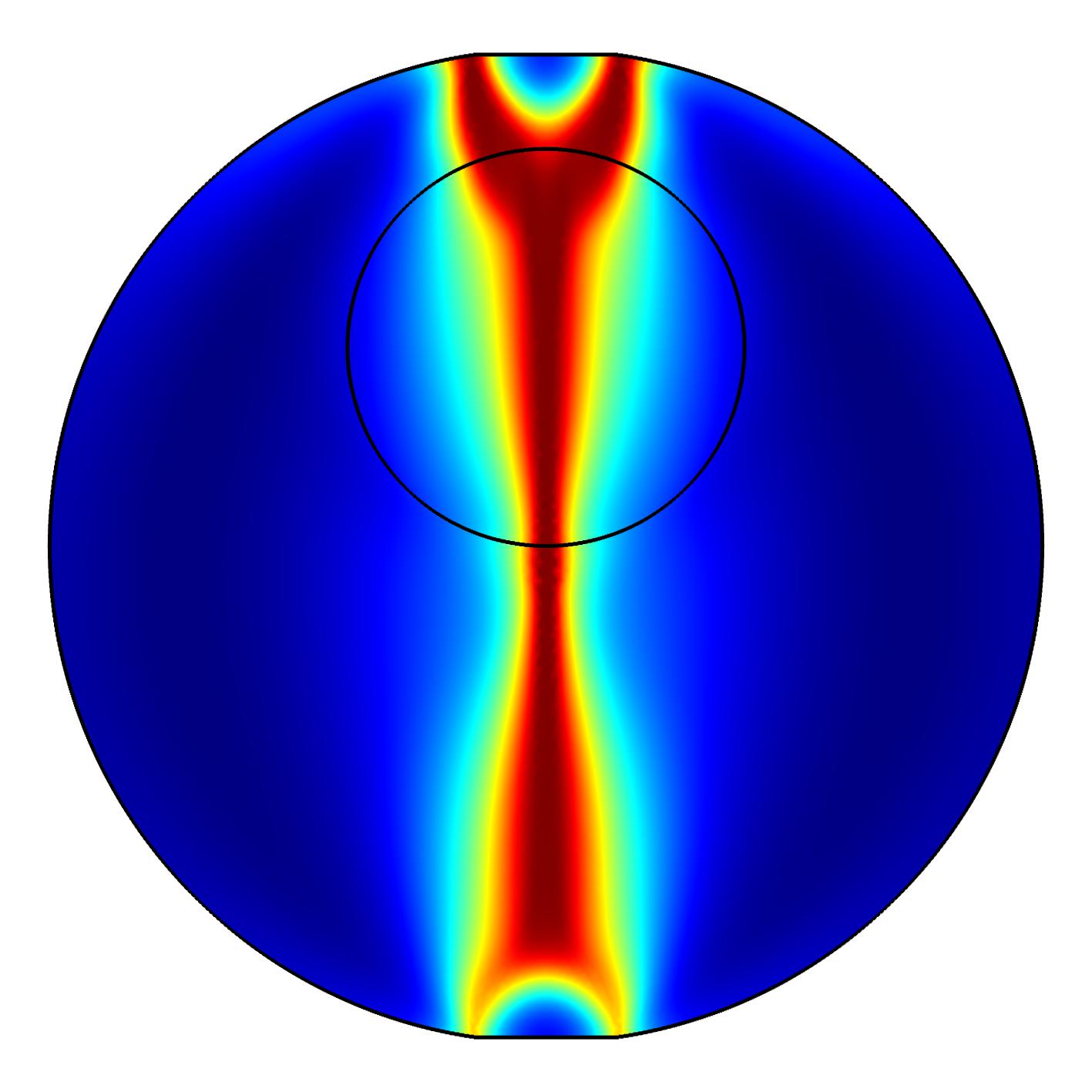}}
	\subfigure[$E_i=15$ GPa, $e = 0.6$]{\includegraphics[height = 3.8cm]{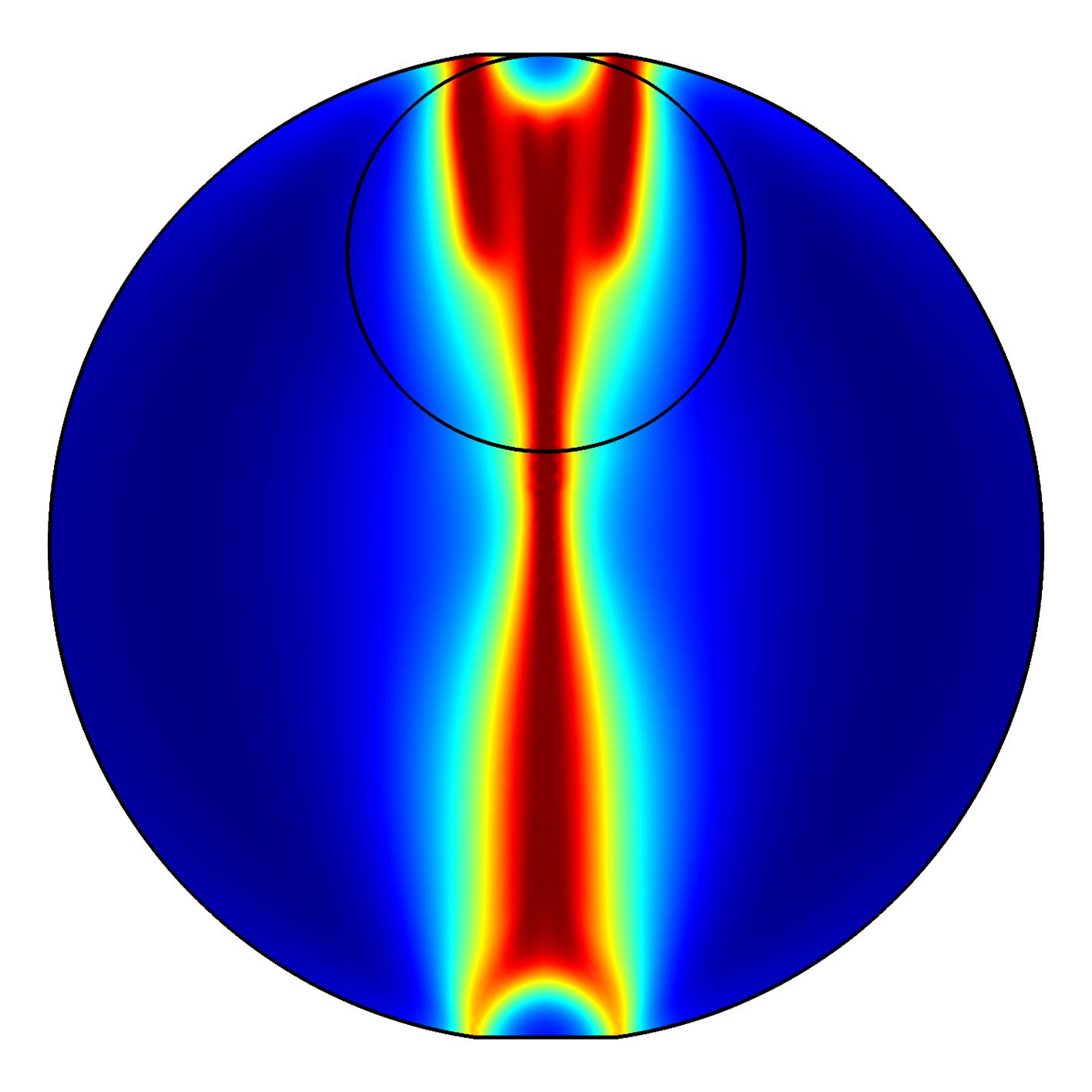}}\\
	\subfigure[$E_i=60$ GPa, $e = 0$]{\includegraphics[height = 3.8cm]{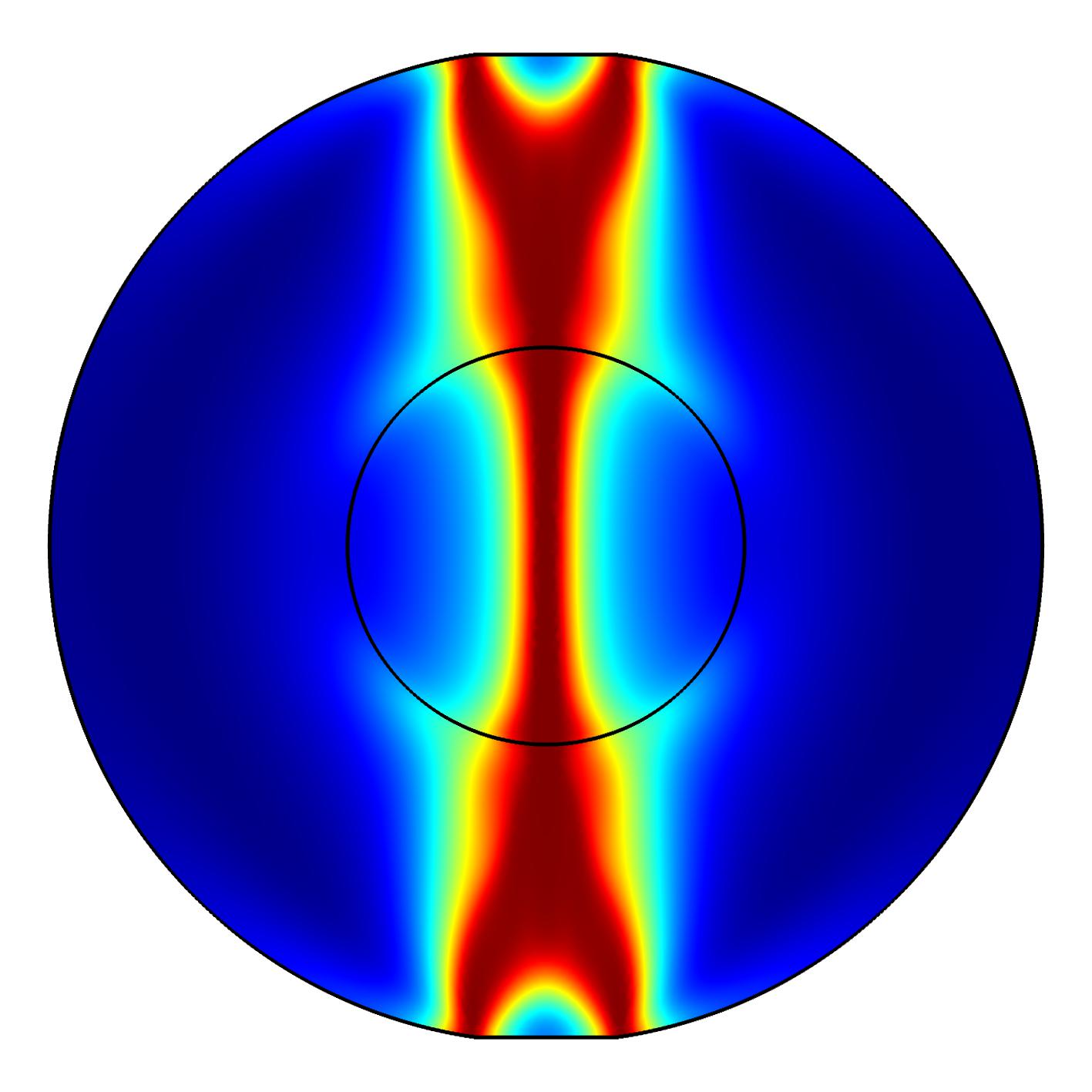}}
	\subfigure[$E_i=60$ GPa, $e = 0.2$]{\includegraphics[height = 3.8cm]{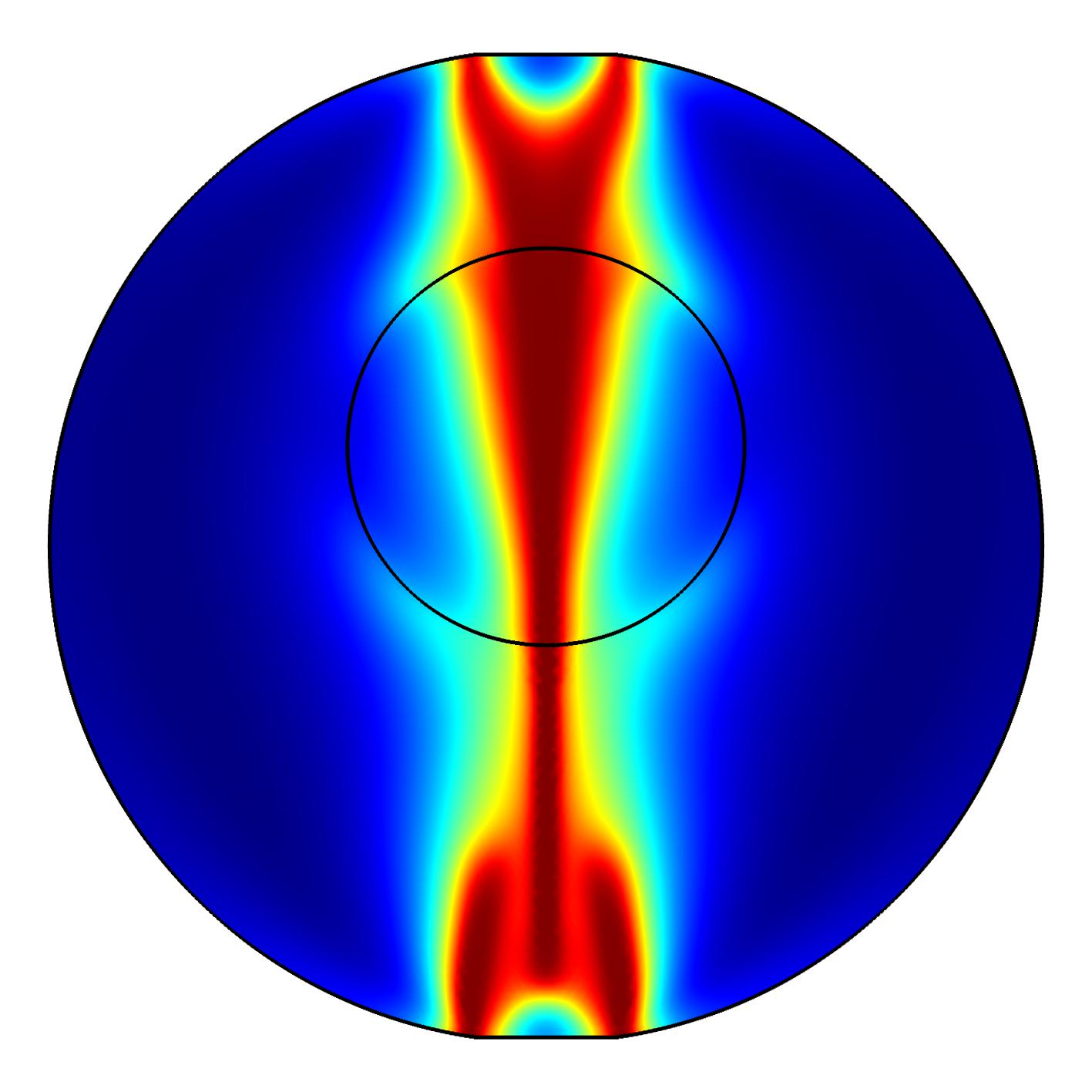}}
	\subfigure[$E_i=60$ GPa, $e = 0.4$]{\includegraphics[height = 3.8cm]{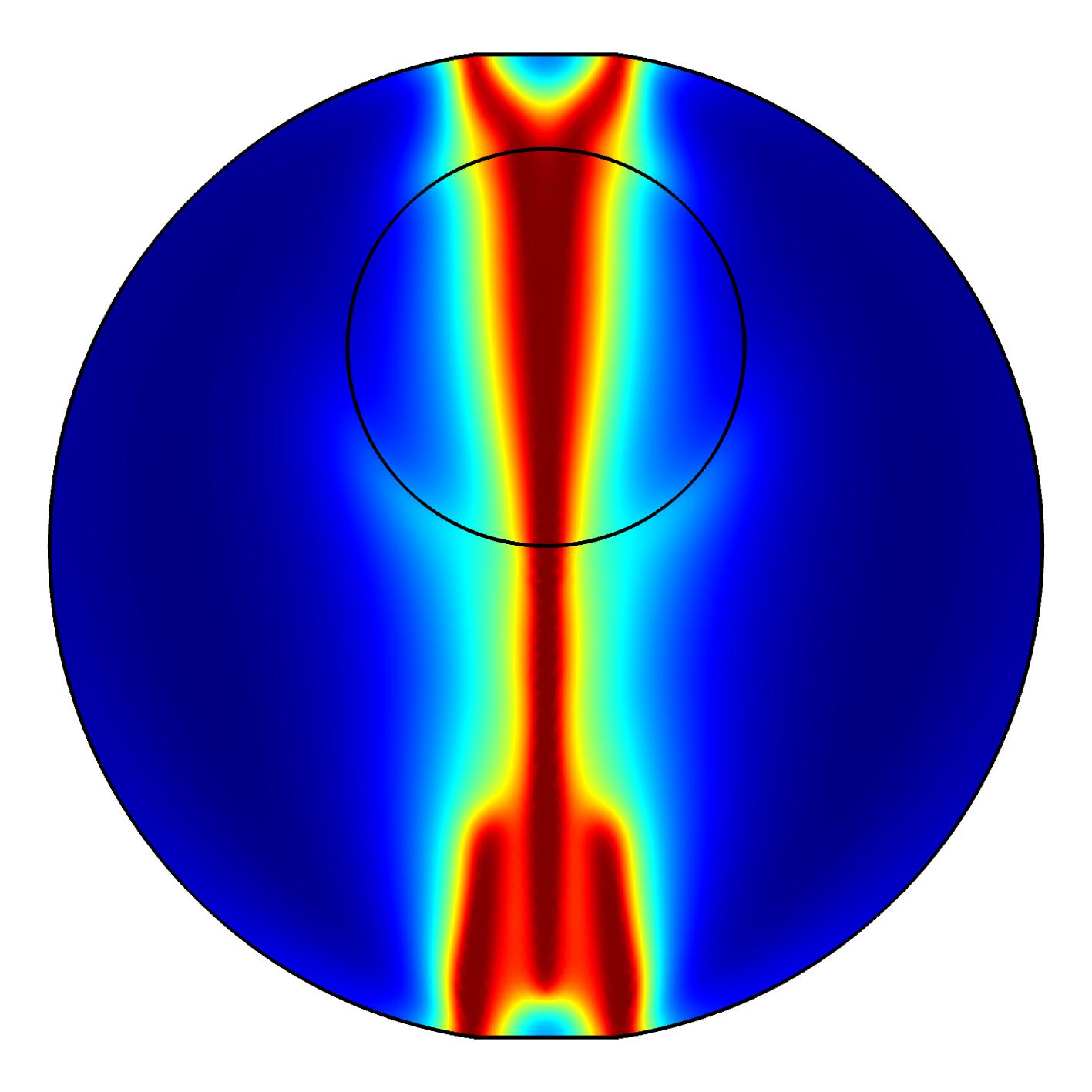}}
	\subfigure[$E_i=60$ GPa, $e = 0.6$]{\includegraphics[height = 3.8cm]{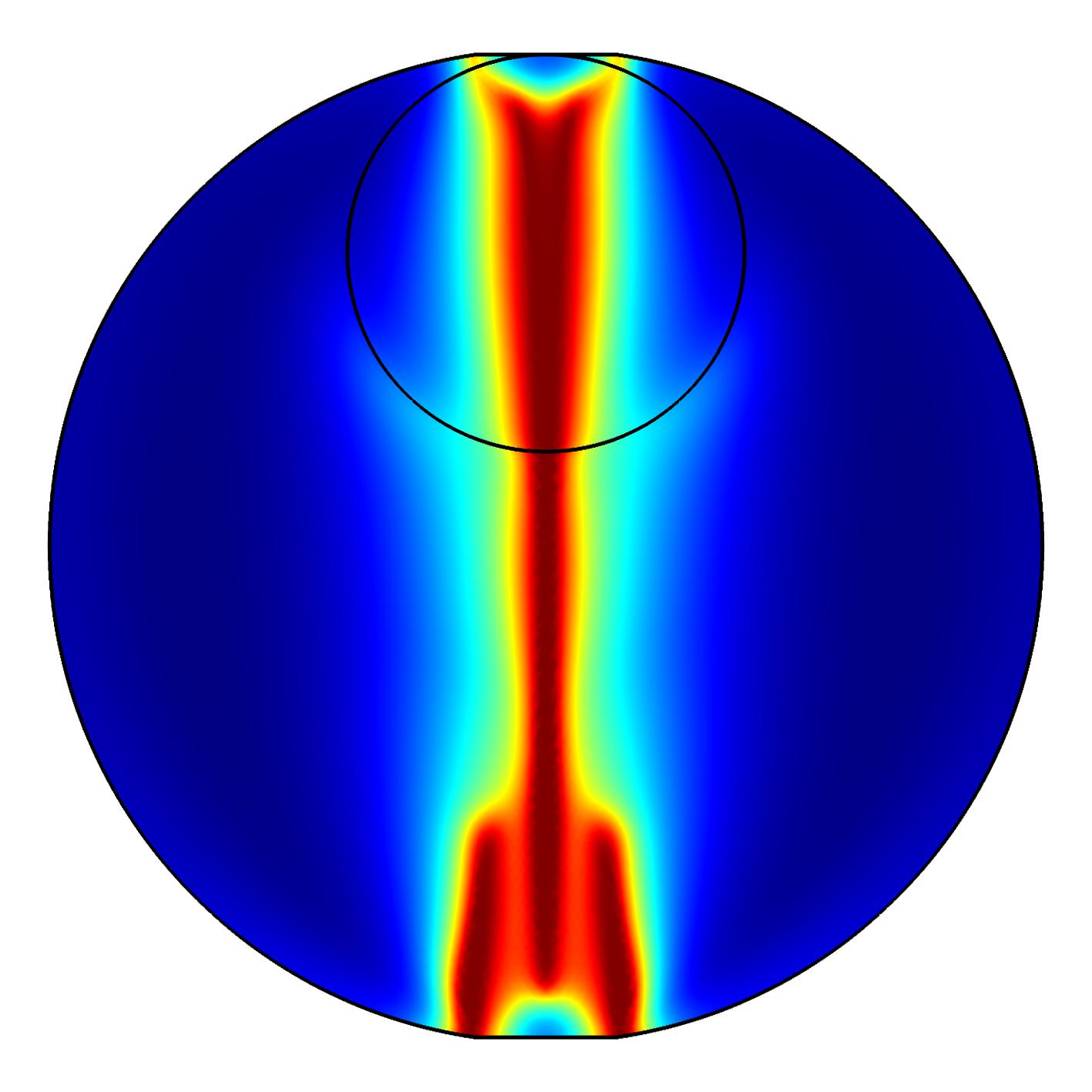}}
	\subfigure{\includegraphics[height = 3cm]{Figure2a_legend.jpg}}
	\caption{Final fracture patterns of the Brazilian discs with an inclusion along the vertical axis}
	\label{Final fracture patterns of the Brazilian discs with an inclusion along the vertical axis}
	\end{figure}

\begin{figure}[htbp]
	\centering
	\subfigure[$E_i=15$ GPa]{\includegraphics[width = 8cm]{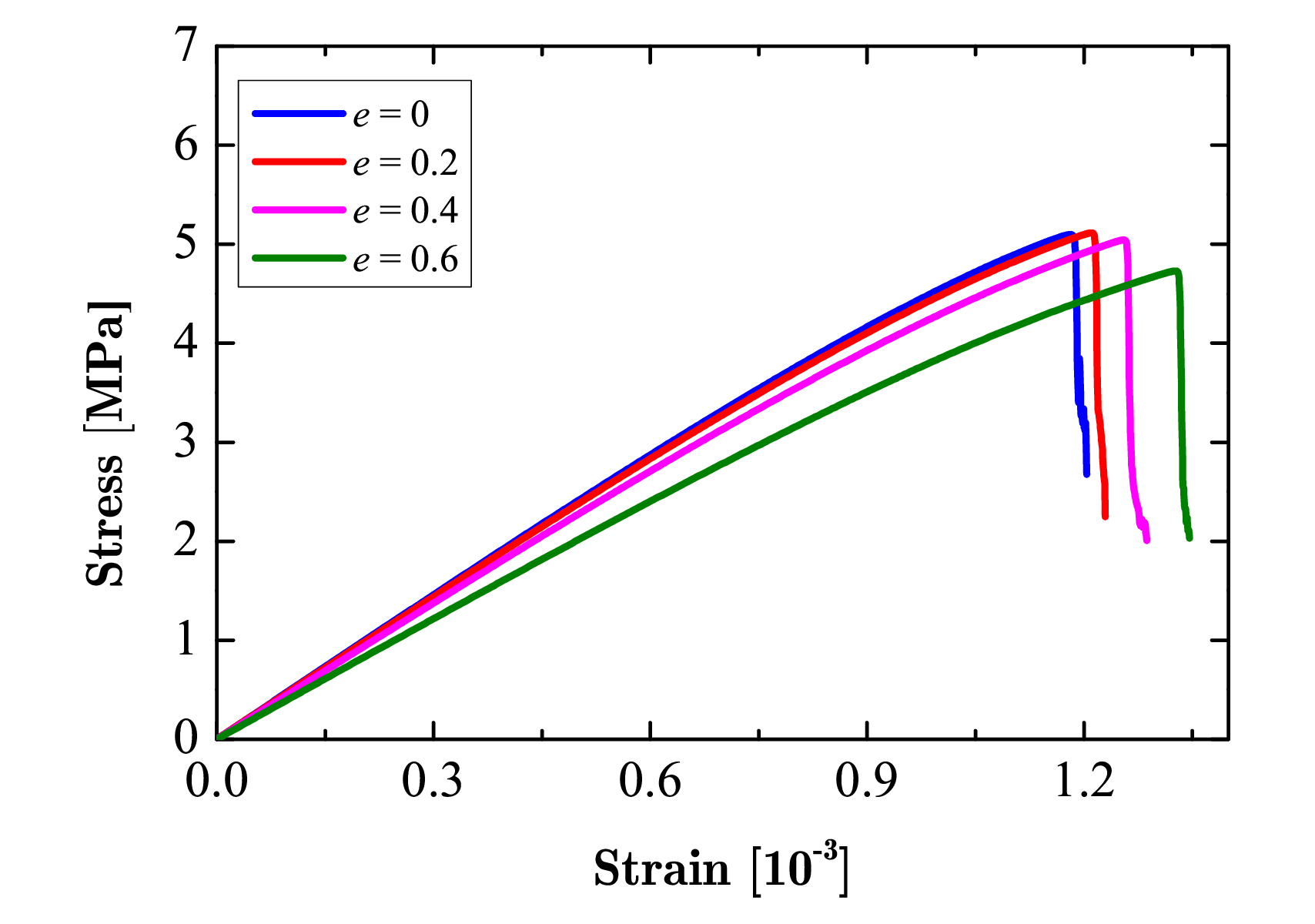}}
	\subfigure[$E_i=60$ GPa]{\includegraphics[width = 8cm]{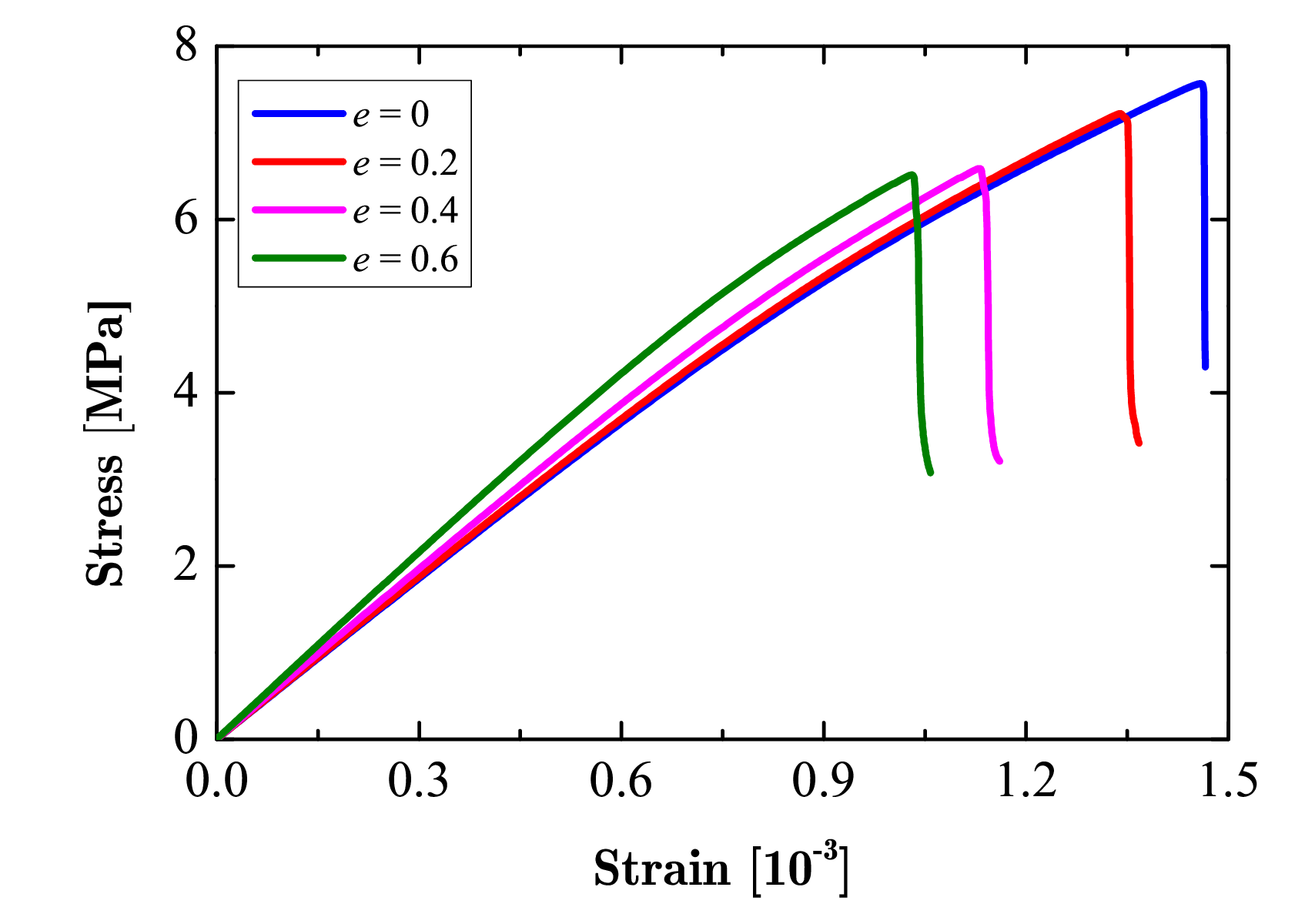}}
	\caption{Stress-strain curves of the Brazilian discs with an inclusion along the vertical axis}
	\label{Stress-strain curves of the Brazilian discs with an inclusion along the vertical axis}
	\end{figure}

\begin{figure}[htbp]
	\centering
	\includegraphics[width = 8cm]{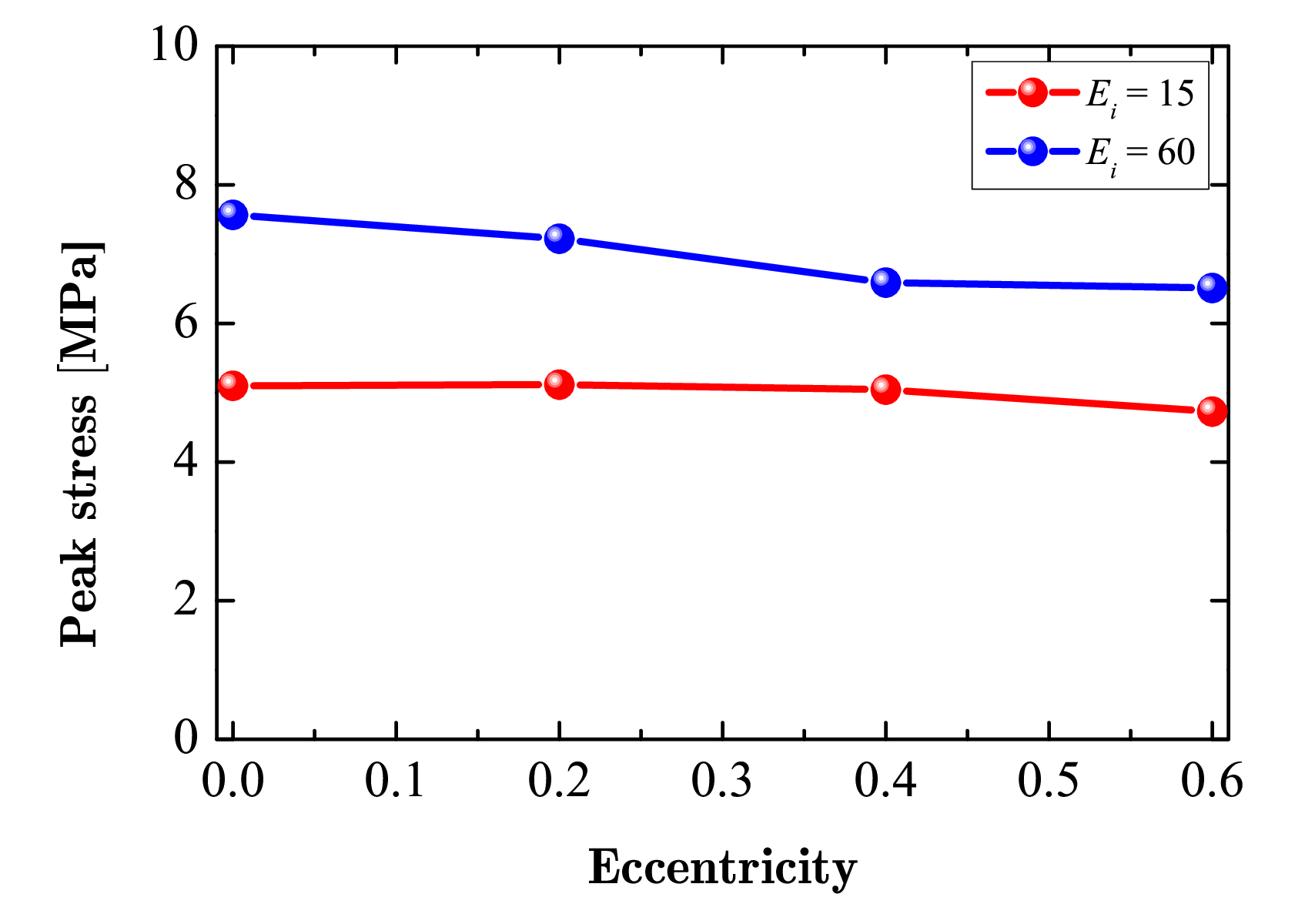}
	\caption{Peak stresses of the Brazilian discs with an inclusion along the vertical axis}
	\label{Peak stresses of the Brazilian discs with an inclusion along the vertical axis}
	\end{figure}

Figure \ref{Final fracture patterns of the Brazilian discs with an inclusion along the vertical axis} shows the final fracture patterns of Brazilian discs with an inclusion along the vertical axis by using the phase field model. The diameter of the inclusion is also 20 mm. The stiffness of the inclusion are $E_i$ = 15 GPa and 60 GPa, respectively. With the varying eccentricity in the vertical direction, the final fractures become more complex. Except the dominant fractures along the vertical axis, some secondary fractures occur around the loading ends. Figure \ref{Stress-strain curves of the Brazilian discs with an inclusion along the vertical axis} shows the stress-strain curves of the Brazilian discs with an inclusion along the vertical axis for different eccentricities. For both softer and stiffer inclusions, the overall strength of the disc decreases with an increasing eccentricity along the vertical axis. Additionally, the overall stiffness of the disc decreases with the increase in the eccentricity $e$. Figure \ref{Peak stresses of the Brazilian discs with an inclusion along the vertical axis} shows the peak stresses of the Brazilian discs with an inclusion along the vertical axis. For $E_i$ = 15 GPa and 60 GPa, the peak stress decreases with the increase in the eccentricity $e$. The decreasing rate of $E_i$ = 60 GPa is larger than that of $E_i$ = 15 GPa. 

\begin{figure}[htbp]
	\centering
	\subfigure[PFM]{\includegraphics[height = 5cm]{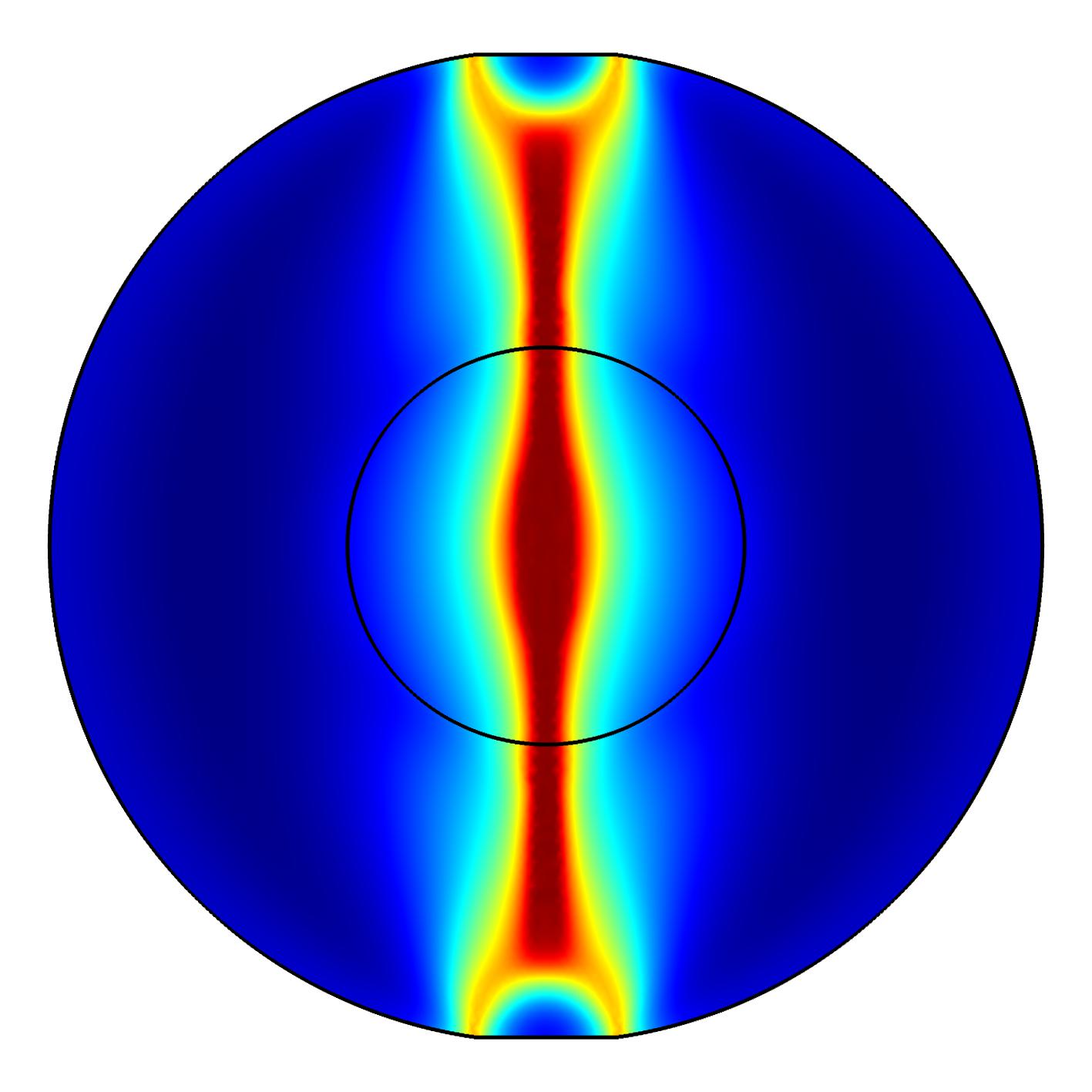}}
	\subfigure[Experimental test \citep{chang2018mechanical}]{\includegraphics[height = 5cm]{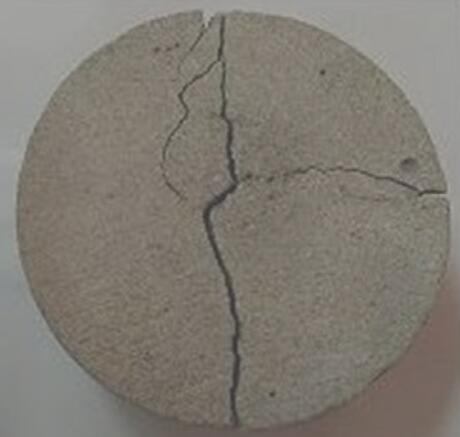}}
	\subfigure[RFPA simulation \citep{chang2018mechanical}]{\includegraphics[height = 5cm]{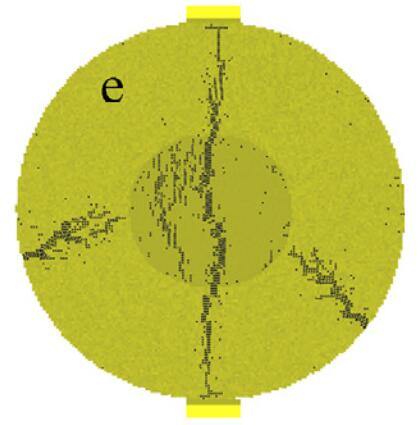}} 
	\caption{Comparison of the fracture patterns in the Brazilian disc with a softer inclusion ($D=20$ mm and $e=0$)}
	\label{Comparison of the fracture patterns in the Brazilian disc with a softer inclusion}
	\end{figure}

\begin{figure}[htbp]
	\centering
	\subfigure[PFM]{\includegraphics[height = 5cm]{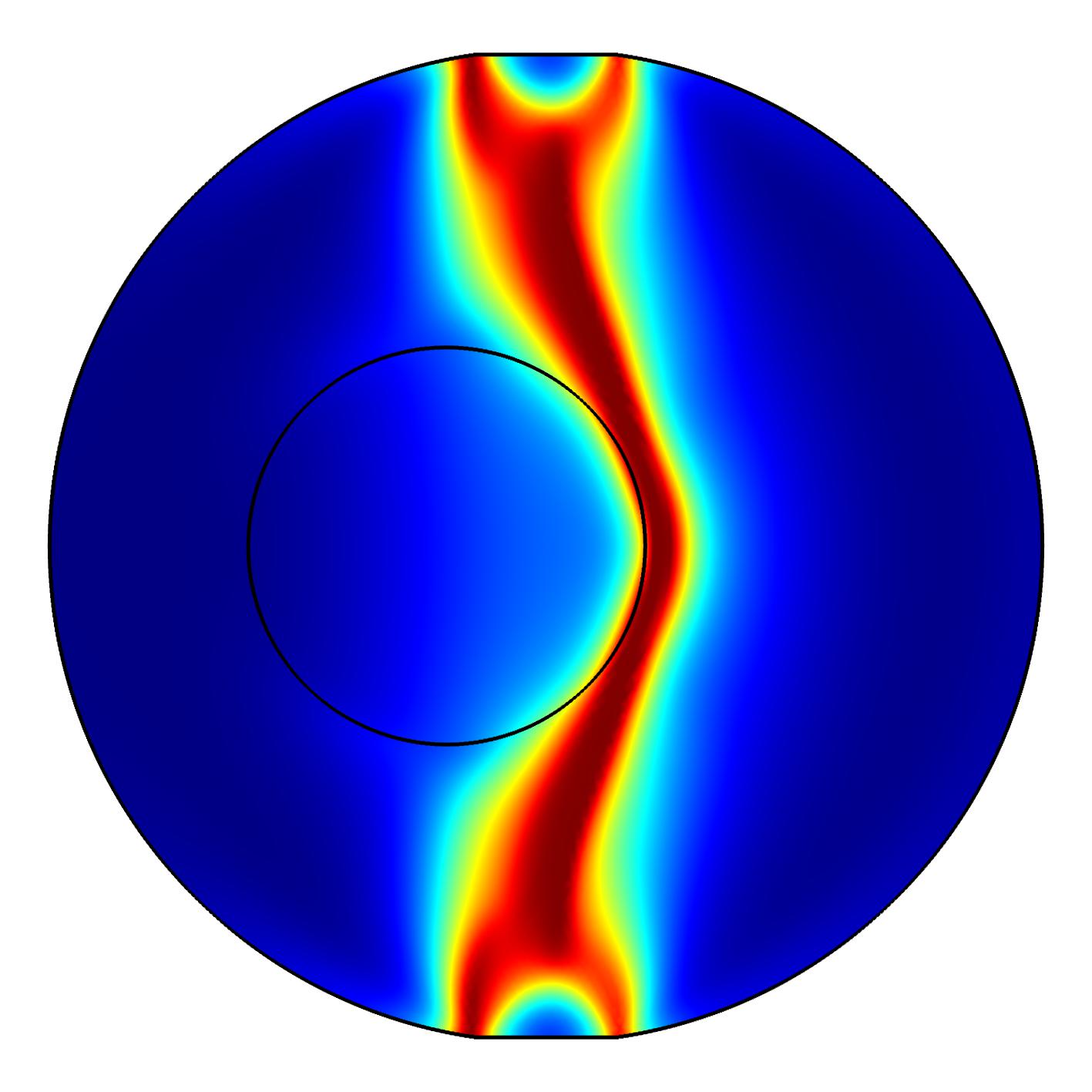}}
	\subfigure[Experimental test \citep{chang2018mechanical}]{\includegraphics[height = 5cm]{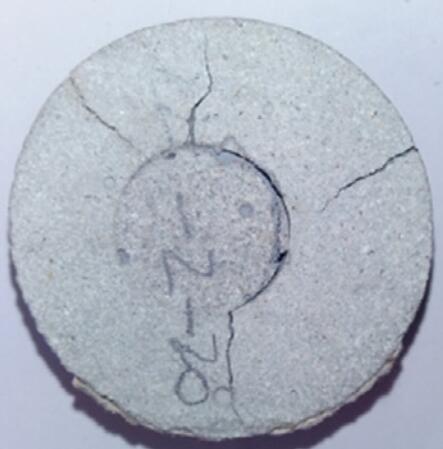}}
	\subfigure[RFPA simulation \citep{chang2018mechanical}]{\includegraphics[height = 5cm]{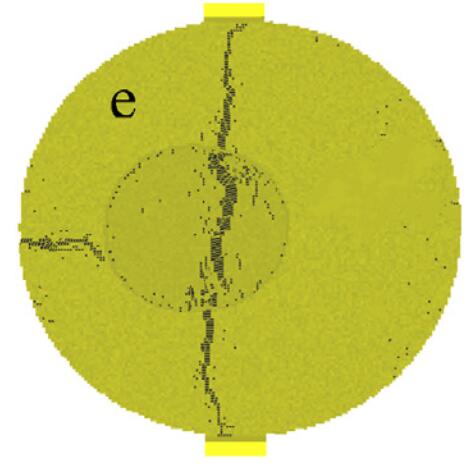}} 
	\caption{Comparison of the fracture patterns in the Brazilian disc with a stiffer inclusion ($D=20$ mm and $e=0.2$)}
	\label{Comparison of the fracture patterns in the Brazilian disc with a stiffer inclusion}
	\end{figure}

In order to investigate the mechanical property of Brazilian discs containing different inclusions, \citet{chang2018mechanical} carried out experimental tests and used RFPA to model fracture patterns. Three cases were considered: $E_i<E_m$, $E_i=E_m$, and $E_i>E_m$ with $E_m$ the Young's modulus of the disc matrix. By comparing \citet{chang2018mechanical} and our simulations, it is found that the phase field prediction of the fractures is in good agreement with the results by using RFPA and experimental tests in most circumstances. Some evidence can be seen in Figs. \ref{Comparison of the fracture patterns in the Brazilian disc with a softer inclusion} and \ref{Comparison of the fracture patterns in the Brazilian disc with a stiffer inclusion} with a little difference between the phase field modeling and RFPA in the case of $E_i>E_m$ (a stiffer inclusion). This difference is mainly because the RFPA is not based on a real approach to fracture and the Young's modulus is assumed to obey the Weibull distribution in the specimen. In the sense of 'fracture', the results obtained by the phase field model seem to be more persuasive because of its strong physical meaning for fracture problems, and the comparisons in Figs. \ref{Comparison of the fracture patterns in the Brazilian disc with a softer inclusion} and \ref{Comparison of the fracture patterns in the Brazilian disc with a stiffer inclusion} also indicate that the phase field simulations are much closer to those experimental observations.

\subsection{Effects of multiple inclusions on the fracture patterns}

\begin{figure}[htbp]
	\centering
	\includegraphics[height = 5cm]{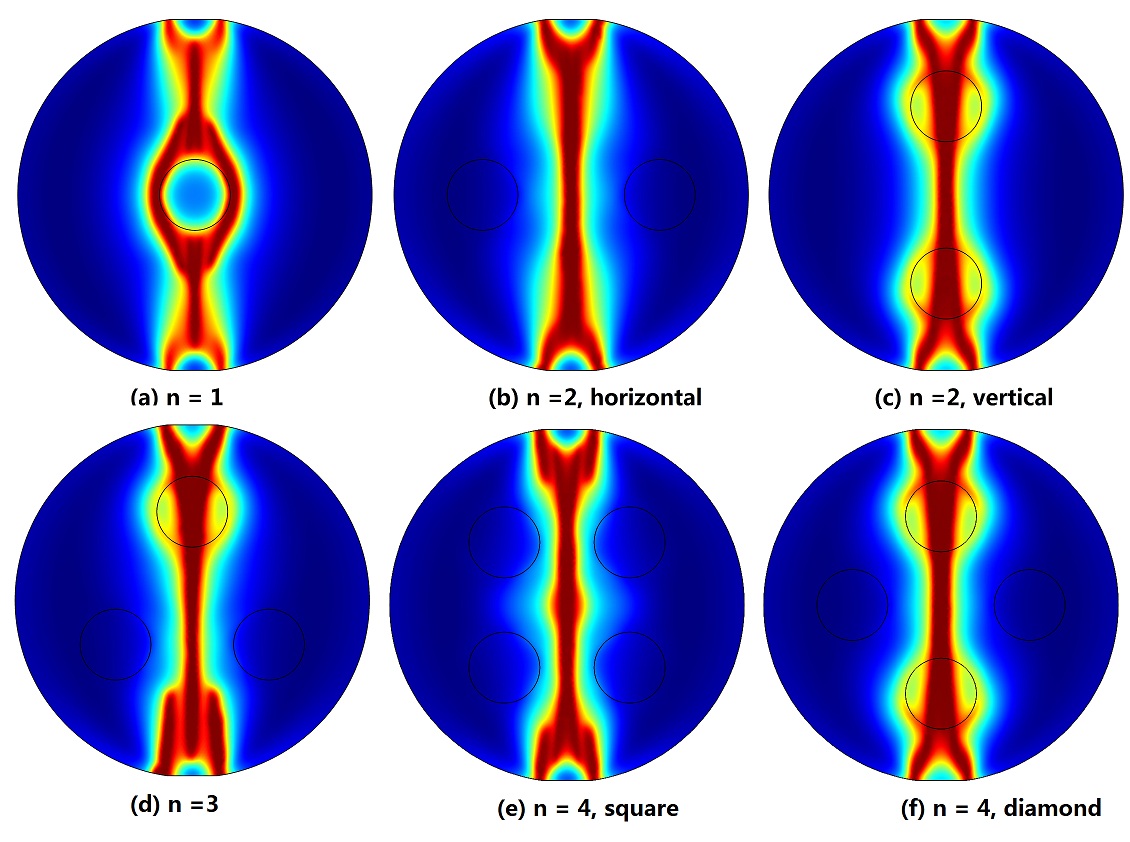}
	\caption{Final fracture patterns of the Brazilian discs with multiple inclusions}
	\label{Final fracture patterns of the Brazilian discs with multiple inclusions}
\end{figure}

\begin{figure}[htbp]
	\centering
	\includegraphics[height = 5cm]{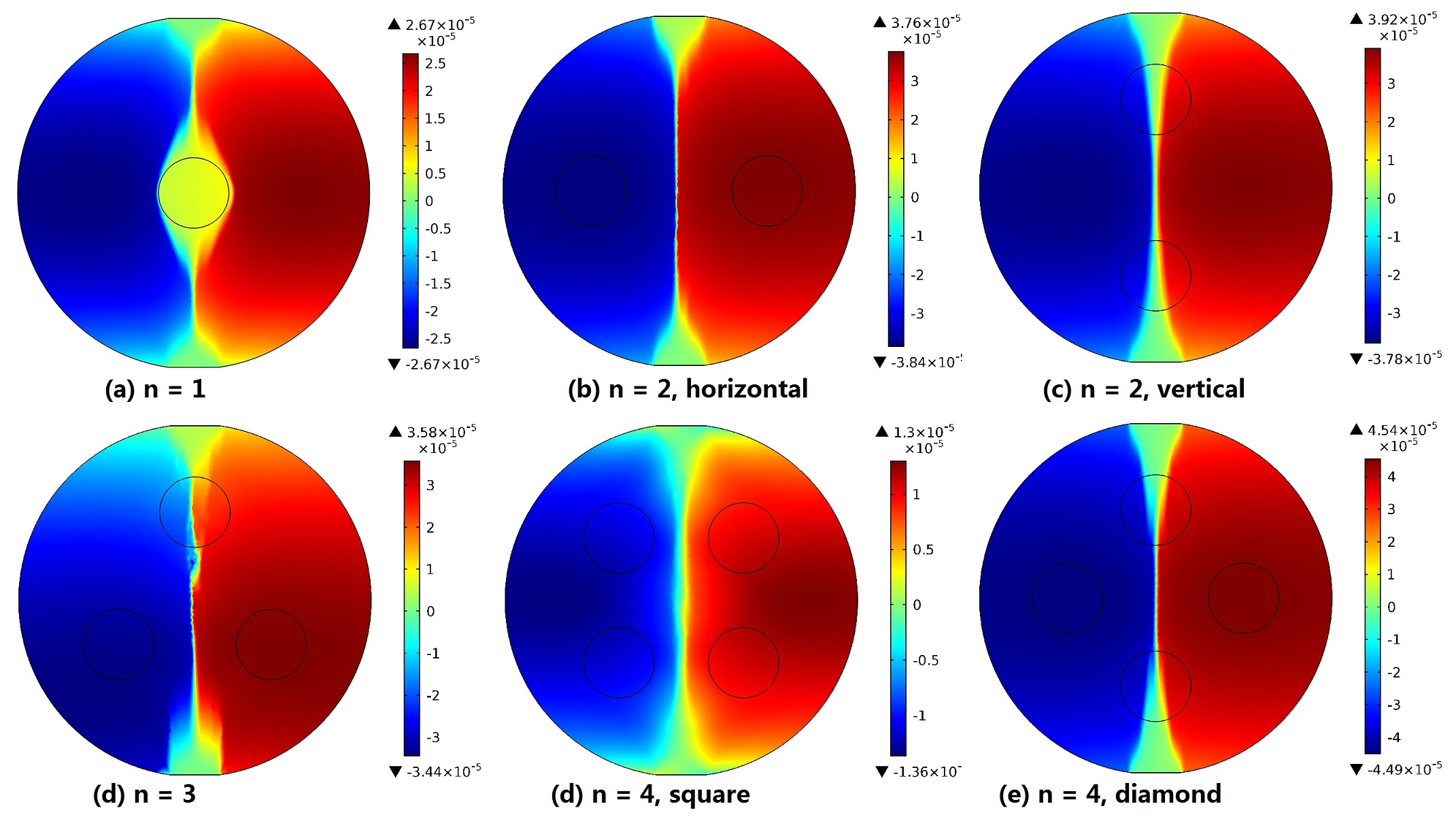}
	\caption{Horizontal displacement of the Brazilian discs with multiple inclusions}
	\label{Horizontal displacement of the Brazilian discs with multiple inclusions}
\end{figure}

This subsection examines the effect of inclusion number on the fracture patterns in the Brazilian disc, and the diameter of the inclusion is fixed to 10 mm. Figure \ref{Final fracture patterns of the Brazilian discs with multiple inclusions} shows the final fracture patterns with different number of inclusions in the Brazilian disc. In most cases, the fracture initiates at center of the disk and propagates through the inclusions. However, for one single inclusion, the fracture propagates along the inclusion boundary. In addition, Fig. \ref{Horizontal displacement of the Brazilian discs with multiple inclusions} shows the horizontal displacement field under different number of inclusions. It is indicated that the horizontal displacement is symmetrical without being affected by the inclusion number.

\section {Conclusions}\label{Conclusions}

A phase field model (PFM) is used to investigate fracture mechanisms in Brazilian discs with circular voids and inclusions. The phase field method describes the fracture behavior in a diffusive manner without any external fracture criteria and we implement the PFM within the framework of the finite element method. The influence of diameter, eccentricity, and quantity of the voids and inclusions on the fracture patterns and stress-strain curves of the discs are subsequently examined. The prediction by using PFM is in good agreement with previous experimental and numerical observations.

The phase field simulations show that the increase in the void diameter reduces the overall strength and stiffness of the Brazilian discs owing to the decrease of load-bearing area. For the disc with a void, the fracture initiates at the top and bottom of the void and propagates towards the two disc ends. If the void number increases, the fracture patterns become more complex with dominant and secondary fractures being observed because of the complex distribution of elastic energy. The strength and stiffness of the discs increase with an increasing eccentricity along the horizontal axis and a decreasing eccentricity along the vertical direction. A softer inclusion has a similar influence with a void on the fracture mechanisms. However, for the discs with a stiffer inclusion, the increase in the inclusion diameter enhances the overall strength and stiffness of the Brazilian discs. The reason is that under the same stress, a larger diameter of stiffer inclusion will have a smaller strain and further a smaller elastic energy, which reduces the possibility of fracture initiation and propagation. The overall strength of the discs increases due to the increase in the eccentricity along the horizontal and vertical directions.

% In the simulations the overall strength and stiffness of the Brazilian discs decrease with an increasing void diameter. For a disc with a void, the fracture initiates at the top and bottom of the void and propagates towards the two disc ends. More complex fractures including the dominant and secondary fractures are observed if the void number increases. For a disc with single inclusion, the peak stress of the disc decreases as the eccentricity along the vertical axis increases. When the eccentricity along the horizontal direction increases, the peak stress of the disc increases for softer inclusion while stiffer inclusion induces an opposite trend.
\section*{Data Availability Statement}
Some or all data, models, or code generated or used during the study are available from the corresponding author by request.

\section*{Acknowledgment}
The authors gratefully acknowledge financial support provided by Deutsche Forschungsgemeinschaft (DFG ZH 459/3-1), and RISE-project BESTOFRAC (734370).
\bibliography{references}
\end{document}